\documentclass[11pt]{article}
\pdfoutput = 1

\usepackage[utf8]{inputenc}
\usepackage{color,graphicx}
\usepackage{epsfig}
\usepackage{amsmath}
\usepackage{verbatim}
\usepackage{amssymb}
\usepackage{mathabx}
\usepackage[mathcal]{eucal}
\usepackage{stmaryrd}
\usepackage{empheq}
\usepackage{esint}
\usepackage{physics}
\usepackage{amsfonts}
\usepackage{cite}
\usepackage{array}
\usepackage{setspace}
\usepackage{braket}
\usepackage{float}
\usepackage{url}
\usepackage{mathtools}
\usepackage{tensor}
\usepackage{bbold}
\usepackage{slashed}
\usepackage{graphicx}
\usepackage{epstopdf}
\usepackage{subcaption}
\usepackage[labelfont=bf,font={sf}]{caption}
\usepackage{mathrsfs}
\usepackage{fancybox}
\usepackage{eurosym}
\usepackage{tensor}
\usepackage[normalem]{ulem}
\allowdisplaybreaks

\usepackage{changepage}

\newcommand{\rev}[1]{#1}

\usepackage[margin = 2.2cm]{geometry}
\usepackage[ragged]{footmisc}
\usepackage[bookmarks=true,bookmarksnumbered=false,
hyperindex=true,bookmarksopen=true,hyperfigures=true,
colorlinks=true,linkcolor=webblue,citecolor=webgreen,urlcolor=webblue,breaklinks]{hyperref}
\definecolor{webgreen}{rgb}{0, 0.5, 0}
\definecolor{webblue}{rgb}{0, 0, 0.5}
\definecolor{webred}{rgb}{0.5, 0, 0}
\definecolor{darkgreen}{rgb}{0,0.5,0}

\def\ben{\begin{equation}}
\def\een{\end{equation}}

     \let\r=v

\def\be{\begin{equation}}
\def\ee{\end{equation}}
\def\ba{\begin{array}}
\def\ea{\end{array}}

\def\dalemb#1#2{{\vbox{\hrule height .#2pt
       \hbox{\vrule width.#2pt height#1pt \kern#1pt
               \vrule width.#2pt}
       \hrule height.#2pt}}}
\def\square{\mathord{\dalemb{6.8}{7}\hbox{\hskip1pt}}}
\newcommand{\bea}{\begin{eqnarray}}
\newcommand{\eea}{\end{eqnarray}}

\let\tilde=\widetilde

\numberwithin{equation}{section}

\begin{document}

\begin{titlepage}
\vspace{1.2cm}
\begin{center}
{\LARGE\bfseries Liouville gravity as a magnetized particle\\[3pt] on the quantum hyperbolic disk\par}
\vspace{1.2cm}
{\large Jacopo Papalini}\\[10pt]
{\small\itshape Department of Physics and Astronomy, Ghent University,\\ Krijgslaan 281-S9, 9000 Gent, Belgium}\\[6pt]
{\small {jacopo.papalini@ugent.be}}
\end{center}
\vspace{1cm}
\begin{abstract}
\noindent We work out a precise duality between Liouville gravity, in its formulation as sinh dilaton gravity with a finite Dirichlet boundary, and \rev{the rotationally symmetric sector of} a charged particle on the quantum hyperbolic disk, in the presence of a magnetic field set by the boundary value of the dilaton. We identify the quantum hyperbolic disk with the quantization of the classical AdS$_2$ saddle of Liouville gravity, the Weyl frame measured by one of the two Liouville fields, and we show that the bulk Hamiltonian reproduces exactly the magnetized quantum-disk Laplacian. The non-commutativity of the disk coordinates is the canonical commutator of the gravitational phase space. Based on this duality, Liouville gravity correlators are rephrased as gravitationally dressed Witten diagrams on the quantum hyperbolic disk, in which bulk matter lines interact with the magnetized boundary particle. We evaluate the matter two-point function at any value of the magnetic field: at fixed Mellin momentum of the matter line, the matrix element is given by a $b$-deformed $6j$ symbol. In the strong-field limit, loops localize close to the boundary and the results reduce to the known Liouville gravity and, upon analytic continuation of the central charge, double-scaled SYK correlators. Away from the strong-field regime, the expressions receive additional non-perturbative contributions, which we characterize in the JT limit as instanton bounces using the method of exact WKB. In the Liouville CFT formulation, we propose to interpret them as the insertion of non-trivial degenerate bulk vertex operators that modify the smooth no-boundary state in the closed channel.
\end{abstract}
\end{titlepage}

\pagebreak
\setcounter{page}{1}
\setcounter{tocdepth}{2}
\tableofcontents
\section{Introduction}
\label{sec:intro}

\begin{figure}[ht]
\centering
\includegraphics{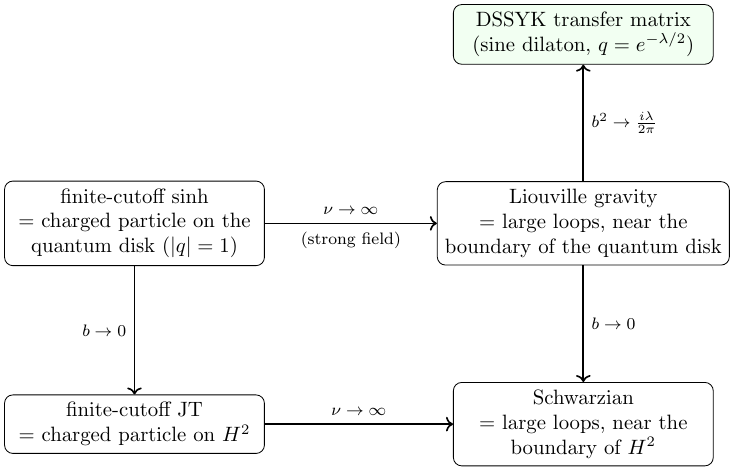}
\caption{The web of limits. Horizontally, the strong-field limit $\nu\to\infty$ removes the cutoff, and the classical loops become large and hug the boundary. Vertically, the JT limit $b\to0$ turns dilations into derivatives and the quantum disk into the hyperbolic plane. At the top, the continuation to sine dilaton gravity turns the Liouville gravity equation into the DSSYK transfer matrix.}
\label{fig:qd_web}
\end{figure}

\rev{Two-dimensional dilaton gravities are the simplest settings in which the gravitational path integral can be computed exactly, and in which one can follow how a bulk geometry emerges from a boundary quantum system. The prototype is Jackiw--Teitelboim (JT) gravity~\cite{teitelboim1983gravitation,jackiw1985lower}, in which the dilaton fixes the curvature to be constant and negative, so that the only dynamical degree of freedom is the shape of the boundary curve. Its dynamics is governed by the Schwarzian action~\cite{Jensen:2016pah,Maldacena:2016upp,Engelsoy:2016xyb}, which also describes the low-energy limit of the Sachdev--Ye--Kitaev (SYK) model~\cite{Sachdev:1992fk,Kitaevtalks,Maldacena:2016hyu,Kitaev:2017awl}, so that JT gravity is the simplest example of a holographic duality between a gravitational theory and a quantum mechanical system. It has been solved exactly with a variety of methods: by localization of the Schwarzian path integral~\cite{Stanford:2017thb}, through Liouville quantum mechanics and the conformal bootstrap~\cite{Bagrets:2016cdf,Mertens:2017mtv,Mertens:2018fds}, by the canonical quantization of its phase space~\cite{henneaux1985quantum,Louis-Martinez:1993bge,Harlow:2018tqv}, in its formulation as a BF theory, where the boundary dynamics becomes that of a particle on the group $SL(2,\mathbb R)$~\cite{Blommaert:2018oro,Blommaert:2018iqz,Iliesiu:2019xuh}, and as a charged particle on the hyperbolic plane in a magnetic field~\cite{Kitaev:2017awl,Kitaev:2018wpr,Yang:2018gdb}. Regarding the latter method, which will be of relevance for this paper, the boundary curve is the worldline of a charged particle on the hyperbolic plane, in a uniform magnetic field set by the boundary value of the dilaton. Boundary correlators then become Witten diagrams in a fixed AdS$_2$, in which gravity only dresses the insertion points with the propagators of the boundary particle~\cite{Yang:2018gdb,Kitaev:2018wpr}. Including all topologies, JT gravity is dual to a random matrix integral~\cite{Saad:2019lba}; see~\cite{Mertens:2022irh} for a review.

Two deformations of JT gravity have recently attracted attention. Sinh dilaton gravity is Liouville gravity, i.e.\ a spacelike Liouville field coupled to a timelike one~\cite{Mertens:2020hbs,Fan:2021bwt}, and its continuation to imaginary $b^2$, sine dilaton gravity~\cite{Blommaert:2024ymv,Blommaert:2024whf,Bossi:2024ffa,Blommaert:2025avl}, is dual to double-scaled SYK (DSSYK)~\cite{Berkooz:2018qkz,Berkooz:2018jqr,Lin:2022rbf,Berkooz:2024lgq}. Both have been studied with two complementary sets of techniques. Their formulation as non-critical string theories, which include the Virasoro minimal string and the complex Liouville string, gives access to the methods of conformal field theory and string theory~\cite{Mertens:2020hbs,Collier:2023cyw,Collier:2024kmo,Collier:2024mlg,Verlinde:2024zrh,Blommaert:2025rgw}. Their exact solutions are instead organized by the quantum group $U_q(\mathfrak{sl}(2,\mathbb R))$, with $q=e^{i\pi b^2}$~\cite{Fan:2021bwt,Mertens:2022aou,Blommaert:2023opb,Blommaert:2023wad,Belaey:2025ijg}: the boundary dynamics is described by the $q$-Schwarzian~\cite{Blommaert:2023wad}, and the exact amplitudes follow from the representation theory of its modular double~\cite{Faddeev:1999fe,Ponsot:2001,Teschner:2012}\rev{, which applies to sinh dilaton gravity, where $q$ is a phase}.

Independently, a non-commutative geometry, the quantum hyperbolic disk~\cite{Shklyarov:1998,Shklyarov:1998b}, whose isometries generate $U_q(\mathfrak{su}(1,1))$, has been proposed as the bulk of DSSYK~\cite{Berkooz:2022mfk,Almheiri:2024ayc}. This symmetry was found in the chord algebra of DSSYK~\cite{Lin:2023trc,Berkooz:2022mfk,vanderHeijden:2025}, where the Hamiltonian was written as that of a particle on a $q$-deformed AdS$_2$~\cite{Berkooz:2022mfk}. On the quantum disk the two coordinates form a $q$-oscillator algebra, which raises and lowers the radial level as the chord operators raise and lower the number of chords~\cite{Almheiri:2024ayc}, and the Laplacian on its line bundles was used to describe a $q$-deformed particle with spin on the hyperbolic plane~\cite{Berkooz:2025classS}. Close to the boundary, the two-point function of sinh dilaton gravity was shown to obey the $q$-Ward identities of the quantum disk~\cite{Belaey:2025}. These two descriptions, a dilaton gravity with a bulk Hamiltonian on one side and a particle on a non-commutative space on the other, have so far been related through their symmetries. What has been missing is a derivation of the particle on the quantum disk from the gravitational theory: which gravitational variables are the coordinates of the disk, where its non-commutativity comes from, and what fixes the magnetic field. The obstacle is that the quantum disk is an intrinsically quantum object. It is defined by an algebra of non-commuting coordinates and by its quantum-group isometries, rather than by a classical action whose quantization would produce it, so that it is not clear which classical geometry it deforms, nor how it could emerge from a gravitational path integral.

}\rev{In this paper we provide this bridge, and with it the classical origin of the quantum disk, for sinh dilaton gravity at finite cutoff, i.e.\ with Dirichlet boundary conditions on the metric and the dilaton at a finite value of the latter. Our starting point is the canonical quantization of~\cite{SinhBox}, which extends that of finite-cutoff JT gravity~\cite{Griguolo:2026eoi}. The reduced phase space is spanned by the length $L$ of the Einstein--Rosen bridge, measured in the Weyl-rescaled metric $g_-=e^{2\pi b^2\Phi}g$, which is locally AdS$_2$, and its conjugate momentum, and the two-sided Hamiltonian becomes a difference operator, with shifts of the length set by $q$. We show that this operator is exactly the Laplacian of the quantum disk, continued to $\abs{q}=1$, acting on the rotationally symmetric sector of a charged particle. Each ingredient of the particle acquires a gravitational meaning:
\begin{itemize}
    \item the radial coordinate of the disk is the length of the Einstein--Rosen bridge, measured from its endpoint, where the two boundaries touch;
    \item the non-commutativity of the disk coordinates is the canonical commutator of the gravitational phase space, with the logarithm of the radial variable conjugate to the angle and the effective Planck constant $2\pi b^2$;
    \item the quantum disk is the quantization of this AdS$_2$ geometry $g_-$, which is the metric measured by one of the two Liouville fields;
    \item the magnetic field is the boundary value of the dilaton in units of the Planck constant, and it is imaginary, $B=i\Phi_b/\hbar$.
\end{itemize}
The finite cutoff is therefore not a technical complication. It makes the field finite, and through the field it controls the size of the loops of the particle. Only when the loops are large, at energies small compared with the cutoff, does the particle stay close to the boundary of the disk, where one recovers Liouville gravity and, after continuation, DSSYK. Small loops probe the full quantum disk, and their dynamics is captured by none of these limits. The resulting web of theories and limits is summarized in Figure~\ref{fig:qd_web}.}

\rev{Already in the JT limit, Section~\ref{sec:jt}, this picture organizes several facts that were so far separate. The boundary particle of~\cite{Kitaev:2018wpr,Yang:2018gdb} has mostly been used at large field, where it reproduces the Schwarzian theory. We show that at finite cutoff the field is finite and imaginary, and that the P\"oschl--Teller Hamiltonian of finite-cutoff JT gravity derived in~\cite{Griguolo:2026eoi} is exactly the rotationally symmetric sector of the magnetic Laplacian of the hyperbolic plane. Its operator ordering, which in~\cite{Griguolo:2026eoi} had to be chosen, is then fixed by the measure of the hyperbolic plane, and the local density of states of the magnetic Laplacian computed in~\cite{Comtet:1986ki} becomes the finite-cutoff density at imaginary field. We also give the two-sided phase space a geometric meaning: the Hartle--Hawking state is the propagator of the particle from the endpoint of the bridge, the length of the bridge is the radius of the disk, and its classical turning point is the diameter of the cyclotron orbit~\cite{Comtet:1986ki,Kitaev:2018wpr} around the horizon. Finally, the particle explains the square root in the Brown--York energy. In the particle description the boundary weight is usually the heat kernel of the Casimir, whereas at finite cutoff the proper length of the boundary is fixed. We show that its conjugate is the momentum of a relativistic particle, which obeys the hyperbolic Larmor relation~\cite{Comtet:1986ki} and gives the $T\bar T$-deformed energy~\cite{Zamolodchikov:2004ce,Smirnov:2016lqw,Cavaglia:2016oda,McGough:2016lol} of finite-cutoff JT gravity~\cite{Gross:2019ach,Aguilar-Gutierrez:2024nst,Chaudhuri:2024yau,Belaey:2026hol,Iliesiu:2020zld,Stanford:2020qhm,Griguolo:2021resurgence,Griguolo:2025kpi,Griguolo:2026eoi}.

Section~\ref{sec:qdisk} contains the main result. We show that, in the absence of the field, the sinh Hamiltonian of~\cite{SinhBox} coincides exactly with the invariant Laplacian of the quantum disk~\cite{Almheiri:2024ayc}, once the radius is measured from the endpoint of the bridge rather than from the centre of the disk, and its spectral parameter is the horizon value of the dilaton. We then show that the field enters as in JT gravity. There the radial operator is a sum of two squares of the Euler operator $\theta=u\partial_u$, with $u=\sinh^2\frac L2$, and the field covariantizes the second one. The sinh operator has exactly the same structure, with each square replaced by its $q$-number, $[X]_q=(q^X-q^{-X})/(q-q^{-1})$,
\begin{equation}\label{eq:intro_squares}
    \theta^2+\frac1u\Big(\theta-\tfrac12-B\Big)^2\quad\longrightarrow\quad[\theta]_q^2+\frac1{q^{2B}u}\Big[\theta-\tfrac12-B\Big]_q^2 .
\end{equation}
We derive this coupling on the quantum disk, without matching it to gravity: we compute the Laplacian of the quantum disk on waves of definite angular momentum, continue the angular momentum to the field and remove the flux at the endpoint of the bridge, a condition that fixes the operator uniquely, up to a sign selected by the JT limit. The result is the little $q$-Jacobi operator with parameters $(0,2B)$, the same as those of the Jacobi operator of JT gravity, and we show that the identification extends to the Hilbert spaces constructed in~\cite{SinhBox}. Continued to real $q$, we find that the eigenfunctions develop poles on the lattice of the disk, with Al-Salam--Chihara polynomials as residues, and the comparison with the line-bundle Laplacian of~\cite{Berkooz:2025classS} suggests the description of sine dilaton gravity at finite cutoff.

In Section~\ref{sec:corr} we use the correspondence to compute correlators, and we show that they become dressed Witten diagrams on the quantum disk: the boundary arcs are propagators of the magnetized particle and the vertices are integrated with the invariant measure. We couple the matter field to the dilaton as was done in sine dilaton gravity~\cite{Blommaert:2024ymv}, so that it propagates on the AdS$_2$ metric $g_-$. On the quantum disk its Green's function is then that of the quantum-disk Laplacian, which we determine exactly. It is a $b$-hypergeometric function~\cite{Nishizawa:1996,Ponsot:2001}, a hyperbolic matrix element of the modular double of $U_q(\mathfrak{sl}(2,\mathbb R))$~\cite{Faddeev:1999fe,SinhBox}, and as such it obeys a second, dual difference equation that, together with the source at the endpoint of the bridge and the decay at the boundary, fixes it uniquely. We then evaluate the two-point function exactly, at any value of the field. We find that, at fixed Mellin momentum of the matter line, the radial integral over the quantum disk is a $b$-$6j$ symbol~\cite{Ponsot:2001,Teschner:2012}, one of whose edges carries the magnetic field, so that the two-point function is a single integral of $b$-$6j$ symbols. In the JT limit it reproduces the two-point function of~\cite{Griguolo:2026eoi}, including the descendants of the primary insertion. In the strong-field limit the loops localize near the boundary, the eigenfunctions become $q$-Bessel functions, and we recover the known two-point functions of Liouville gravity~\cite{Mertens:2020hbs,Fan:2021bwt} and, after continuation, of DSSYK~\cite{Berkooz:2018jqr} as limits of this exact answer. In the semiclassical limit we show that the radial integral localizes on the length of the bridge continued to Euclidean time, and the boundary is a Larmor orbit around the horizon in the AdS$_2$ metric $g_-$, with a deformed Larmor relation.

Away from the strong-field regime the density of states receives corrections that are exponentially suppressed in the boundary dilaton, and that in finite-cutoff JT gravity were associated with Euclidean configurations in which the length of the bridge returns to zero~\cite{Griguolo:2026eoi}. In Section~\ref{sec:nonpert} we derive their complete structure, mostly in the JT limit, where the spectral problem is the P\"oschl--Teller equation. We formulate the density as a connection problem between the endpoint of the bridge and the boundary, and solve it with exact WKB, in which the Borel-resummed WKB series are glued across Stokes curves and the spectrum is encoded in the quantum periods~\cite{Dunham:1932,Voros:1983,Silverstone:1985,Delabaere:1997,Delabaere:1999,Kawai:2005,Zinn-Justin:2004vcw,Gaiotto:2009hg,Iwaki:2014vad,Dunne:2014bca,Ito:2018eon,Sueishi:2020yux}. Exact WKB turns the density into a statement about geometry. We show that it is generated by exactly three cycles, whose quantum periods receive no corrections in $\hbar$, so that the density is one-loop exact, and we identify its sectors as bounces in which the boundary winds several times around the outer or the inner horizon, with a conical excess at the horizon, the analogue of the conical defects of JT gravity~\cite{Mertens:2019tcm,Maxfield:2020ale,Witten:2020wvy,Turiaci:2020fjj,Eberhardt:2023rzz,Kruthoff:2024gxc}. We find that their alternating sign has two independent origins, the reflection phases of the bounces and the negative modes of the wound boundaries, and that it relies on the ordering term, which makes the endpoint of the bridge reflect like a smooth turning point. For a real field, the same periods give the Landau levels of the hyperbolic plane~\cite{Comtet:1986ki} through an exact Bohr--Sommerfeld condition. At finite $b$ each sector becomes the difference of two defects, with angles split by $\pm2\pi b^2$ around the classical excess. In Section~\ref{subsec:zz} we propose an interpretation of these sectors in the Liouville CFT formulation, as degenerate bulk vertex operators that replace the smooth cap of the disk.}

The paper is organized as follows. Section~\ref{sec:sinh} collects the results on finite-cutoff sinh dilaton gravity that we need. Section~\ref{sec:jt} treats the JT limit, Section~\ref{sec:qdisk} establishes the correspondence with the magnetized quantum disk, Section~\ref{sec:corr} computes correlators, Section~\ref{sec:nonpert} analyses the non-perturbative structure and Section~\ref{subsec:zz} its CFT interpretation. We conclude in Section~\ref{sec:discussion}. Technical derivations are collected in the appendices.

\section{Review: canonical quantization of sinh dilaton gravity}
\label{sec:sinh}

In this section we collect the results on the canonical quantization of sinh dilaton gravity at finite cutoff that will be needed in the rest of the paper. We refer to \cite{SinhBox} for the details, and to \cite{Griguolo:2026eoi} for the analogous treatment of JT gravity. We focus on the elements that will acquire a geometric meaning later on: the AdS$_2$ frame in which the length variable is defined, the canonical pair of the reduced phase space, and the difference equation obeyed by the wavefunctions. \rev{Throughout, we work on the disk topology, although formulations of these theories on higher topologies exist, both through their description as non-critical string theories~\cite{Collier:2023cyw,Collier:2024kmo,Collier:2024mlg,Blommaert:2025avl} and, for DSSYK, through its matrix model for eigenstate thermalization and chord diagrams~\cite{Jafferis:2022uhu,Jafferis:2022wez,Okuyama:2023kdo,Okuyama:2023wormhole,Berkooz:2020fvm,Raz:2025ramp}.}

The Lorentzian action of sinh dilaton gravity is
\begin{equation}\label{eq:sinh_action}
    S[g,\Phi]=\frac12\int_{\mathcal M}\dd^2x\sqrt{-g}\,\Big(\Phi R+W(\Phi)\Big)+\int_{\partial\mathcal M}\dd y\sqrt{\abs{h}}\,\bigg(\Phi K-\sqrt{\frac{\cosh(2\pi b^2\Phi)-1}{2\pi^2b^4}}\bigg),
\end{equation}
with the dilaton potential
\begin{equation}\label{eq:sinh_W}
    W(\Phi)=\frac{\sinh(2\pi b^2\Phi)}{\pi b^2}.
\end{equation}
As $b$ goes to zero the potential becomes linear in the dilaton and the action reduces to that of JT gravity; we will refer to this regime as the JT limit. We impose Dirichlet boundary conditions both on the dilaton and on the induced metric at a timelike boundary,
\begin{equation}
    \Phi\big|_{\partial\mathcal M}=\Phi_b,\qquad \dd s^2\big|_{\partial\mathcal M}=-\dd\tau^2 ,
\end{equation}
so that the boundary sits at a finite value $\Phi_b$ of the dilaton. The equations of motion read
\begin{equation}\label{eq:sinh_eom}
    R=-W'(\Phi)=-2\cosh(2\pi b^2\Phi),\qquad \nabla_\mu\nabla_\nu\Phi-g_{\mu\nu}\Box\Phi+\tfrac12g_{\mu\nu}W(\Phi)=0 ,
\end{equation}
and taking the trace of the second one we obtain
\begin{equation}\label{eq:sinh_box}
    \Box\Phi=W(\Phi).
\end{equation}
The static solutions are black holes,
\begin{equation}\label{eq:sinh_bh}
    \dd s^2=-\frac{\cosh(2\pi b^2r)-\cosh(2\pi b^2\Phi_h)}{2\pi^2b^4}\,\dd t^2+\frac{2\pi^2b^4\,\dd r^2}{\cosh(2\pi b^2r)-\cosh(2\pi b^2\Phi_h)},\qquad \Phi=r,
\end{equation}
labelled by the value $\Phi_h$ of the dilaton at the horizon, and cut off at the boundary $\Phi=\Phi_b$. The quasi-local Brown--York energy \rev{of each boundary, which is the same for the two boundaries of the symmetric two-sided geometry,} is\rev{\footnote{\rev{It is $E=F(\Phi_b)-\partial_n\Phi$, with the counterterm $F(\Phi_b)=\sqrt{(\cosh(2\pi b^2\Phi_b)-1)/2\pi^2b^4}$ of \eqref{eq:sinh_action} and $\partial_n\Phi=\sqrt{f(\Phi_b)}$, with $f$ the $\dd t^2$ coefficient of \eqref{eq:sinh_bh}. With this counterterm the energy vanishes at $\Phi_h=0$ and reduces to the energy of finite-cutoff JT gravity \eqref{eq:JTspectral}.}}}
\begin{equation}\label{eq:sinh_quasilocal}
    H(\Phi_h)=\rev{\frac{1}{2\pi b^2}\Big[\sqrt{2\cosh(2\pi b^2\Phi_b)-2}-\sqrt{2\cosh(2\pi b^2\Phi_b)-2\cosh(2\pi b^2\Phi_h)}\,\Big]}.
\end{equation}

The first equation in \eqref{eq:sinh_eom} shows that, unlike in JT gravity, the physical metric is not AdS$_2$: its curvature depends on the dilaton. However, it becomes AdS$_2$ after a Weyl rescaling by the dilaton itself. To see this, recall that under a Weyl rescaling $\tilde g=e^{2\sigma}g$ the curvature transforms as
\begin{equation}
    \tilde R=e^{-2\sigma}\big(R-2\Box\sigma\big).
\end{equation}
Choosing $\sigma=\mp\pi b^2\Phi$ and using \eqref{eq:sinh_eom} and \eqref{eq:sinh_box}, we find
\begin{equation}
    R-2\Box\sigma=-2\cosh(2\pi b^2\Phi)\pm2\sinh(2\pi b^2\Phi)=-2\,e^{\mp2\pi b^2\Phi}=-2\,e^{2\sigma},
\end{equation}
so that both rescaled metrics
\begin{equation}\label{eq:sinh_weyl}
    g_\pm\equiv e^{\mp2\pi b^2\Phi}\,g
\end{equation}
have constant curvature $\tilde R=-2$ on shell. If we further require that the asymptotic boundary of the physical spacetime be mapped to the asymptotic boundary of AdS$_2$, we select $g_-=e^{2\pi b^2\Phi}g$~\cite{SinhBox,Blommaert:2024ymv}. In this frame the black hole \eqref{eq:sinh_bh} becomes the AdS$_2$ black hole
\begin{equation}\label{eq:mappings}
    \dd s^2_{\rm AdS}=-(\rho^2-\rho_h^2)\,\dd t^2+\frac{\dd\rho^2}{\rho^2-\rho_h^2},\qquad \rho_h=\frac{\sinh(2\pi b^2\Phi_h)}{2\pi b^2},\qquad \rho_b=\frac{e^{2\pi b^2\Phi_b}-\cosh(2\pi b^2\Phi_h)}{2\pi b^2}.
\end{equation}
As $b$ goes to zero at fixed dilaton the Weyl factor tends to one, and we recover the JT statement that the metric is AdS$_2$.

This Weyl frame is the geometry measured by one of the two Liouville fields of the Liouville formulation of the theory, as we will see in Section~\ref{subsec:qd_meaning}, and it is the space that becomes the quantum hyperbolic disk upon quantization.

The auxiliary AdS$_2$ is also where the length variable of the reduced phase space is defined. Following \cite{Harlow:2018tqv,SinhBox}, one foliates the two-sided geometry with the images of the boundary-to-boundary geodesics of the AdS$_2$ frame, and takes as first canonical variable their length $L$, measured in the AdS$_2$ metric. \rev{On the black hole \eqref{eq:mappings}, the geodesic joining the two boundaries at times $t_L$ and $t_R$ has length
\begin{equation}\label{eq:sinh_ERB}
    \sinh\frac L2=\sqrt{\frac{\rho_b^2}{\rho_h^2}-1}\;\cosh\frac{\rho_h(t_L+t_R)}{2} .
\end{equation}} The conjugate momentum $P$ follows from Hamilton's equations, and in terms of these variables the conserved quantity $\cosh(2\pi b^2\Phi_h)$ reads~\cite{SinhBox}
\begin{equation}\label{eq:sinh_cosh_hamiltonian}
    \cosh(2\pi b^2\Phi_h)=\sqrt{1+\frac{e^{4\pi b^2\Phi_b}}{\sinh^2(L/2)}}\;\frac{\cosh\!\big(2\pi b^2P\tanh(L/2)\big)}{\tanh(L/2)}-\frac{e^{2\pi b^2\Phi_b}}{\sinh^2(L/2)} .
\end{equation}
The energy is then obtained from \eqref{eq:sinh_quasilocal}. Two limits provide useful checks of this expression. In the JT limit it reduces to the classical P\"oschl--Teller Hamiltonian of finite-cutoff JT gravity~\cite{Griguolo:2026eoi},
\begin{equation}\label{eq:sinh_PTclassical}
    \Phi_h^2=P^2+\frac{\Phi_b^2}{\cosh^2(L/2)} ,
\end{equation}
while removing the cutoff at fixed renormalized length one recovers the Hamiltonian of Liouville gravity~\cite{Mertens:2020hbs},
\begin{equation}
    \cosh(2\pi b^2\Phi_h)\simeq\sqrt{1+(2\pi b^2)^2e^{-L_{\rm ren}}}\,\cosh(2\pi b^2P).
\end{equation}

Upon quantization the canonical variables obey $[\hat L,\hat P]=i\hbar$. It is convenient to trade them for the pair
\begin{equation}\label{eq:sinh_LambdaPi}
    \hat\Lambda=2\log\sinh\frac{\hat L}2,\qquad \hat\Pi=\frac12\Big(\hat P\tanh\frac{\hat L}2+\tanh\frac{\hat L}2\,\hat P\Big),
\end{equation}
which is again canonical, $[\hat\Lambda,\hat\Pi]=i\hbar$. The key feature of \eqref{eq:sinh_cosh_hamiltonian} is that the momentum appears only through hyperbolic functions, so that after quantization it acts through exponentials of $\hat\Pi$. These are not derivatives but finite translations of $\Lambda$ in the imaginary direction. It is convenient to rescale $\Phi_h\to\hbar s$, $\Phi_b\to\hbar\nu$ and $b^2\to b^2/\hbar$. In these units the translations have size $2\pi b^2$, and the pair formed by $\Lambda$ and the rescaled momentum $p\equiv2\pi b^2\Pi$ satisfies
\begin{equation}\label{eq:sinh_hbar}
    [\hat\Lambda,\hat p]=i\hbar_{\rm eff},\qquad \hbar_{\rm eff}=2\pi b^2 .
\end{equation}
This effective Planck constant will reappear in Section \ref{sec:qdisk} as the parameter that controls the non-commutativity of the quantum disk.

With a symmetric operator ordering, chosen so as to reproduce the ordering of JT gravity in the limit $b\to0$, the eigenvalue problem for $\cosh(2\pi b^2\hat\Phi_h)$ becomes a second-order difference equation in $\Lambda$~\cite{SinhBox}. It takes its simplest form after a similarity transformation by a ratio of Faddeev quantum dilogarithms,
\begin{equation}\label{eq:sinh_Phinu}
    \psi_s(\Lambda)=\Phi_\nu(\Lambda)\,\chi_s(\Lambda),\qquad \Phi_\nu(\Lambda)=\bigg[\frac{\phi_b\big(-\frac{\Lambda}{2\pi b}\big)}{\phi_b\big(\frac{-\Lambda+4\pi b^2\nu}{2\pi b}\big)}\bigg]^{1/2}.
\end{equation}
The function $\Phi_\nu$ has unit modulus on the real line, and in the JT limit it reduces to
\begin{equation}
    \Phi_\nu(\Lambda)\to\big(1+e^{-\Lambda}\big)^{i\nu}.
\end{equation}
In terms of $\chi_s$ the equation reads~\cite{SinhBox}
\begin{equation}\label{eq:sinh_difference}
\begin{aligned}
    \cosh(2\pi b^2s)\,\chi_s(\Lambda)={}&\tfrac12\big(1+e^{-\Lambda+i\pi b^2}\big)\chi_s(\Lambda-2\pi ib^2)+\tfrac12\big(1+e^{4\pi b^2\nu}e^{-\Lambda-i\pi b^2}\big)\chi_s(\Lambda+2\pi ib^2)\\
    &-e^{2\pi b^2\nu}e^{-\Lambda}\chi_s(\Lambda).
\end{aligned}
\end{equation}
Its coefficients are polynomial in $e^{-\Lambda}$, and this equation will be the starting point of Section~\ref{sec:qdisk}.

Equation \eqref{eq:sinh_difference} can be solved by a Mellin transform, which turns it into a first-order recursion solved by double sine functions $S_b$ (see Appendix~\ref{app:doublesine}). The solution regular at $L=0$ is~\cite{SinhBox}
\begin{equation}\label{eq:sinh_eigenfunction}
    \chi_s(\Lambda)=N_s\int_{c+i\mathbb R}\frac{\dd t}{2\pi i}\;e^{(\Lambda-2\pi b^2\nu)t}\;\frac{S_b(bt\pm ibs)\,S_b(\frac b2+ib\nu-bt)}{S_b(\frac b2-ib\nu+bt)},\qquad N_s=\frac{2\pi b^2}{S_b(\frac b2+ib\nu\pm ibs)},
\end{equation}
where $0<c<\frac12$ and we use the shorthand $S_b(x\pm y)=S_b(x+y)S_b(x-y)$. \rev{In terms of the $b$-hypergeometric function~\cite{Nishizawa:1996,Ponsot:1999uf,Ponsot:2001,Ip,SinhBox}
\begin{equation}\label{eq:sinh_Fb}
    F_b(\alpha,\beta,\gamma;z)=\frac{2\pi b\,S_b(\gamma)}{S_b(\alpha)S_b(\beta)}\int\frac{\dd t}{2\pi i}\,(-z)^t\,e^{i\pi bt(\alpha+\beta-\gamma)}\,\frac{S_b(\alpha+bt)\,S_b(\beta+bt)\,S_b(-bt)}{S_b(\gamma+bt)},
\end{equation}
which tends to ${}_2F_1(A,B;C;z)$ for $(\alpha,\beta,\gamma)=b(A,B,C)$ and $b\to0$, the eigenfunction \eqref{eq:sinh_eigenfunction} reads
\begin{equation}\label{eq:sinh_chiFb}
    \chi_s(\Lambda)=e^{(\Lambda-2\pi b^2\nu)(\frac12+i\nu)}\,F_b\Big(\frac b2+ib\nu+ibs,\frac b2+ib\nu-ibs,b;-e^{\Lambda}\Big).
\end{equation}
With this normalization $\chi_s(\Lambda)\simeq e^{(\Lambda-2\pi b^2\nu)(\frac12+i\nu)}$ as $\Lambda\to-\infty$, independently of the energy.} In the JT limit it reduces to the hypergeometric wavefunction of finite-cutoff JT gravity~\cite{Griguolo:2026eoi},
\begin{equation}
    \psi_s\propto\sinh\frac L2\,\cosh^{2i\nu}\frac L2\;{}_2F_1\Big(\tfrac12+i\nu+is,\tfrac12+i\nu-is;1;-\sinh^2\frac L2\Big).
\end{equation}
At large $\Lambda$ the eigenfunctions behave as plane waves, and their normalization fixes the density of states~\cite{SinhBox},
\begin{equation}\label{eq:sinh_density}
    \rho_b(s)=\frac{1}{\pi b^2}\,\frac{\sinh(2\pi b^2s)\,\sinh(2\pi s)}{\cosh(2\pi\nu)+\cosh(2\pi s)} .
\end{equation}
In the rescaled units the energy \eqref{eq:sinh_quasilocal} becomes
\begin{equation}\label{eq:sinh_energy}
    E_b(s)=\rev{\frac{1}{2\pi b^2}\Big[\sqrt{2\cosh2\pi b^2\nu-2}-\sqrt{2\cosh2\pi b^2\nu-2\cosh2\pi b^2s}\,\Big]},
\end{equation}
and the disk partition function is the transition amplitude between states of vanishing length,
\begin{equation}
    Z_b(\beta)=\bra{L=0}e^{-\beta\hat H}\ket{L=0}\propto\int\rev{_0^\nu}\dd s\;\rho_b(s)\,e^{-\beta E_b(s)}.
\end{equation}

Two limits of these results will be used throughout the paper. In the JT limit the density \rev{\eqref{eq:sinh_density}} and the energy \rev{\eqref{eq:sinh_energy}} become those of finite-cutoff JT gravity~\cite{Griguolo:2026eoi},
\begin{equation}\label{eq:JTspectral}
    \rho(s)=\frac{2s\,\sinh(2\pi s)}{\cosh(2\pi\nu)+\cosh(2\pi s)},\qquad E(s)=\nu-\sqrt{\nu^2-s^2},
\end{equation}
and the difference equation becomes the P\"oschl--Teller equation
\begin{equation}\label{eq:sinh_PT}
    -\psi''(L)+\Big[\nu^2\,\mathrm{sech}^2\frac L2-\frac{1}{4\sinh^2L}\Big]\psi(L)=s^2\psi(L),\qquad L\ge0 .
\end{equation}
In the opposite direction, removing the cutoff at fixed renormalized coordinate
\begin{equation}\label{eq:renorm}
    \Lambda_{\rm ren}=\Lambda-4\pi b^2\nu,
\end{equation}
the difference equation \eqref{eq:sinh_difference} becomes the equation of Liouville gravity,
\begin{equation}\label{eq:sinh_liouville}
    \tfrac12\chi(\Lambda_{\rm ren}-2\pi ib^2)+\tfrac12\big(1+e^{-i\pi b^2}e^{-\Lambda_{\rm ren}}\big)\chi(\Lambda_{\rm ren}+2\pi ib^2)=\cosh(2\pi b^2s)\,\chi(\Lambda_{\rm ren}),
\end{equation}
and, up to a cutoff-dependent factor, the density \rev{\eqref{eq:sinh_density}} becomes the Liouville gravity density~\cite{Mertens:2020hbs,Fan:2021bwt}, proportional to $\sinh(2\pi b^2s)\sinh(2\pi s)$. The two limits commute. \rev{An interesting comment is that the large-cutoff limit fixes the form of the Liouville gravity equation. In the quantum-group derivation of~\cite{Fan:2021bwt}, the eigenfunctions of \eqref{eq:sinh_liouville} arise as Whittaker functions, and the equation comes in a family labelled by an integer $\epsilon$, which determines how the Liouville potential is ordered with respect to the shifts; the amplitudes of Liouville gravity require $\epsilon=\pm1$, a choice that has to be imposed by hand.\footnote{\rev{Here $\epsilon$ is the difference of the twists of the two Whittaker vectors by the Cartan generator, and \eqref{eq:sinh_liouville} is the case $\epsilon=-1$ in the conventions of~\cite{Fan:2021bwt}.}} Starting from the finite-cutoff difference equation \eqref{eq:sinh_difference} and its eigenfunctions \eqref{eq:sinh_eigenfunction}, the limit selects this value automatically, without requiring a twisted diagonalization of the parabolic vectors.}

\section{The JT limit: a charged particle on the hyperbolic plane}
\label{sec:jt}

Before turning to the deformed theory, we show that finite-cutoff JT gravity is exactly the rotationally symmetric sector of a charged particle moving on the hyperbolic plane, in a constant and imaginary magnetic field set by the boundary dilaton. This is the classical structure that will be $q$-deformed in Section \ref{sec:qdisk}, and we will present it in a form adapted to that purpose.

Consider the hyperbolic plane in geodesic polar coordinates around a point $O$,
\begin{equation}\label{eq:mag_polar}
    \dd s^2=\dd L^2+\sinh^2L\,\dd\vartheta^2,\qquad L\ge0,\quad\vartheta\sim\vartheta+2\pi .
\end{equation}
These are the hyperbolic analogue of ordinary polar coordinates in the plane. Walking straight out of $O$ at fixed angle $\vartheta$ is the shortest way to reach a point, so the radial lines are geodesics and $L$ is simply the distance from $O$ measured along them. 

A uniform magnetic field has a field strength proportional to the area form, and in the symmetric gauge, which is regular at $O$, it is described by the potential
\begin{equation}\label{eq:mag_A}
    A=B\,(\cosh L-1)\,\dd\vartheta .
\end{equation}
The Hamiltonian of a particle of unit charge is the magnetic Laplacian,\rev{\footnote{\rev{\label{fn:maglap}It is $\hat H_B=-\frac{1}{\sqrt g}D_\mu\big(\sqrt g\,g^{\mu\nu}D_\nu\big)$, with the covariant derivative $D_\mu=\partial_\mu-iA_\mu$. In the coordinates \eqref{eq:mag_polar}, $\sqrt g=\sinh L$, $g^{\vartheta\vartheta}=1/\sinh^2L$, and since $A_L=0$ one finds $\hat H_B=-\frac{1}{\sinh L}\partial_L\big(\sinh L\,\partial_L\big)-\frac{1}{\sinh^2L}\big(\partial_\vartheta-iA_\vartheta\big)^2$. On $e^{im\vartheta}R(L)$ the angular derivative becomes $im$, which gives \eqref{eq:mag_radial_m}.}}} and on a wavefunction of angular momentum $m$, $\Psi=e^{im\vartheta}R(L)$, it acts as
\begin{equation}\label{eq:mag_radial_m}
    \hat H_B\big|_m=-R''-\coth L\,R'+\frac{\big(m-B(\cosh L-1)\big)^2}{\sinh^2L}\,R .
\end{equation}
In the rotationally symmetric sector the magnetic term reduces to a P\"oschl--Teller potential. Indeed, for $m=0$ we have
\begin{equation}
    \frac{B^2(\cosh L-1)^2}{\sinh^2L}=B^2\tanh^2\frac L2=B^2-B^2\,\mathrm{sech}^2\frac L2 ,
\end{equation}
and after removing the first-derivative term with $R=\psi/\sqrt{\sinh L}$ the eigenvalue equation $\hat H_BR=\mathcal ER$ becomes
\begin{equation}\label{eq:mag_PT}
    -\psi''+\Big[-B^2\,\mathrm{sech}^2\frac L2-\frac{1}{4\sinh^2L}\Big]\psi=\Big(\mathcal E-\frac14-B^2\Big)\psi .
\end{equation}
This is precisely the equation \eqref{eq:sinh_PT} of finite-cutoff JT gravity, including the ordering term, provided we identify
\begin{equation}\label{eq:mag_dictionary}
    B=i\nu=\frac{i\Phi_b}{\hbar},\qquad \mathcal E=\frac14+B^2+s^2 ,
\end{equation}
and we identify the length of the Einstein--Rosen bridge in the two-sided canonical description with the radius of the hyperbolic disk.
It is worth stressing that the ordering term, which in \cite{Griguolo:2026eoi} was a choice, is here forced by the measure of the hyperbolic plane: the JT ordering is the one that turns the problem into the Laplacian of a two-dimensional space. The Hilbert spaces match as well, since the norm of $\psi$ on the half-line equals the norm of $R$ with the measure $\sinh L\,\dd L$, and regularity of $R$ at the centre is exactly the boundary condition $\psi\simeq\sqrt L$ imposed in~\cite{Griguolo:2026eoi}.\footnote{Finally, the combination $\mathcal E-\frac14-B^2=s^2$ is, up to normalization, the $\mathfrak{sl}(2,\mathbb R)$ Casimir in the principal series with $j=\frac12+is$, and the horizon value of the dilaton is $\Phi_h=\hbar s$.}

The identification passes a non-trivial spectral check. For real field, the local density of continuum states of the magnetic Laplacian at $O$ was computed in~\cite{Comtet:1986ki}, and in our normalization it reads
\begin{equation}\label{eq:mag_comtet}
    \rho_B(s)=\frac{2s\,\sinh(2\pi s)}{\cosh(2\pi s)+\cos(2\pi B)} .
\end{equation}
At vanishing field this is the Plancherel density of the hyperbolic plane, while at $B=i\nu$ it coincides exactly with the density \eqref{eq:JTspectral} of finite-cutoff JT gravity. For real $B>\frac12$ the magnetic Laplacian also has bound states, the Landau levels of the hyperbolic plane, located at $s=\pm i(B-n-\frac12)$.  We will meet them again in Section \ref{sec:nonpert}.

\subsection{A geometric picture: why the two systems are equivalent}

The reason why the two problems coincide becomes clear in the Euclidean path integral. Since the dilaton imposes constant curvature, the bulk is a region of the hyperbolic plane bounded by a closed curve $\gamma$, and by the Gauss--Bonnet theorem the boundary term can be rewritten as
\begin{equation}
    \Phi_b\oint_\gamma K=2\pi\Phi_b\,T+\Phi_b\,\mathrm{Area}(\gamma),
\end{equation}
where $T$ is the turning number of the curve. The boundary weight therefore contains the exponential of $\Phi_b/\hbar$ times the enclosed area. This is the Aharonov--Bohm phase of a unit charge in a uniform field, once the field is taken to be imaginary and equal to the boundary dilaton in units of $\hbar$. 
This is indeed the boundary-particle description of JT gravity~\cite{Kitaev:2018wpr,Yang:2018gdb}: after integrating out the dilaton, what remains is a particle on the hyperbolic plane tracing the boundary curve, charged under a field whose strength is the boundary value of the dilaton. The topological term $2\pi\Phi_bT$ will play an important role in Section \ref{sec:nonpert}\rev{: the particle is weighted by the enclosed area alone, so its path integral differs from the gravitational one by the factor $e^{-2\pi\Phi_bT/\hbar}$. For curves with $T=1$ this is an overall normalization, while it weights the curves that wind several times}.

The same identification can be understood from the point of view of symmetry, explaining why the length of the Einstein--Rosen bridge can be identified with a radius in a Euclidean setting. The two-sided Hartle--Hawking state is prepared by a Euclidean path integral over half a disk, and a two-sided configuration is the analytic continuation of a pair of points on the same Euclidean boundary curve, the bridge continuing to the geodesic chord that joins them. The boundary curve is defined up to isometries of the hyperbolic plane, so we can use this freedom to place one of the two points at the centre $O$. The only gauge-invariant information left about the other point is its distance from $O$, which is its radial coordinate $L$. The residual gauge group is the group of rotations around $O$, and invariance under it is the condition that we are in the rotationally symmetric sector, i.e.\ $m=0$.\footnote{In Lorentzian signature the stabilizer of a point is a boost, which is why the canonical analysis finds the boost reduction. A rotation by an imaginary angle is a boost: the magnetic field of the Euclidean one-particle picture and the boost charge of the Lorentzian two-sided picture are the same quantity, the boundary dilaton, seen through a Wick rotation. This connects with the analysis performed in~\cite{Griguolo:2026eoi,SinhBox}.} 

\begin{figure}[t]
\centering
\includegraphics{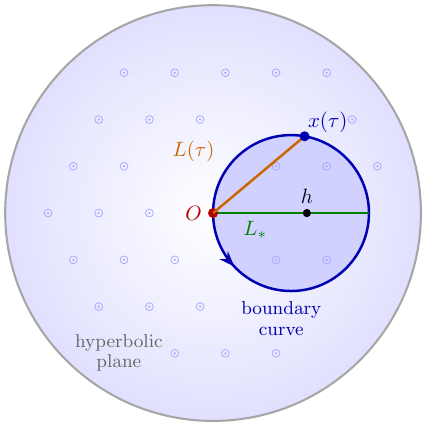}
\caption{Euclidean picture of the length variable in the Poincar\'e disk. The boundary curve is classically a circle around the horizon $h$. Using the isometries, the boundary point $x(0)$ is placed at the centre $O$, and $L(\tau)$ is the geodesic distance from $O$ to $x(\tau)$, i.e.\ the radial coordinate of the hyperbolic plane. Its maximum is the diameter $L_*$ through the horizon, the Einstein--Rosen bridge at $t=0$. \rev{The boundary curve is the trajectory of the charged particle, while the large disk is the whole hyperbolic plane, threaded by the uniform field $B$ ($\odot$); the shaded region is the Euclidean geometry bounded by the curve.}}
\label{fig:mag_disk}
\end{figure}

Every element of the semiclassical analysis now acquires a geometric meaning, as illustrated in Figure \ref{fig:mag_disk}. A vanishing length corresponds to two coincident boundary points, i.e.\ to the closing of the boundary curve, and the partition function is the heat kernel at coincident points. On a smooth classical disk the boundary is a circle around the horizon, and the longest chord from $O$ is the diameter through the horizon,
\begin{equation}\label{eq:mag_Lstar}
    L_*=2\,\mathrm{arccosh}\frac{\Phi_b}{\Phi_h} ,
\end{equation}
which is at the same time the turning point of the P\"oschl--Teller barrier and the length of the bridge at $t=0$\rev{, i.e.\ the JT limit of \eqref{eq:sinh_ERB} at $t_L+t_R=0$}. The classically forbidden region consists of Euclidean chords, and a round trip from $O$ to $L_*$ and back is one revolution of the boundary around the horizon.

We now rewrite the radial operator in the form that will be useful in the $q$-deformed setting. Let us introduce the variable
\begin{equation}\label{eq:jt_u}
    u=\sinh^2\frac L2=e^\Lambda,\qquad \theta=u\,\partial_u=\partial_\Lambda ,
\end{equation}
in terms of which the radial Laplacian of the hyperbolic plane is $\partial_u\,u(1+u)\,\partial_u$. In the symmetric gauge \eqref{eq:mag_A} the angular potential is $A_\vartheta^2/\sinh^2L=B^2u/(1+u)$, and the radial operator \rev{\eqref{eq:mag_radial_m} at $m=0$}, acting on $R(u)$, reads
\begin{equation}\label{eq:qd_LB}
    \mathcal L_B=\partial_u\,u(1+u)\,\partial_u-B^2\frac{u}{1+u} .
\end{equation}
The frame in which the quantization of Section \ref{sec:sinh} is performed is related to this one by
\begin{equation}\label{eq:qd_chiR}
    \chi=g\,R,\qquad g=u^{\frac12}\Big(\frac{u}{1+u}\Big)^{i\nu}.
\end{equation}
The factor $u^{1/2}$ accounts for the change of measure, from the measure $\sinh L\,\dd L$ of the hyperbolic plane to the flat measure in $\Lambda$. The second factor is a phase that depends only on the radius. Such a phase is a radial gauge transformation: it adds to the connection an exact radial component, which does not change the field strength and therefore carries no physics. Explicitly, in the symmetric gauge the connection is purely angular as in \eqref{eq:mag_A}, while the connection in the new gauge associated to \eqref{eq:qd_chiR} is
\begin{equation}\label{eq:jt_radialgauge}
    A=2Bu\;\dd\vartheta+\frac{B}{1+u}\;\dd\Lambda=B(\cosh L-1)\,\dd\vartheta+\frac{B}{\cosh^2\frac L2}\,\dd\Lambda .
\end{equation}
In this frame the operator \rev{\eqref{eq:qd_LB}} takes the remarkably simple form
\begin{equation}\label{eq:qd_gaugeid}
    g\,\mathcal L_B\,g^{-1}+\tfrac14+B^2=\theta^2+\frac1u\Big(\theta-\tfrac12-B\Big)^2 ,
\end{equation}
and the radial equation becomes
\begin{equation}\label{eq:qd_classical2}
    \theta^2\chi+\frac1u\Big(\theta-\tfrac12-B\Big)^2\chi+s^2\chi=0 .
\end{equation}
The operator is a sum of two squares of radial derivatives. The first, which we call the boundary term, dominates as $u\to\infty$; the second, weighted by $1/u$ and which we call the centre term, dominates near $O$. Notice that the magnetic field appears only in the centre term, through the replacement of $\theta$ by $\theta-B$. We stress again that this should not be read as a radial magnetic field: it is the angular potential rewritten in the radial gauge \eqref{eq:jt_radialgauge}. The solution of \eqref{eq:qd_classical2} regular at $O$ is
\begin{equation}\label{eq:solution}
    \chi=u^{\frac12+B}\,{}_2F_1\Big(\tfrac12+B+is,\tfrac12+B-is;1;-u\Big).
\end{equation}

The structure of the equation \rev{\eqref{eq:qd_classical2}} is best summarized in terms of exponents, which characterize the power law behavior of the solution close to its singular points. The equation has three special points: the centre $O$ at $u=0$, the point $L=i\pi$ at $u=-1$ and the boundary at $u=\infty$. A Fuchsian equation of this type is completely determined by the exponent differences at its singular points, which encode gauge-invariant information\footnote{They do not change under rescalings $\chi\to u^{\alpha}(1+u)^{\beta}\chi$.} and have a clear physical interpretation. These are read off directly from \eqref{eq:qd_classical2} and from the explicit solution \eqref{eq:solution} above:
\begin{itemize}
    \item At $O$ ($u=0$) the centre term dominates, and $(\theta-\frac12-B)^2\chi=0$ gives the double root $\rho=\frac12+B$, with solutions $u^{\frac12+B}$ and $u^{\frac12+B}\log u$. The exponent difference is $0$: the wave is rotationally symmetric and the field carries no flux at $O$.
    \item At $L=i\pi$ ($u=-1$) the hypergeometric function has exponents $0$ and $-2B$ in the variable $1+u$, i.e.\ exponents $\pm B$ in the symmetric gauge, where $R\propto(1+u)^{B}\chi$. The exponent difference is $2B$: this is the magnetic field.
    \item At the boundary ($u=\infty$) the boundary term dominates, and $\theta^2\chi=-s^2\chi$ gives $\chi\sim u^{\pm is}$. The exponent difference is $2is$: this is the energy.
\end{itemize}
In the language of exponents, the magnetic field is nothing but the exponent difference at $L=i\pi$. A useful way to see this is to compare with a neutral particle of angular momentum $m$. Its radial operator, \rev{i.e.\ \eqref{eq:mag_radial_m} at $B=0$ written in the variable $u$ (see footnote~\ref{fn:maglap}),}
\begin{equation}\label{eq:qd_Lm}
    \mathcal L_m=\partial_u\,u(1+u)\,\partial_u-\frac{m^2}{4u(1+u)}=\partial_u\,u(1+u)\,\partial_u-\frac{m^2}{4u}+\frac{m^2}{4(1+u)},
\end{equation}
has a centrifugal barrier with poles both at $O$ and at $L=i\pi$, and exponent difference $m$ at each: the angle rotates around both poles of the complexified geometry. For the charged particle, the effective angular momentum is $-A_\vartheta=-B(\cosh L-1)$, which vanishes at $O$ and equals $2B$ at $L=i\pi$. Correspondingly, the angular potential has no pole at $O$, and at $L=i\pi$ it has exactly the pole of a neutral wave with $m=2B$. Comparing \eqref{eq:qd_LB} with \eqref{eq:qd_Lm} we find
\begin{equation}\label{eq:qd_LBLm}
    \mathcal L_B+B^2=\mathcal L_{m=2B}+\frac{B^2}{u}:
\end{equation}
the charged rotationally symmetric wave is the neutral wave of angular momentum $2B$ with its barrier at $O$ removed. In other words, seen from $L=i\pi$ the magnetic field is angular momentum. This simple observation will be the key to the quantum-disk coupling of the field in Section \ref{sec:qdisk}.

\subsection{The \texorpdfstring{$T\bar T$}{TTbar} structure from the relativistic particle}
\label{sec:ttbar}

The classical Hamiltonian of finite-cutoff JT gravity is
\begin{equation}\label{eq:tt_H}
    H=\Phi_b-\sqrt{\Phi_b^2-\Phi_h^2},\qquad \Phi_h^2=P^2+\Phi_b^2\,\mathrm{sech}^2\frac L2,
\end{equation}
where $\Phi_h^2$ is the conserved $SL(2,\mathbb R)$ Casimir \cite{Griguolo:2026eoi}. On the other hand, the particle introduced above has as Hamiltonian the magnetic Laplacian whose eigenvalue is the Casimir itself. The Boltzmann factors of gravity therefore involve a square root of the particle Hamiltonian, which is the $T\bar T$ structure of finite-cutoff JT gravity~\cite{Gross:2019ach,Iliesiu:2020zld}. In this section we show that the square root has a simple origin in the particle picture. The Dirichlet condition fixes the length of the worldline and the quantity conjugate to the length of a worldline is the momentum of a relativistic particle. We will now derive an expression for it in terms of the gravitational quantities and show it indeed encodes the $T\bar{T}$ structure \eqref{eq:tt_H}.

In order to do so, we first recall the classical motion~\rev{\cite{Comtet:1986ki,Kitaev:2018wpr}}. A charged particle on the hyperbolic plane in a constant field moves along a geodesic bent by the Lorentz force. The geodesic equation of a unit mass particle on the hyperboloid, described in embedding coordinates $X\cdot X=-1$, is given by
\begin{equation}\label{eq:tt_eom}
    \ddot X=(\dot X\cdot\dot X)\,X+B\,X\times\dot X,
\end{equation}
where the first term is the force that keeps the particle on the hyperboloid while the second is the Lorentz force. Solving \eqref{eq:tt_eom} one finds that the magnitude $\abs{P}=\abs{\dot X}$ of the momentum is conserved and that for $\abs{P}<B$ the particle moves on a circle, the Larmor orbit, of radius
\begin{equation}\label{eq:tt_larmor}
    \tanh R=\frac{\abs{P}}{B},\qquad\text{i.e.}\qquad \sinh R=\frac{\abs{P}}{\sqrt{B^2-\abs{P}^2}} .
\end{equation}
This is the hyperbolic version of the cyclotron radius $R=\abs{P}/B$, which is the standard flat space relation, recovered from \eqref{eq:tt_larmor} when $R$ is small and the curvature is negligible.

The same orbits follow from a variational principle which is directly the gravitational one. In the Euclidean path integral discussed above, the boundary term weighs the boundary curve by $\Phi_b$ times the enclosed area. Imposing that the boundary has a fixed length $\beta$ with a Lagrange multiplier $\lambda$, the boundary curve $\gamma$ extremizes
\begin{equation}\label{eq:tt_action}
    I[\gamma]=-B\,A[\gamma]+\lambda\big(\ell[\gamma]-\beta\big),\qquad B=\Phi_b ,
\end{equation}
where we can interpret the term $\lambda\ell$ as the action of a relativistic particle of mass $\lambda$. Varying the shape of $\gamma$ balances the tension against the Lorentz force and the extremal curves are circles of area $A=2\pi(\cosh R-1)$ and length $\ell=2\pi\sinh R$. Extremizing the action with respect to the radius $R$ we get:
\begin{equation}\label{eq:tt_extremize}
    \frac{\dd I}{\dd R}=-2\pi B\sinh R+2\pi\lambda\cosh R=0\qquad\Longrightarrow\qquad \lambda=B\tanh R ,
\end{equation}
which is the Larmor relation \eqref{eq:tt_larmor} with $\abs{P}=\lambda$, as expected.

So the momentum of the particle is fixed by the Larmor orbit relation \eqref{eq:tt_larmor}, once the radius of the orbit is known. The latter can be found by matching with the gravitational solution. The boundary is the circle around the horizon of Figure \ref{fig:mag_disk} and its diameter is the Einstein--Rosen bridge at $t=0$. Therefore one has $L_*=2R$ and by \eqref{eq:mag_Lstar} $\cosh R=\Phi_b/\Phi_h$.\footnote{Another perspective on the same identification is found by computing the guiding centre of the magnetic orbit. Using the geodesic equation \eqref{eq:tt_eom}, one can show that the vector $\mu=X\times P+BX$ is conserved: it is the charge of the magnetic translations and $\mu/\abs{\mu}$ is the centre of the orbit. The scalar product $\Phi(x)=-\mu\cdot X(x)=\abs{\mu}\cosh\left(d(x,\mu/\abs{\mu})\right)$~\cite{Kitaev:2018wpr,Yang:2018gdb} measures the distance from the centre of the orbit and corresponds precisely to the classical dilaton solution $\Phi=\Phi_h \cosh(d(x,h))$ which measures the distance from the black hole horizon. One therefore infers $\abs{\mu}=\Phi_h$ and from the norm of $\mu$,  $\mu\cdot\mu=\abs{P}^2-B^2$, recovers precisely the relation \eqref{eq:tt_P}.} From \eqref{eq:tt_larmor}, the momentum of the particle is then given by:
\begin{equation}\label{eq:tt_P}
    \abs{P}=\lambda=\Phi_b\tanh R=\sqrt{\Phi_b^2-\Phi_h^2} .
\end{equation}
We notice this value of $P$ satisfies the constraint as well, since from \eqref{eq:tt_P} one finds $\beta=2\pi\sinh R=2\pi\sqrt{\Phi_b^2-\Phi_h^2}/\Phi_h$, which is indeed the inverse Tolman temperature of the JT black hole. The momentum \eqref{eq:tt_P} also has a direct gravitational meaning: it is the normal derivative of the dilaton at the boundary,\footnote{As one can check using that for the Euclidean black hole in the Schwarzschild gauge $\dd s^2=(r^2-\Phi_h^2)\dd\tau^2+\dd r^2/(r^2-\Phi_h^2)$ with $\Phi=r$, the unit normal is $n=\sqrt{r^2-\Phi_h^2}\,\partial_r$. }
\begin{equation}\label{eq:tt_dn}
    \abs{P}=\partial_n\Phi .
\end{equation}

We are now ready to show \eqref{eq:tt_P} is the quantity that enters the Boltzmann factors and, to see why, we consider the path integral associated with the constrained action \eqref{eq:tt_action}. Including the counterterm, the full boundary action \rev{of a curve with turning number one} is $-\Phi_b\oint(K-1)=-2\pi\Phi_b-\Phi_bA+\Phi_b\beta$. Summing over all boundary curves and imposing their length $\beta$ with the Lagrange multiplier $\lambda$ of \eqref{eq:tt_action}, the partition function schematically reads 
\begin{equation}\label{eq:tt_laplace}
    Z(\beta)=e^{2\pi\Phi_b}\int\dd\lambda\;e^{-\beta(\Phi_b-\lambda)}\,G_\lambda,\qquad G_\lambda=\int\mathcal D\gamma\;e^{-\lambda\ell[\gamma]+\Phi_bA[\gamma]} .
\end{equation}
Here $G_\lambda$ is the propagator of a relativistic particle of mass $\lambda$ in the field $\Phi_b$. On the classical saddle computed above, the multiplier is the momentum of the Larmor orbit, $\lambda=\abs{P}$ in~\eqref{eq:tt_P}, and from \eqref{eq:tt_laplace} the Boltzmann factor is $e^{-\beta E}$ with
\begin{equation}\label{eq:tt_E}
    E=\Phi_b-\abs{P}=\Phi_b-\partial_n\Phi=\Phi_b-\sqrt{\Phi_b^2-\Phi_h^2},
\end{equation}
which is indeed the Brown--York energy \eqref{eq:tt_H}. In short, the boundary length is conjugate to the mass of the relativistic particle, which on the Larmor orbit is the momentum $\abs{P}$. From the boundary particle perspective, this is why the square root appears in the Boltzmann factors.

The quantum statement follows from the same structure. The $T\bar T$ flow of a one-dimensional system maps each energy level according to $E_0\mapsto\frac{1}{4\lambda}(1-\sqrt{1-8\lambda E_0})$ without changing the eigenstates~\cite{Gross:2019ach,Iliesiu:2020zld}, and by the identification \eqref{eq:mag_dictionary} we obtain the operator statement
\begin{equation}\label{eq:mag_ttbar}
    \hat H=\Phi_b-\sqrt{\Phi_b^2-2\Phi_b\,\hat H_0},\qquad \hat H_0=\frac{\hbar^2}{2\Phi_b}\Big(\hat H_B-\tfrac14-B^2\Big)\Big|_{B=i\Phi_b/\hbar} .
\end{equation}
Finite-cutoff JT gravity is therefore the $T\bar T$ deformation of the charged particle at finite field, rather than of Liouville quantum mechanics. The two seeds have the same eigenvalue function, hence the same energies, but they act on different Hilbert spaces, the half-line with a regular origin in one case and the full line of the renormalized length in the other, and they have different spectral measures. The cutoff enters twice, in the $T\bar T$ coupling and in the magnetic field of the seed, and replacing the seed by its large-cutoff limit before deforming loses the non-perturbative corrections discussed in Section~\ref{sec:nonpert}.

Since $\hat H$ is a function of $\hat H_0$, the two operators have the same eigenfunctions and only the eigenvalue is deformed. Throughout the paper we therefore work with the seed, the magnetic Laplacian inside the square root, both in JT and in sinh dilaton gravity: wavefunctions, propagators and densities of states are those of the seed. The square root enters only through the Boltzmann factors, where the eigenvalue of the seed is replaced by its $T\bar T$ image. In sinh dilaton gravity the energy \eqref{eq:sinh_quasilocal} still has the Brown--York form,
\begin{equation}
    E=\rev{\frac{\sqrt{2\cosh(2\pi b^2\Phi_b)-2}}{2\pi b^2}}-\partial_n\Phi,\qquad \partial_n\Phi=\frac{\sqrt{2\cosh(2\pi b^2\Phi_b)-2\cosh(2\pi b^2\Phi_h)}}{2\pi b^2},
\end{equation}
and the square root is again the momentum conjugate to the boundary length. However, the boundary contains other terms besides the flux, and the Larmor relation \eqref{eq:tt_larmor} is deformed, as we will see in Section~\ref{subsec:corr_semiclassical}.

The particle picture also suggests how to treat the range of the spectrum. The particle energy \eqref{eq:tt_E} is real only for $\Phi_h<\Phi_b$, i.e.\ $s<\nu$, and has a square-root branch point at $s=\nu$. In~\cite{Griguolo:2026eoi} two options were considered: projecting the Hilbert space onto the states with $s<\nu$ or extending the theory beyond the horizon, where the cutoff curve becomes spacelike. From the point of view of the boundary particle, the classical saddles of the Boltzmann weight $e^{-\beta E}$ are the closed Larmor orbits of length $\beta$, with real tension $\lambda=\abs{P}$. These orbits exist for $\Phi_h<\Phi_b$. As $\Phi_h\to\Phi_b$ the momentum vanishes and the orbit shrinks to a point. For $\Phi_h>\Phi_b$ the mass shell gives $\abs{P}^2=\Phi_b^2-\Phi_h^2<0$: the momentum and the radius of the would-be orbit are imaginary, and the relativistic particle of \eqref{eq:tt_action} becomes tachyonic. 

The particle picture, at least according to this classical argument, points toward projecting the spectrum onto $0<s<\nu$, and we adopt it throughout both in JT and sinh dilaton gravity. \rev{Accordingly, whenever a sum over energy eigenstates is inserted, the resolution of the identity is replaced by the projector $\int_0^\nu\dd s\,\rho(s)\ket{s}\bra{s}$, and all spectral integrals run over $0<s<\nu$.} 
\section{Sinh dilaton gravity as a charged particle on the quantum disk}
\label{sec:qdisk}

In JT gravity the dilaton can be integrated out, which sets the local geometry to be a piece of the hyperbolic disk, and the identification of Section \ref{sec:jt} follows from a classical geometric picture. In sinh dilaton gravity the dilaton cannot be integrated out, and indeed the quantum hyperbolic disk is a quantum object that cannot be obtained from any classical solution.

\rev{Nevertheless, there is} a good reason to expect the charged-particle picture to persist, and it comes from the same boundary term that produced the magnetic field in Section \ref{sec:jt}. The same rewriting can be performed in sinh dilaton gravity, expressing the boundary term in \eqref{eq:sinh_action} in terms of the AdS$_2$ frame $g_-=e^{2\sigma}g$, $\sigma=\pi b^2\Phi$, of Section~\ref{sec:sinh}. Then, applying Gauss--Bonnet to the boundary curve in the effective AdS$_2$ frame, where $R_-=-2$ on shell, the boundary term becomes
\begin{equation}\label{eq:qd_GBsinh}
    \Phi_b\oint K\,\dd s=2\pi\Phi_b\,T+\Phi_b\,\mathrm{Area}_-(\gamma)-\pi b^2\,\Phi_b\oint\partial_n\Phi\,\dd s ,
\end{equation}
where $T$ is the turning number of the boundary curve and $\mathrm{Area}_-$ is the area it encloses in the AdS$_2$ frame. The first two terms have exactly the structure found in JT gravity: a topological term, and the flux of a uniform magnetic field, proportional to the boundary dilaton, through the region enclosed by the boundary in the AdS$_2$ frame.\footnote{The last term comes from the change of the extrinsic curvature under a Weyl rescaling, i.e.\ $K_-\,\dd s_-=(K+\partial_n\sigma)\,\dd s$. We will see in Section~\ref{subsec:corr_semiclassical} that this term combines with the on-shell bulk term to provide a semiclassical magnetic field description in the effective geometry.} We therefore expect the boundary to behave again as a particle charged under a uniform field set by $\Phi_b$, now moving in the AdS$_2$ frame. The difference with JT gravity is that the AdS$_2$ frame is constant-curvature only on shell, and its quantum fluctuations are not suppressed. The space on which the particle moves is then not the classical hyperbolic plane, but its quantization, the quantum hyperbolic disk\rev{, which is the quantization of the AdS$_2$ frame $g_-$, the metric measured by one of the two Liouville fields, as we explain in Section~\ref{subsec:qd_meaning}}. Since we cannot derive this picture by integrating out the bulk, we establish it by matching the dynamics: we show that the difference equation \eqref{eq:sinh_difference} is exactly the Schr\"odinger equation of such a particle.

This is the main result of this paper: the difference equation \eqref{eq:sinh_difference} describes the rotationally symmetric sector of a charged particle on the quantum hyperbolic disk, analytically continued to $\abs{q}=1$, in the magnetic field $B=i\nu$ set by the boundary dilaton. We will establish this in three steps:
\begin{itemize}
    \item we switch the field off, and show that the sinh Hamiltonian is then exactly the invariant Laplacian of the quantum disk. Here we use the semiclassical intuition developed in Section \ref{sec:jt}. The length of the Einstein--Rosen bridge is the radial coordinate of one of its endpoints, measured from the other one, placed at a reference point $O$, and the two-sided dynamics is the rotationally symmetric sector around $O$. On the quantum disk we keep this identification, with the length variable $\Lambda$ playing the role of the radial coordinate;
    \item we switch the field on by covariantizing the $q$-derivative, following the sum-of-squares structure of the JT operator \rev{\eqref{eq:qd_gaugeid}} derived in Section~\ref{sec:jt}; this reproduces the sinh Hamiltonian exactly;
    \item we show that this coupling is the one dictated by the quantum disk itself, by computing its Laplacian on wavefunctions of definite angular momentum and continuing the angular momentum to the magnetic field.
\end{itemize}

Let us first rewrite the sinh equation \rev{\eqref{eq:sinh_difference}} in the variable $u=e^\Lambda=\sinh^2\frac L2$ used in the JT analysis. We set
\begin{equation}
    q=e^{i\pi b^2},\qquad q^2=e^{\varepsilon},\qquad \varepsilon=2\pi ib^2,\qquad q^{-2B}=e^{2\pi b^2\nu},
\end{equation}
where in the last relation we used $B=i\nu$. The shifts of $\Lambda$ by $\pm2\pi ib^2$ appearing in \eqref{eq:sinh_difference} are then dilations of $u$ by $q^{\pm2}$, and the equation becomes
\begin{equation}\label{eq:qd_H}
    \cosh(2\pi b^2s)\,\chi(u)=\tfrac12\Big(1+\frac qu\Big)\chi(q^{-2}u)+\tfrac12\Big(1+\frac{q^{-1-4B}}{u}\Big)\chi(q^{2}u)-\frac{q^{-2B}}{u}\,\chi(u)\equiv\mathcal H_{\mathrm{sinh}}\chi .
\end{equation}
This is a second-order $q$-difference operator: derivatives have been replaced by finite dilations. Since $\varepsilon$ is imaginary, these dilations are translations of $\Lambda$ in the imaginary direction, while $\Lambda$ itself remains a continuous real variable: nothing is discretized. The structure of the operator becomes transparent once we write the dilations as exponentials of the Euler operator $\theta=u\partial_u$, namely $\chi(q^{\pm2}u)=e^{\pm\varepsilon\theta}\chi(u)$. A short computation shows that \eqref{eq:qd_H} can be rewritten exactly as
\begin{equation}\label{eq:qd_opform}
    \mathcal H_{\mathrm{sinh}}=\cosh(\varepsilon\theta)+\frac{q^{-2B}}{u}\Big[\cosh\!\big(\varepsilon(\theta-\tfrac12-B)\big)-1\Big] .
\end{equation}
Expanding to second order in $\varepsilon$, and using $\cosh(2\pi b^2s)=\cos(\varepsilon s)$, the eigenvalue equation reduces precisely to the JT equation \eqref{eq:qd_classical2}. The sinh operator is thus a term-by-term deformation of the sum-of-squares form \rev{\eqref{eq:qd_gaugeid}} of the JT operator: this is the structure that we want to understand geometrically.

\subsection{The quantum hyperbolic disk and the neutral particle}
\label{subsec:qd_neutral}
\label{subsec:qd_disk}

The space that will play the role of the hyperbolic plane is the quantum hyperbolic disk. Recall first that the Poincar\'e disk, with metric $4\dd z\dd\bar z/(1-\abs{z}^2)^2$, is related to the polar coordinates \eqref{eq:mag_polar} by
\begin{equation}\label{eq:qd_poincare}
    z=\tanh\frac L2\;e^{i\vartheta},
\end{equation}
so that all radial variables are functions of $\abs{z}^2$. In particular, the variable
\begin{equation}\label{eq:radiusqdisk}
    y\equiv1-\abs{z}^2=\mathrm{sech}^2\frac L2=\frac{1}{1+u}
\end{equation}
will be the natural radial variable on the quantum disk. The quantum disk keeps the complex coordinate, but makes $z$ and its conjugate $z^*$ non-commuting~\cite{Shklyarov:1998,Shklyarov:1998b,Almheiri:2024ayc}:
\begin{equation}\label{eq:qd_relation}
    z^*z=q^2\,zz^*+1-q^2 .
\end{equation}
In the limit $q\to1$ the two coordinates commute and the classical disk is recovered. The relation is covariant under the quantum group $U_q(\mathfrak{su}(1,1))$, the deformation of the M\"obius isometries, and it is invariant under rotations $z\to e^{i\alpha}z$, which are not deformed. If we define the quantum radial variable as $y=1-zz^*$\rev{, the quantum version of \eqref{eq:radiusqdisk}}, the relation \rev{\eqref{eq:qd_relation}} implies that multiplication by $z$ dilates it, $yz=q^2zy$. Functions on the quantum disk are therefore naturally acted upon by $q$-dilations of the radial variable, which is exactly the kind of operator we found in \eqref{eq:qd_H}.

In particular, two regimes are relevant for us. For real $0<q<1$ the algebra has a Hilbert-space representation in which $y$ is diagonal with eigenvalues $q^{2n}$, $n=0,1,2,\dots$, and $z$ raises $n$ by one~\cite{Shklyarov:1998,Shklyarov:1998b,Almheiri:2024ayc}.
Defining $L$ by the classical relation \eqref{eq:radiusqdisk} between $y$ and the radius, the radii become quantized and the disk is a set of concentric circles. Moreover, the circles become equally spaced near the boundary; this is the regime of sine dilaton gravity and DSSYK \cite{Blommaert:2024ymv}. For $q=e^{i\pi b^2}$ instead, which is the case of sinh dilaton gravity, the dilations of $y$ are imaginary translations of $\log y$ and the radius stays continuous. Physically, this is the analytic continuation of the quantum disk to $\abs{q}=1$, with the same algebra and the same operators acting on functions of a continuous radius.

There is one more feature of $q$-deformed operators that will be crucial. Dilations of a radial variable have exactly two fixed points, which we call the anchors of the deformation. These are the only points where the deformed operator retains the local structure of the classical one. From \eqref{eq:radiusqdisk} it is clear that the natural dilations of the quantum disk act on $x=1/y=\cosh^2\frac L2$ and their fixed points are the boundary, $x=\infty$, and $x=0$, i.e.\ $L=i\pi$. The centre of the disk, at $x=1$, is not an anchor. Instead the sinh operator \eqref{eq:qd_H} dilates $u=\sinh^2\frac L2$ and is anchored at the boundary and at the endpoint $O$ of the Einstein--Rosen bridge, at $u=0$. We will see in a moment that this mismatch has a simple and important resolution.

The invariant Laplacian of the quantum disk is given in~\cite{Almheiri:2024ayc} as
\begin{equation}\label{eq:qd_AP}
    \Box=q^{-1}(1-zz^*)^2\,\partial_z\partial_{z^*},
\end{equation}
where the $q$-derivatives are defined by their commutation relations with the coordinates,
\begin{equation}\label{eq:qd_qder}
    \partial_zz=1+q^{-2}z\partial_z,\qquad \partial_zz^*=q^2z^*\partial_z,\qquad \partial_{z^*}z^*=1+q^2z^*\partial_{z^*},\qquad \partial_{z^*}z=q^{-2}z\partial_{z^*}.
\end{equation}
We denote by $\Box_0$ its restriction to radial functions, i.e.\ functions of $x=\cosh^2\frac L2$. In Appendix \ref{app:qdlap} we show that
\begin{equation}\label{eq:qd_SSV}
    \Box_0=q^{-3}\,\frac1x\,[\theta]_q\,\big(q^{-1}x-1\big)\,[\theta]_q ,
\end{equation}
where $[\theta]_q$ is the symmetric $q$-derivative, i.e.\ the $q$-number of the Euler operator $\theta=x\partial_x$,
\begin{equation}\label{eq:qd_jackson}
    [\theta]_q\,f(x)=\frac{f(qx)-f(q^{-1}x)}{q-q^{-1}},\qquad [X]_q\equiv\frac{q^X-q^{-X}}{q-q^{-1}} ,
\end{equation}
which tends to $\theta$ as $q\to1$. Since $\frac1x\,\theta\,(x-1)\,\theta=\partial_x\,x(x-1)\,\partial_x$, \eqref{eq:qd_SSV} is the natural $q$-deformation of the radial Laplacian of the hyperbolic plane, in which the derivatives are replaced by symmetric $q$-derivatives. We also checked that it coincides, up to the normalization $q^{-3}$, with the radial Laplacian of the quantum disk given in \cite{Shklyarov:1998,Shklyarov:1998b}.

We can now compare the sinh operator \rev{\eqref{eq:qd_H}} with the quantum-disk Laplacian \rev{\eqref{eq:qd_SSV}} in the absence of the magnetic field. Setting $B=0$ in \eqref{eq:qd_H}, and changing variable in the Laplacian \eqref{eq:qd_SSV} according to
\begin{equation}\label{eq:qd_swapvar}
    x=-q\,u ,
\end{equation}
we find the exact identity
\begin{equation}\label{eq:qd_vaksman}
    \mathcal H_{\mathrm{sinh}}\big|_{B=0}=\frac{q^4}{2}\,(q^{-1}-q)^2\;u^{\frac12}\,\Box_0\,u^{-\frac12}+\frac{q+q^{-1}}{2} .
\end{equation}
We provide more details in Appendix \ref{app:qdlap}. The conjugation by $u^{1/2}$ is the same change of measure that we encountered in JT gravity. Moreover, in \cite{Almheiri:2024ayc} the eigenfunctions of the quantum-disk Laplacian are labeled by the principal series $\Delta=\frac12+i\rho$, with eigenvalues
\begin{equation}\label{eq:qd_APeigen}
    \lambda_\rho=\frac{1+q^2-2q\cos(2\rho\log q)}{(1-q^2)^2}.
\end{equation}
For $q=e^{i\pi b^2}$ the numerator becomes $2q\big[\cos(\pi b^2)-\cosh(2\pi b^2\rho)\big]$, and by \eqref{eq:qd_vaksman} the eigenvalue $\cosh(2\pi b^2s)$ of the neutral sinh operator corresponds precisely to $\rho=s$. The spectral parameter of the quantum disk is the horizon value of the dilaton and the neutral sinh Hamiltonian is therefore exactly the Laplacian of the quantum disk. 

To understand the meaning of the change of variable \eqref{eq:qd_swapvar} we can consider its classical limit, which is solved by
\begin{equation}\label{eq:shift}
    \cosh^2\frac{L_{\rm disk}}2=-\sinh^2\frac L2 \qquad \Longrightarrow \qquad  L=L_{\rm disk}+i\pi .
\end{equation}
\rev{The sinh length is therefore not measured from the centre of the disk, $x=1$, but from its second anchor $x=0$, i.e.\ $L_{\rm disk}=i\pi$: the shift \eqref{eq:shift} maps the anchor $O$ of the sinh operator to this anchor of the disk.} Classically the shift is allowed at $B=0$, since the neutral rotationally symmetric problem is invariant under $L\to L+i\pi$, which exchanges $u$ with $-1-u$ and leaves the Laplacian $\partial_u\,u(1+u)\,\partial_u$ invariant. 

However, for the charged particle the two choices are not equivalent because the gauge field is not invariant under the relabelling: the field regular around the centre of the disk becomes, when seen from $O$, the field regular at $O$ plus an Aharonov--Bohm flux. The JT problem is the rotationally symmetric wave of the field regular at $O$, and therefore the field has to be coupled after performing the shift \eqref{eq:shift}.

The relabelling also has a natural interpretation in terms of the eigenfunction \eqref{eq:sinh_eigenfunction}, which is defined by regularity at $O$. We show in Appendix \ref{app:realq} that the eigenfunction \eqref{eq:sinh_eigenfunction}, when continued to real $q$ where the quantum disk is originally defined, becomes a Heine series ${}_2\varphi_1(\ldots;q^2,x)$ in precisely the variable $x=-qu$, centred at $O$\rev{, up to corrections that are non-perturbative in $\log q$}. The radius of convergence of the Heine series \rev{\eqref{eq:rq_heine}} ends precisely at the centre of the disk, at $L=i\pi$. Beyond the centre it develops simple poles exactly on the Fock lattice of the disk, anchored at the centre, with Al-Salam--Chihara polynomials as residues, as we show in Appendix \ref{app:realq}. The relabelling therefore does not discard any structure: the description anchored at $O$ contains the discrete one anchored at the centre and the swap of anchors appears as a natural step once $q$ is analytically continued.

\subsection{Coupling the magnetic field}
\label{subsec:qd_field}

After the relabelling, the neutral operator is \eqref{eq:qd_opform} at $B=0$. Its structure becomes clearer in terms of the symmetric $q$-derivative \eqref{eq:qd_jackson}, now acting on the sinh variable, $\theta=u\partial_u$. Using the identity $\cosh(\varepsilon X)-1=\frac12(q-q^{-1})^2[X]_q^2$, the neutral operator can be written as
\begin{equation}\label{eq:qd_qsquares}
    \mathcal H_0=1+\frac{(q-q^{-1})^2}{2}\Big\{[\theta]_q^2+\frac1u\,[\theta-\tfrac12]_q^2\Big\}.
\end{equation}
This is, term by term, the $q$-number version of the neutral JT operator \eqref{eq:qd_gaugeid} when $B=0$: a boundary term and a centre term, each the square of a radial $q$-derivative. In JT gravity, the magnetic field enters this structure as a covariantization of the radial derivative in the centre term, $\theta\to\theta-B$, as shown in \eqref{eq:qd_classical2}. The natural $q$-version of this is to covariantize the $q$-derivative in the centre term in \eqref{eq:qd_qsquares}, i.e.
\begin{equation}\label{eq:qd_qsquaresB}
    \mathcal H_B=1+\frac{(q-q^{-1})^2}{2}\Big\{[\theta]_q^2+\frac1w\,[\theta-\tfrac12-B]_q^2\Big\},\qquad w=q^{2B}u .
\end{equation}
This operator coincides exactly with the sinh operator \eqref{eq:qd_opform}.
The only subtlety is the radial variable in which the coupling is performed, $w$ rather than $u$, which corresponds to a shift of the origin of $\Lambda$ by $2\pi b^2\nu$. This shift vanishes in the JT limit and it is fixed by charge conjugation: classically, the rotationally symmetric wave depends on the field only through $B^2$, and $w$ is the variable in which the deformed operator preserves this symmetry. 

The agreement \eqref{eq:qd_qsquaresB} shows that the sinh Hamiltonian is obtained from the quantum-disk Laplacian by the most natural minimal coupling suggested by the JT limit. It does not yet show that this is the coupling of a magnetic field on the quantum disk, since a differential operator admits infinitely many $q$-deformations with the same classical limit. 

To settle this question we turn to the quantum disk itself, and follow the classical logic of \eqref{eq:qd_LBLm}: the charged rotationally symmetric wave is the neutral wave of angular momentum $2B$, with the centrifugal barrier at $O$ removed. Our plan is therefore to compute the Laplacian \rev{\eqref{eq:qd_AP}} of the quantum disk on waves of angular momentum $m$, to continue $m$ to the magnetic field and then to erase the flux at $O$ by a suitable modification of the equation. 

We first ask how a wavefunction of definite angular momentum looks on the quantum disk. Since rotations are an exact symmetry of the relation \eqref{eq:qd_relation}, angular momentum is well defined and a wavefunction of angular momentum $m\ge0$ is, exactly as classically, $z^m$ times a radial function (with $z^m$ written on the left). In Appendix \ref{app:mwave} we compute the action of the quantum-disk Laplacian \eqref{eq:qd_AP} on these wavefunctions in closed form, finding that it preserves the angular momentum and acts on the radial function as
\begin{equation}\label{eq:qd_boxm}
\begin{aligned}
    &\Box\big(z^mf\big)=z^m\,\Box_mf,\\
    &\Box_mf(y)=\frac{q^{-1}}{(1-q^2)^2}\Big\{\big(q^{-2}-q^{2m}y\big)\big[f(q^2y)-f(y)\big]-(1-y)\big[f(y)-f(q^{-2}y)\big]\Big\}.
\end{aligned}
\end{equation}
At $m=0$ this is the radial Laplacian \eqref{eq:qd_SSV}. Performing the same relabelling \eqref{eq:qd_swapvar} as in the neutral case, $y=-1/(qu)$, \rev{and a radial gauge transformation that distributes the hopping coefficients as in the sinh operator (Appendix~\ref{app:mwave}),} we obtain the operator acting on the wave of angular momentum $m$ in the sinh variable,
\begin{equation}\label{eq:qd_Hm}
    \mathcal H_m\chi=\frac12\Big(1+\frac qu\Big)\chi(q^{-2}u)+\frac12\Big(1+\frac{q^{2m-1}}u\Big)\chi(q^2u)-\frac{1+q^{2m}}{2u}\,\chi(u).
\end{equation}

Before continuing $m$, we need to understand how the classical exponents studied in Section \ref{sec:jt} are encoded in a $q$-difference equation, since these are the data on which we impose the physical conditions. We write the operator as
\begin{equation}
A_+(u)\chi(q^2u)+A_-(u)\chi(q^{-2}u)+A_0(u)\chi(u),
\end{equation}
with hopping coefficients $A_\pm$ and a diagonal term $A_0$. As shown in Appendix \ref{app:exponents}, the two singular points are encoded in different places:
\begin{itemize}
    \item the exponents at $O$, i.e.\ the angular momentum around $O$ shifted by any flux through it, are fixed by the diagonal term;
    \item the exponent difference at $L=i\pi$, i.e.\ the holonomy of the gauge field around it, is fixed by the hopping coefficients, through the splitting of the zeros of their gauge-invariant product.
\end{itemize}
For the wave \eqref{eq:qd_Hm} both are equal to $m$, as for the two poles of the centrifugal barrier in \eqref{eq:qd_Lm}.\footnote{Around $u=0$ \rev{the solutions of \eqref{eq:qd_Hm} behave as $u^{1/2}$ and $u^{1/2-m}$}, i.e.\ the exponent difference is $m$, as for the classical case \eqref{eq:qd_Lm}. Moreover the ratio of the zeros of the hopping product is $q^{2m}=q^{2(\gamma_1-\gamma_2)}$, which encodes the exponent difference $\gamma_1-\gamma_2=m$ at $u=-1$ or $L=i\pi$.}

We can now carry out the plan. We continue \eqref{eq:qd_Hm} to $m=-2B$, the sign being a choice of orientation. We keep its hopping coefficients intact since they encode the holonomy $2B$ of the field around $L=i\pi$. They are
\begin{equation}\label{eq:hoppingterms}
    A_-=\frac12\Big(1+\frac qu\Big),\qquad A_+=\frac12\Big(1+\frac{q^{-1-4B}}u\Big),
\end{equation}
exactly the hopping coefficients of the sinh operator \eqref{eq:qd_H}. We then modify the diagonal term, so as to remove the flux at $O$, i.e.\ we replace it with the unique term of the form $-a/u$ for which the two exponents at $O$ coincide. The indicial equation at $O$,
\begin{equation}
    \frac12q^{-1-4B}X+\frac{q}{2X}-a=0,\qquad X=q^{2\rho},
\end{equation}
has a double root if and only if $a^2=q^{-4B}$ and the existence of the JT limit selects $a=q^{-2B}$. This is exactly the diagonal term of the sinh operator \rev{\eqref{eq:qd_H}}. Altogether,
\begin{equation}\label{eq:qd_HsinhHm}
    \mathcal H_{\mathrm{sinh}}=\mathcal H_{m=-2B}+\frac{(1-q^{-2B})^2}{2u} ,
\end{equation}
which is the exact quantum-disk version of the classical relation \eqref{eq:qd_LBLm}.\footnote{Indeed, the classical limit of the last term is $\frac{\varepsilon^2}2\frac{B^2}u$.}

\rev{The sinh operator therefore follows from two conditions, with no coefficient adjusted to it. The hopping coefficients are those of the Laplacian of the quantum disk on waves of angular momentum $-2B$, \eqref{eq:qd_Hm}, and the diagonal term is fixed, up to a sign selected by the JT limit, by requiring that the field carry no flux at the endpoint of the Einstein--Rosen bridge. The only other choice is the radial gauge of \eqref{eq:qd_Hm}, a rescaling of the wavefunction by a function of $u$ which leaves the spectrum unchanged (Appendix \ref{app:mwave}).} Since rotations are not deformed, the angular momentum is not renormalized, which explains why the magnetic field enters the sinh operator through $q^{-4B}$ exactly, with $B=i\nu$ to all orders in $b$. \rev{Since \eqref{eq:qd_HsinhHm} is the same operator as \eqref{eq:qd_qsquaresB}, the covariantization $\theta\to\theta-B$ of the centre term is not an additional assumption: it is the coupling derived here from the quantum disk, written in the radial gauge in which the field enters only the centre term, as in \eqref{eq:qd_classical2}.}

As a final confirmation, the resulting operator is the little $q$-Jacobi operator, the $q$-analogue of the Jacobi operator that arises from matrix elements of representations of $U_q(\mathfrak{su}(1,1))$~\cite{Kakehi:1995,Koelink:2000}. As we show in Appendix \ref{app:qjacobi}, after a conjugation by a power of the variable and a rescaling of $u$, one has that
\begin{equation}\label{eq:qd_identity}
    \mathcal H_{\mathrm{sinh}}=X^{-\gamma}\,\mathcal L_{(\alpha,\beta)}\,X^\gamma,\qquad (\alpha,\beta)=(0,2B),\qquad X=q^{1+4B}u,\qquad\gamma=-\tfrac12-B,
\end{equation}
i.e.\ the magnetized quantum disk Laplacian is mapped to the $\mathcal L_{(0,2B)}$ little $q$-Jacobi operator. In the JT limit the radial equation \eqref{eq:sinh_PT} is the Jacobi equation with the very same parameters, $\alpha=0$ corresponding to the rotationally symmetric wave at $O$ and $\beta=2B$ to the magnetic field. 
For real $q$, its solution regular at $O$ is expressed in terms of a Heine series,
\begin{equation}\label{eq:qd_heinesol}
    \chi_s\propto u^{\frac12+B}\,{}_2\varphi_1\Big(q^{1+2B-2is},q^{1+2B+2is};q^2;q^2;-qu\Big),
\end{equation}
the $q$-analogue of the hypergeometric JT wavefunction.

At $\abs{q}=1$, the regime of sinh dilaton gravity, the Heine series is no longer available and its role is played by the $b$-hypergeometric function \eqref{eq:sinh_eigenfunction}. This is the modular-double version of \eqref{eq:qd_heinesol}, which fixes the periodic ambiguity of the difference equation \rev{\eqref{eq:qd_H}}~\cite{SinhBox}. The two descriptions are consistent. In fact in Appendix \ref{app:realq} we continue \eqref{eq:sinh_eigenfunction} to real $q$ and show that it reduces to the Heine series \eqref{eq:qd_heinesol}, up to an infinite tower of corrections suppressed by powers of the dual base $\tilde q=e^{-2\pi i/b^2}$, which are non-perturbative in $\log q$ and come from the dual series. 

\subsection{Phase-space non-commutativity as spacetime non-commutativity}
\label{subsec:qd_meaning}

Having identified the space on which sinh dilaton gravity describes a particle, we can ask what this space is from the gravitational point of view. The answer is clearest in the formulation of sinh dilaton gravity as Liouville gravity, i.e.\ as a spacelike Liouville field $\phi$, with background charge $Q=b+b^{-1}$, coupled to a timelike Liouville field $\tilde\phi$, with background charge $\tilde Q=b^{-1}-b$~\cite{Mertens:2020hbs}. We will come back to the double Liouville CFT formulation in Section \ref{subsec:zz}.

Choosing a flat fiducial metric $\hat g$, the action of the two Liouville theories is
\begin{equation}\label{eq:qd_liouvilleaction}
    S=\frac1{4\pi}\int\dd^2x\Big[(\hat\nabla\phi)^2+4\pi\mu\,e^{2b\phi}\Big]+\frac1{4\pi}\int\dd^2x\Big[-(\hat\nabla\tilde\phi)^2+4\pi\tilde\mu\,e^{2b\tilde\phi}\Big],
\end{equation}
up to the curvature couplings, which only contribute at the boundary. In conformal gauge $g=e^{2\rho}\hat g$, the two fields are related to the gravitational variables by
\begin{equation}\label{eq:zz_map}
    \phi=\frac\rho b-\pi b\,\Phi,\qquad \tilde\phi=\frac\rho b+\pi b\,\Phi .
\end{equation}
Inserting \eqref{eq:zz_map} in \eqref{eq:qd_liouvilleaction}, the two kinetic terms combine into $-\int\partial\Phi\cdot\partial\rho$, which is the coupling $\Phi R$ of dilaton gravity, while the two cosmological terms become $\sqrt g\,(\mu\,e^{-2\pi b^2\Phi}+\tilde\mu\,e^{2\pi b^2\Phi})$. For $\tilde\mu=-\mu=-\frac{1}{4 \pi b^2}$ this is the sinh potential~\cite{Mertens:2020hbs}.

The two Liouville fields are decoupled, and each of them obeys its own Liouville equation,
\begin{equation}\label{eq:qd_liouvilleeom}
    \hat\nabla^2\phi=4\pi\mu b\,e^{2b\phi},\qquad \hat\nabla^2\tilde\phi=4\pi\mu b\,e^{2b\tilde\phi} ,
\end{equation}
where we used $\tilde\mu=-\mu$. A Liouville field is the logarithm of the Weyl factor of a metric, and the Liouville equation states that this metric has constant negative curvature, $R=-8\pi\mu b^2=-2$. The two metrics
\begin{equation}\label{eq:sinh_liouvillemetrics}
    e^{2b\phi}\,\hat g=g_+,\qquad e^{2b\tilde\phi}\,\hat g=g_-
\end{equation}
are thus the two AdS$_2$ geometries \eqref{eq:sinh_weyl} found in Section~\ref{sec:sinh}. The canonical quantization of Section~\ref{sec:sinh} is carried out in the AdS$_2$ frame $g_-$, where the length $L$ of the Einstein--Rosen bridge is defined, and this is the geometry measured by a Liouville field. Having matched the two-sided dynamics of this effective AdS$_2$ geometry with those of a particle on a quantum hyperbolic disk, we infer that the quantum hyperbolic disk corresponds to the quantization of this geometry. 

We now show that the non-commutativity of the quantum disk is nothing but the canonical commutator of the gravitational phase space. On the gravitational side, in terms of the canonical pair \eqref{eq:sinh_hbar}, the Euler operator is $\varepsilon\theta=-p$ and the operator form \eqref{eq:qd_opform} becomes
\begin{equation}\label{eq:qd_Hclassical}
    \mathcal H_{\mathrm{sinh}}=\cosh p+e^{2\pi b^2\nu}\,e^{-\Lambda}\Big[\cosh\big(p-2\pi b^2\nu+\tfrac{i}{2}\hbar_{\rm eff}\big)-1\Big].
\end{equation}
Before quantization this is an ordinary function on a commutative phase space. Once the canonical commutator $[\Lambda,p]=i\hbar_{\rm eff}$ is imposed, the exponentials of the momentum become finite translations of the length,
\begin{equation}\label{eq:qd_shiftPi}
    e^{\pm p}\,\chi(\Lambda)=\chi(\Lambda\mp i\hbar_{\rm eff}),\qquad q^2=e^{i\hbar_{\rm eff}} .
\end{equation}
The $q$-difference structure of the sinh equation \rev{\eqref{eq:qd_H}} is the canonical commutator.

On the side of the quantum disk, we write the complex coordinate in polar form, as a radius times a phase,
\begin{equation}
    z=\sqrt{1-y}\;V ,
\end{equation}
where $\sqrt{1-y}$ is the quantum version of $\tanh\frac{L_{\rm disk}}2$ and $V$ the quantum version of $e^{i\vartheta}$. The defining relation \eqref{eq:qd_relation} is then equivalent to
\begin{equation}\label{eq:qd_radang}
    V^\dagger\,y\,V=q^2\,y\qquad\Longleftrightarrow\qquad [\log y,\hat\vartheta]=-2i\log q=\hbar_{\rm eff} .
\end{equation}
The non-commutativity of $z$ and $z^*$ is a non-commutativity between the radius and the angle: the logarithm of the radial variable and the angle form a Heisenberg pair. After the relabelling \eqref{eq:qd_swapvar}, the logarithm of the radial variable is, up to a constant, minus the length $\Lambda$, and the conjugation by the phase, which dilates the radial variable, is the shift \eqref{eq:qd_shiftPi} produced by the momentum. The dictionary is therefore
\begin{equation}\label{eq:qd_dictionaryPi}
    e^{-\Lambda}\;\longleftrightarrow\;-q\,y,\qquad [\Lambda,p]=i\hbar_{\rm eff}\;\longleftrightarrow\;[\log y,\hat\vartheta]=\hbar_{\rm eff},
\end{equation}
and the two commutators are equal, with the angle playing the role of $i$ times the momentum. This identification also explains why $\Lambda=2\log\sinh\frac L2$, rather than the length $L$ itself, is the correct gravitational variable on the phase space of sinh dilaton gravity: it is the variable conjugate to the angle of the quantum hyperbolic disk. Through this dictionary, the sinh Hamiltonian \eqref{eq:qd_Hclassical} is the Laplacian of the quantum disk.

The logic of the identification can then be summarized as follows. Classically, \eqref{eq:qd_Hclassical} is a function on a commutative phase space. Imposing the canonical commutator turns the exponentials of the momentum into finite shifts and only this quantum operator coincides with the Laplacian of a non-commutative disk. The quantization of the gravitational phase space appears geometrically as the non-commutativity of the disk: a Planck cell of the phase space is mapped to an elementary cell of the quantum disk and the fuzziness of the phase space becomes the geometric fuzziness of the effective AdS$_2$. \rev{This is consistent with what was found in~\cite{Belaey:2025} through the analysis of the two-point function close to the boundary, where its $U_q(\mathfrak{su}(1,1))$ covariance was related to a $q$-deformation of the effective AdS$_2$ geometry.} 

Two qualifications are in order. For sinh dilaton gravity, $q$ has unit modulus and the Planck constant of the disk is imaginary, so that the angle of the disk corresponds to $i$ times the momentum, and the cells describe the algebra rather than an actual lattice. For sine dilaton gravity, instead, $q$ is real and the quantum disk is the Hermitian one. The angle is then a genuine periodic variable, $\hat\vartheta\sim\hat\vartheta+2\pi$, and its periodicity discretizes the conjugate variable $\log y$, so that the radial levels are discrete, $y=q^{2n}$. This is precisely the logic by which, in sine dilaton gravity, the periodicity of the momentum $p$ discretizes the length of the Einstein--Rosen bridge~\cite{Blommaert:2024whf}: as we will see in Section~\ref{subsec:corr_limits}, the radial levels are the chord numbers of DSSYK in the regime of large loops.

\section{Correlators as dressed Witten diagrams on the quantum disk}
\label{sec:corr}
\rev{Having established the duality between sinh dilaton gravity with a finite Dirichlet boundary and a magnetized particle on the quantum disk, we now use it to compute correlators.}

In JT gravity, it was observed in~\cite{Yang:2018gdb} that a boundary correlator is a Witten diagram in a fixed AdS$_2$. Matter lines run through the bulk between insertion points, but the insertion points sit on a fluctuating boundary curve, which is the worldline of the charged particle of Section \ref{sec:jt}. Gravity enters only through the propagators of this particle along the boundary arcs between the insertions. Since we have shown that finite-cutoff sinh dilaton gravity is the same charged particle on the quantum hyperbolic disk, its correlators should be dressed Witten diagrams on the quantum disk. In this section we construct these diagrams and evaluate the two-point function exactly.

\begin{figure}[t]
\centering
\includegraphics{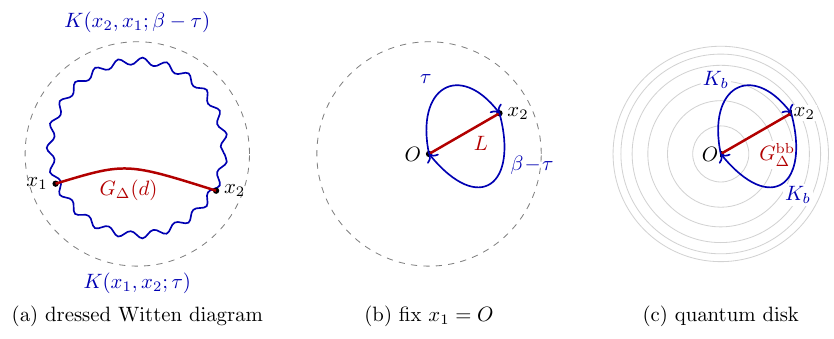}
\caption{(a) The dressed Witten diagram for the two-point function. The wiggly curve is the fluctuating boundary, the worldline of the charged particle; the two insertions split it into arcs of proper time $\tau$ and $\beta-\tau$, which are particle propagators between the same two bulk points. The red line is the bulk matter propagator. (b) After placing $x_1$ at the centre $O$, only the radius $L$ of $x_2$, the length of the matter line, is integrated over. (c) On the quantum disk the arcs become $q$-propagators $K_b$ and the matter line becomes the Green's function $G^{\rm bb}_\Delta$ of the quantum-disk Laplacian; the circles depict the discrete radii of the disk at real $q$.}
\label{fig:corr_witten}
\end{figure}

Let us first recall the construction in JT gravity. A free matter field of mass $m^2=\Delta(\Delta-1)$ propagates on the fixed hyperbolic background. Inserting two boundary operators at times $0$ and $\tau$ splits the closed boundary curve into two arcs, of proper lengths $\tau$ and $\beta-\tau$, and the matter field connects the two insertion points $x_1$ and $x_2$ through the bulk, as shown in Figure \ref{fig:corr_witten}a. Each ingredient of the diagram has a simple meaning. The boundary particle has to go from $x_1$ to $x_2$ in a time $\tau$ and then come back from $x_2$ to $x_1$ in the remaining time $\beta-\tau$: this produces two particle propagators, which together make up the closed boundary curve. The matter field, on the other hand, propagates from $x_1$ to $x_2$ through the bulk and by the isometries its propagator depends only on the geodesic distance between the two points. Finally, since we do not know where the boundary curve is at time $\tau$, we must integrate over the position of $x_2$. Putting these ingredients together, the two-point function is
\begin{equation}\label{eq:corr_yang}
    \langle\mathcal O_\Delta(\tau)\mathcal O_\Delta(0)\rangle_\beta=\frac{1}{Z(\beta)}\int\dd^2x_2\;K(x_1,x_2;\tau)\;G_\Delta\big(d(x_1,x_2)\big)\;K(x_2,x_1;\beta-\tau).
\end{equation}
The two propagators carry opposite Wilson lines of the gauge field, whose phases cancel. Using the isometries we can then place $x_1$ at the centre $O$, as we did in Section \ref{sec:jt}, so that the distance between the two insertions becomes the radius $L$ of $x_2$ (see Figure \ref{fig:corr_witten}b) and the integral over $x_2$ becomes an integral over all lengths of the matter line, each weighted by the amplitude for the boundary to have its two insertions that far apart. With the spectral representation of the propagators, we obtain
\begin{equation}\label{eq:corr_2pt}
\begin{aligned}
    \langle\mathcal O_\Delta(\tau)\mathcal O_\Delta(0)\rangle_\beta&=\frac{1}{Z(\beta)}\int\dd s_1\dd s_2\,\rho(s_1)\rho(s_2)\,e^{-\tau E(s_1)-(\beta-\tau)E(s_2)}\,M(s_1,s_2),\\
    M(s_1,s_2)&=\int\dd L\;\psi^*_{s_1}(L)\,G_\Delta(L)\,\psi_{s_2}(L).
\end{aligned}
\end{equation}
At large cutoff the matter line becomes the boundary-to-boundary propagator, the exponential of the renormalized length, and the matrix element reduces to the well-known ratio of Gamma functions~\cite{Yang:2018gdb}. At finite cutoff, \rev{the full AdS$_2$ propagator was used as matter insertion in~\cite{Griguolo:2026eoi}.}

The formula \eqref{eq:corr_2pt} can also be written canonically, as the insertion of an operator $G_\Delta(\hat L)$ between two Euclidean evolutions in the two-sided quantum mechanics~\cite{Griguolo:2026eoi}. The diagram adds two things. The function of the length is not chosen but derived as the bulk propagator of the matter field, fixed by the bulk symmetry. Moreover the integral over $L$ acquires a geometric meaning as the integral over the position of an insertion point on the fluctuating boundary. At strong field the same diagram will produce the Liouville gravity amplitudes as Witten diagrams on the quantum disk, independently of the Liouville CFT~\cite{Mertens:2020hbs} and quantum-group~\cite{Fan:2021bwt} derivations.

\subsection{Matter line and boundary particle propagators on the quantum disk}
\label{subsec:corr_qdisk}
\label{subsec:corr_yang}
\label{subsec:corr_vsQM}

Each ingredient of the diagram has a natural counterpart on the quantum disk, as depicted in Figure \ref{fig:corr_witten}c. We begin with the boundary particle. Its propagator from the centre $O$, i.e.\ from $L=0$, to the radial position $\Lambda$ in the Euclidean time $\tau$ is the matrix element of the Euclidean evolution of the sinh dilaton gravity Hamiltonian of Section \ref{sec:sinh},
\begin{equation}\label{eq:corr_qprop_def}
    K_b(\Lambda;\tau)=\bra{\Lambda}e^{-\tau\hat H}\ket{L=0} .
\end{equation}
To evaluate it we insert \rev{the projector on the physical spectrum $0<s<\nu$, $\int_0^\nu\dd s\,\rho_b(s)\ket{s}\bra{s}$, as discussed in Section \ref{sec:ttbar}}, and this gives $K_b=\int\rev{_0^\nu}\dd s\,\rho_b(s)\,e^{-\tau E_b(s)}\,\psi_s(\Lambda)\,\overline{\psi_s(L=0)}$. The value of the eigenfunctions at $O$ is independent of the energy, and we absorb it in the definition of $\ket{L=0}$ as for the partition function \rev{of Section~\ref{sec:sinh}}. We obtain
\begin{equation}\label{eq:corr_qprop}
\begin{aligned}
    K_b(\Lambda;\tau)&=\int_0^{\rev{\nu}}\dd s\;\rho_b(s)\,e^{-\tau E_b(s)}\;\psi_s(\Lambda),\\
    \psi_s(\Lambda)&=\Phi_\nu(\Lambda)\,N_s\int_{c+i\mathbb R}\frac{\dd t}{2\pi i}\;e^{(\Lambda-2\pi b^2\nu)t}\;\frac{S_b(bt\pm ibs)\,S_b(\frac b2+ib\nu-bt)}{S_b(\frac b2-ib\nu+bt)},
\end{aligned}
\end{equation}
with the density \eqref{eq:sinh_density}, the energy \eqref{eq:sinh_energy}, and the normalization and phase of \eqref{eq:sinh_eigenfunction} and \eqref{eq:sinh_Phinu}. The return arc is the conjugate amplitude, $\bra{L=0}e^{-(\beta-\tau)\hat H}\ket{\Lambda}=\overline{K_b(\Lambda;\beta-\tau)}$.

The propagator has two readings. On the quantum disk, it is the amplitude for the magnetized boundary particle to travel from the centre $O$ to a point at radial distance $L$ in the Euclidean time $\tau$. In the two-sided channel, it is the wavefunction of the state prepared by a Euclidean evolution of duration $\tau$ from the state of vanishing length,
\begin{equation}
    \Psi_\tau(L)=\bra{L}e^{-\tau\hat H}\ket{L=0} ,
\end{equation}
i.e.\ the amplitude for a universe of zero size to evolve into a universe with an Einstein--Rosen bridge of length $L$.\footnote{This is the Hartle--Hawking state at Euclidean preparation time $\tau$, and the two-point function \eqref{eq:corr_2pt} is the expectation value of the matter line between two such states, $\bra{\Psi_{\beta-\tau}}G_\Delta(\hat L)\ket{\Psi_\tau}$.} 
In the JT limit\rev{, where the density, the energy and the eigenfunctions of Section~\ref{sec:sinh} reduce to those of finite-cutoff JT gravity,} the propagator \eqref{eq:corr_qprop} becomes that of the charged particle on the hyperbolic plane\rev{~\cite{Yang:2018gdb}}, while at strong field, at fixed renormalized coordinate, it becomes the propagator of Liouville gravity.

\rev{We now turn to the matter line. On the gravity side we consider the same type of matter as in sine dilaton gravity, where it was chosen to reproduce the correlators of DSSYK~\cite{Blommaert:2024ymv}: a scalar field coupled non-minimally to the metric, i.e.\ coupled to the dilaton as well. Its action is
\begin{equation}\label{eq:corr_matteraction}
    I_{\rm m}=\frac12\int\dd^2x\sqrt g\,\Big(g^{\mu\nu}\partial_\mu\phi\,\partial_\nu\phi+m^2\,e^{2\pi b^2\Phi}\phi^2\Big)=\frac12\int\dd^2x\sqrt{g_-}\,\Big(g_-^{\mu\nu}\partial_\mu\phi\,\partial_\nu\phi+m^2\phi^2\Big),
\end{equation}
where in the second step we used that the kinetic term is Weyl invariant in two dimensions. The coupling to the dilaton is thus absorbed into the metric $g_-=e^{2\pi b^2\Phi}g$ of \eqref{eq:sinh_weyl}, and at the semiclassical level the field propagates on this effective AdS$_2$ geometry rather than on the physical metric $g$. Upon quantization of the phase space, $g_-$ becomes the quantum disk of Section \ref{sec:qdisk}, whose isometries generate $U_q(\mathfrak{su}(1,1))$. The simplest wave equation for a scalar field which is invariant under these isometries and linear in the field is built from the quantum-disk Laplacian, $(-\square_q+m^2)\phi=0$~\cite{Almheiri:2024ayc}. It is the quantum deformation of the Klein--Gordon equation on $g_-$ that follows from \eqref{eq:corr_matteraction}, to which it reduces for $q\to1$, when the quantum disk becomes the classical disk with metric $g_-$. This is the semiclassical limit of Section \ref{subsec:corr_semiclassical}, at fixed $2\pi b^2\Phi$.\footnote{\rev{In the JT limit, at fixed $\Phi$, $g_-$ coincides with $g$, and one recovers a minimally coupled scalar on the AdS$_2$ geometry of JT gravity.}} The Green's function of $\square_q$ depends only on the invariant distance from the source.} The matter field is neutral and its bulk-to-bulk propagator is therefore the Green's function of the quantum disk Laplacian without magnetic field, i.e.\ of the neutral operator \rev{\eqref{eq:qd_qsquares}} of Section \ref{sec:qdisk} after the relabelling \eqref{eq:qd_swapvar}. We write it as a function of the radial coordinate $\Lambda$ of the insertion point\rev{, which on the quantum disk measures its invariant distance from $O$}. Away from the source, the propagator solves the neutral equation, i.e.\ \eqref{eq:sinh_difference} at $\nu=0$, at the imaginary energy $s=i(\frac12-\Delta)$. Written for $G_\Delta(\Lambda)=e^{-\Lambda/2}\chi(\Lambda)$, which removes the factor $e^{\Lambda/2}$ of the eigenfunctions,\footnote{The eigenfunctions of the difference operator are normalized with respect to the flat measure $\dd\Lambda$. The corresponding scalar functions on the disk are $f=e^{-\Lambda/2}\chi$, since $\int\dd\Lambda\,\abs{\chi}^2=\int\dd u\,\abs{f}^2$ with the invariant radial measure $\dd u=e^\Lambda\dd\Lambda$, which becomes the hyperbolic area element $\dd u\propto\sinh L\,\dd L$ in the JT limit.} this equation reads
\begin{equation}\label{eq:corr_neutral}
\begin{aligned}
    \cos\big(\pi b^2(2\Delta-1)\big)\,G_\Delta(\Lambda)={}&\tfrac12\big(e^{-i\pi b^2}+e^{-\Lambda}\big)G_\Delta(\Lambda-2\pi ib^2)+\tfrac12\big(e^{i\pi b^2}+e^{-\Lambda}\big)G_\Delta(\Lambda+2\pi ib^2)\\
    &-e^{-\Lambda}G_\Delta(\Lambda) ,
\end{aligned}
\end{equation}
and in the JT limit it becomes the Klein--Gordon equation $(\nabla^2-m^2)G_\Delta=0$ on AdS$_2$, with $m^2=\Delta(\Delta-1)$. The propagator is the solution that decays at the boundary and has a source singularity at $O$. In the classical case this is the familiar AdS$_2$ propagator, a hypergeometric function, which admits the Mellin representation
\begin{equation}\label{eq:corr_GJT}
    G^{\rm JT}_\Delta=u^{-\Delta}{}_2F_1\big(\Delta,\Delta;2\Delta;-\tfrac1u\big)=\frac{\Gamma(2\Delta)}{\Gamma(\Delta)^2}\int\frac{\dd k}{2\pi i}\,u^k\,\frac{\Gamma(\Delta+k)\Gamma(-k)^2}{\Gamma(\Delta-k)} .
\end{equation}
On the quantum disk we write $G_\Delta(\Lambda)=\int\frac{\dd k}{2\pi i}\,e^{k\Lambda}g_\Delta(k)$ so that \eqref{eq:corr_neutral} becomes a first-order recursion for the Mellin kernel:
\begin{equation}\label{eq:corr_recursion}
    \frac{g_\Delta(k+1)}{g_\Delta(k)}=\frac{\cos\big(\pi b^2(2\Delta-1)\big)-\cos\big(\pi b^2(2k+1)\big)}{\cos\big(2\pi b^2(k+1)\big)-1}=\frac{\sin\big(\pi b^2(\Delta+k)\big)\sin\big(\pi b^2(\Delta-1-k)\big)}{\sin^2\big(\pi b^2(k+1)\big)} .
\end{equation}
By the shift relation $S_b(x+b)=2\sin(\pi bx)S_b(x)$, it is solved by double sine functions and the solution with the required decay and source is given by the following $b$-deformed hypergeometric function
\begin{equation}\label{eq:corr_gbb}
\begin{aligned}
    G^{\rm bb}_\Delta(\Lambda)&=\int_{c_k+i\mathbb R}\frac{\dd k}{2\pi i}\,e^{k\Lambda}\,g^{\rm bb}_\Delta(k),\qquad -\Delta<c_k<0,\\
    g^{\rm bb}_\Delta(k)&=2\pi b\,\frac{S_b(2b\Delta)}{S_b(b\Delta)^2}\,\frac{S_b(b\Delta+bk)\,S_b(-bk)^2}{S_b(b\Delta-bk)}.
\end{aligned}
\end{equation}
\rev{In terms of the $b$-hypergeometric function \eqref{eq:sinh_Fb}, the substitution $k=-\Delta-t$ gives
\begin{equation}\label{eq:corr_GFb}
    G^{\rm bb}_\Delta(\Lambda)=e^{-\Delta\Lambda}\,F_b\big(b\Delta,b\Delta,2b\Delta;-e^{-\Lambda}\big),
\end{equation}
the $b$-deformation of $u^{-\Delta}{}_2F_1(\Delta,\Delta;2\Delta;-\frac1u)$ in \eqref{eq:corr_GJT}. This form is not a choice: it is a hyperbolic matrix element of the modular double of $U_q(\mathfrak{sl}(2,\mathbb R))$. In~\cite{SinhBox} the solutions of the sinh difference equation were identified with such matrix elements. The one that decays at infinity is $\langle h|g|h\rangle$ with hyperbolic label $h=\nu$, which reduces to the $SL(2,\mathbb R)$ element $K_{++}$~\cite{Vilenkin:1968,Vilenkin:1991} in the JT limit, while the universe wavefunction \eqref{eq:sinh_chiFb} is the combination of two matrix elements that is regular at $O$.\footnote{\rev{Its second difference equation, with shifts $\Lambda\to\Lambda\pm2\pi i$, is the dual Casimir equation of the modular double, twisted by an operator that maps the branch point at $O$ into a regular point~\cite{SinhBox}.}} For the neutral matter field $h=0$, and at the representation label $\frac b2+ibs=b\Delta$, i.e.\ at the energy $s=i(\frac12-\Delta)$ of \eqref{eq:corr_neutral}, the decaying element is exactly $e^{\Lambda/2}G^{\rm bb}_\Delta(\Lambda)$. The matter line is thus a single matrix element, whose branch point at $O$ is the source. As a matrix element of the modular double, and not only of $U_q(\mathfrak{sl}(2,\mathbb R))$, $G^{\rm bb}_\Delta$ solves both \eqref{eq:corr_neutral} and its dual under $b\to1/b$, with shifts $\Lambda\to\Lambda\pm2\pi i$. The first equation determines its solutions only up to factors periodic under $\Lambda\to\Lambda+2\pi ib^2$, and the second removes them. Together with the decay at the boundary and the source at $O$, this fixes the propagator uniquely.} It is pleasing that the matter line, like each gravitational propagator \eqref{eq:corr_qprop}, is a $b$-hypergeometric function with four double sines in its Mellin kernel. Let us now check some important properties of the solution just found.
Using the property \eqref{eq:app_Sbclassical} where the $S_b$ function reduces to a standard $\Gamma$ function in the $b \rightarrow 0$ limit, one can check that \eqref{eq:corr_gbb} reduces to the Mellin representation \eqref{eq:corr_GJT} of the AdS$_2$ propagator in the JT limit\rev{, consistently with the fact that the latter is the matrix element $K_{++}$ at $\frac12+is=\Delta$}. 

Another important property is that it decays with unit coefficient as $e^{-\Delta \Lambda}$ at large distance as long as $b\Delta<\frac Q2$.
In order to show this, as $\Lambda \to+\infty$ one can close the contour in \eqref{eq:corr_gbb} to the left, picking up the poles of $g^{\rm bb}_\Delta(k)$ located at:\footnote{The double sine has poles at $x=-mb-nb^{-1}$ and zeros at $x=Q+mb+nb^{-1}$, with $m,n\ge0$. The factors $S_b(-bk)^2$ only have poles at $k\ge0$, to the right of the contour.}
\begin{equation}
    k=-\Delta-m-\frac{n}{b^2}, \qquad k=-\Big(\frac Qb-\Delta\Big)-m-\frac{n}{b^2}.
\end{equation}
The propagator therefore has the asymptotic expansion
\begin{equation}\label{eq:tower}
    G^{\rm bb}_\Delta(\Lambda)=\sum_{m,n\ge0}c_{mn}\,e^{-(\Delta+m+n/b^2)\Lambda}+\sum_{m,n\ge0}\tilde c_{mn}\,e^{-(Q/b-\Delta+m+n/b^2)\Lambda} .
\end{equation}
The first tower contains the classical descendants $e^{-(\Delta+m)\Lambda}$ of the JT propagator, together with their duals under $b\to1/b$. The second tower is generated by the reflected weight $Q/b-\Delta$, and is pushed to infinity in the JT limit. The leading term is $e^{-\Delta \Lambda}$ provided $b\Delta<\frac Q2$ and the coefficient follows from the residue at $k=-\Delta$.\footnote{Near $k=-\Delta$, $S_b(b\Delta+bk)\simeq1/(2\pi b(k+\Delta))$ so that $\mathrm{Res}_{k=-\Delta}\,g^{\rm bb}_\Delta(k)=2\pi b\,\frac{S_b(2b\Delta)}{S_b(b\Delta)^2}\cdot\frac{1}{2\pi b}\cdot\frac{S_b(b\Delta)^2}{S_b(2b\Delta)}=1$.}
Therefore we conclude that
\begin{equation}\label{eq:explimit}
    G^{\rm bb}_\Delta(\Lambda)\simeq e^{-\Delta \Lambda} ,\qquad \Lambda \gg 1 ,
\end{equation}
i.e.\ the required decay of the propagator.

Let us now check instead the form of the propagator close to $O$. The point $O$ is at $\Lambda\to-\infty$ and therefore we can close the contour in \eqref{eq:corr_gbb} to the right, where the dominant contribution comes from the double pole at $k=0$. Its residue is linear in $\Lambda$ and yields \rev{$G^{\rm bb}_\Delta\simeq-c_\Delta\,\Lambda+\text{const}$, with $c_\Delta=\frac{S_{b}(2b\Delta)}{2\pi b\,S_{b}(b\Delta)^2}$, which tends to $\frac{\Gamma(2\Delta)}{\Gamma(\Delta)^2}$ in the JT limit,} and since $e^\Lambda=\sinh^2\frac L2\simeq L^2/4$, this is a logarithm of the geodesic distance from $O$. In two dimensions $\log L$ is the Green's function of the Laplacian, $\nabla^2\log L=2\pi\,\delta_O$, so this is consistent with a $\delta$-function source in $O$.\footnote{\rev{In the JT limit $(\nabla^2-m^2)G_\Delta=-4\pi c_\Delta\,\delta_O$. The factor $4\pi c_\Delta$ reflects the normalization \eqref{eq:explimit} at infinity instead of a unit source, and only rescales the boundary operator.}} 

\rev{In the JT limit this is the matter insertion used for the two-point function of finite-cutoff JT gravity in~\cite{Griguolo:2026eoi}, where the full AdS$_2$ propagator \eqref{eq:corr_GJT} is inserted between the Pöschl--Teller eigenfunctions. Its structure is transparent when it is} expanded in powers of $\mathrm{sech}^2\frac L2$,
\begin{equation}\label{eq:desce}
    G^{\rm JT}_\Delta=\sum_{n\ge0}\frac{(\Delta)_n^2}{n!\,(2\Delta)_n}\,\cosh\Big(\frac L2\Big)^{-2(\Delta+n)} .
\end{equation}
\rev{The leading term, $\cosh(L/2)^{-2\Delta}$, behaves like a primary insertion, while the subleading terms have weights $\Delta+n$, like descendants. At large separation, which is the regime that dominates at large cutoff, the primary alone reproduces the two-point function, while at finite cutoff the descendants contribute as well~\cite{Griguolo:2025kpi}. The propagator \eqref{eq:corr_gbb} is the $b$-deformation of the full insertion, and this structure of a primary with its descendants survives at finite $b$, as we show in Section~\ref{sec:discussion}.}

\subsection{Exact two-point function on the quantum disk}
\label{subsec:corr_exact}

The last ingredient is the integral over the position of the second insertion, which on the quantum disk becomes the invariant integral. On our rotationally symmetric configurations it reduces to the radial integral over $\Lambda$, with the flat measure of the symmetric ordering. Putting everything together, the two-point function on the quantum disk reads
\begin{equation}\label{eq:corr_qWitten}
    \langle\mathcal O_\Delta(\tau)\mathcal O_\Delta(0)\rangle_\beta=\frac{1}{Z_b(\beta)}\int_{-\infty}^{\infty}\dd\Lambda\;\overline{K_b(\Lambda;\beta-\tau)}\;G^{\rm bb}_\Delta(\rev{\Lambda})\;K_b(\Lambda;\tau).
\end{equation}
It has the same reading as in JT gravity: the boundary particle propagates on the quantum disk from $O$ to radial position $\Lambda$ in time $\tau$ and back in time $\beta-\tau$, while the matter field propagates from $O$ to $\Lambda$ through the quantum bulk and the radial position of the second insertion is integrated over. Inserting the spectral representation \eqref{eq:corr_qprop} of the two propagators, the two-point function takes the form
\begin{equation}\label{eq:corr_spectral}
    \langle\mathcal O_\Delta(\tau)\mathcal O_\Delta(0)\rangle_\beta=\frac{1}{Z_b(\beta)}\int_0^{\rev{\nu}}\dd s_1\,\dd s_2\;\rho_b(s_1)\,\rho_b(s_2)\;e^{-(\beta-\tau)E_b(s_1)-\tau E_b(s_2)}\;M_b(s_1,s_2),
\end{equation}
where
\begin{equation}\label{eq:corr_Mbdef}
    M_b(s_1,s_2)=\int_{-\infty}^{\infty}\dd\Lambda\;\overline{\psi_{s_1}(\Lambda)}\;G^{\rm bb}_\Delta(\rev{\Lambda})\;\psi_{s_2}(\Lambda)
\end{equation}
is the matrix element of the matter line \rev{\eqref{eq:corr_gbb}} between \rev{the} energy eigenstates \rev{\eqref{eq:corr_qprop}}, which we now evaluate.

The computation is most conveniently performed in Mellin space, since each of the three lines of the diagram is a Mellin integral.
We denote by
\begin{equation}
    A_{s,\nu}(t)=\frac{S_b(bt\pm ibs)\,S_b(\frac b2+ib\nu-bt)}{S_b(\frac b2-ib\nu+bt)}
\end{equation}
the Mellin kernel of the eigenfunctions in \eqref{eq:corr_qprop}. On a vertical contour, complex conjugation flips the sign of the field in the kernel, so the bra contains $A_{s_1,-\nu}(t')$.\footnote{For real $b$ the double sine is real on the real axis, so $\overline{S_b(x)}=S_b(\bar x)$.} The matter line is $G^{\rm bb}_\Delta(\rev{\Lambda})=\int\frac{\dd k}{2\pi i}\,e^{\rev{\Lambda k}}g^{\rm bb}_\Delta(k)$ by \eqref{eq:corr_gbb}.

We now focus on the radial integral over $\Lambda$. The Faddeev phases of bra and ket are complex conjugate and cancel $\overline{\Phi_\nu}\Phi_\nu=1$, so the only dependence on $\Lambda$ left is $e^{(t'+t+k)\Lambda}$.\footnote{For real $b$ Faddeev's quantum dilogarithm satisfies $\overline{\phi_b(x)}=1/\phi_b(\bar x)$.} On imaginary contours the radial integral 
\begin{equation}
\int\dd\Lambda\,e^{(t'+t+k)\Lambda}=2\pi i\,\delta(t'+t+k)
\end{equation}
yields a delta function which conserves the Mellin momenta of the three lines at the insertion point. \rev{On its support the exponentials of the field in the two eigenfunctions become $e^{-2\pi b^2\nu(t'+t)}=e^{2\pi b^2\nu k}$, which multiplies the Mellin kernel of the matter line.} We thus obtain\footnote{With contours $0<c<\frac12$, $-\Delta<c_k<0$ and $0<-c-c_k<\frac12$.}
\begin{equation}\label{eq:corr_Mb}
    M_b(s_1,s_2)=\overline{N_{s_1}}N_{s_2}\int_{c_k+i\mathbb R}\frac{\dd k}{2\pi i}\,\rev{e^{2\pi b^2\nu k}}\,g^{\rm bb}_\Delta(k)\;\mathcal I(k),\qquad
    \mathcal I(k)=\int_{c+i\mathbb R}\frac{\dd t}{2\pi i}\,A_{s_1,-\nu}(-t-k)\,A_{s_2,\nu}(t).
\end{equation}
At fixed Mellin momentum $k$ of the matter line, the inner integral has a remarkable structure. Using the reflection relation $S_b(x)S_b(Q-x)=1$ to bring all double sines to the numerator, its integrand contains eight of them, four depending on $t$ through $+bt$ with arguments 
\begin{equation}
\pm ibs_2 ,\qquad \frac b2-ib\nu+bk ,\qquad Q-\frac b2-ib\nu+bk ,
\end{equation}
and four through $-bt$ with arguments 
\begin{equation}
\frac b2+ib\nu ,\qquad Q-\frac b2+ib\nu ,\qquad -bk\pm ibs_1 .
\end{equation}
These arguments add up to $2Q$, which is exactly the balancing condition of the integral representation \eqref{eq:corr_TV} of the $b$-$6j$ symbol of the modular double of $U_q(\mathfrak{sl}(2,\mathbb R))$~\cite{Ponsot:2001,Teschner:2012}. To compare with it we change the integration variable to
\begin{equation}\label{eq:corr_ushift}
    u=bt+Q+\tfrac b2-ib\nu ,
\end{equation}
so that the four factors with $+bt$ take the form $S_b(u-U_k)$ and the four with $-bt$ the form $S_b(V_k-u)$ and the vertical contour in $t$ becomes a vertical contour in $u$ that separates the two families of poles. One can easily check that with the following choice of labels
\begin{equation}\label{eq:corr_labels}
\begin{aligned}
    &\alpha_1=\tfrac b2+i \frac{b}{2}(s_1-s_2),\qquad \alpha_2=\tfrac Q2-i\frac{b}{2}(s_1+s_2),\qquad \alpha_s=\tfrac Q2-ib\nu,\\
    &\alpha_3=\tfrac Q2+i \frac{b}{2}(s_1+s_2),\qquad \alpha_4=\tfrac b2-i\frac{b}{2}(s_1-s_2),\qquad \alpha_t=-bk ,
\end{aligned}
\end{equation}
the integrand for $\mathcal I(k)$ is identical to the integrand of \eqref{eq:corr_TV} so that
\begin{equation}\label{eq:corr_I6j}
    \mathcal I(k)=\frac{1}{2\pi i\,b\;D(\underline\alpha)}\Big\{\begin{matrix}\alpha_1&\alpha_2&\alpha_s\\[2pt] \alpha_3&\alpha_4&\alpha_t\end{matrix}\Big\}_b,
\end{equation}
where $D$ is the product of triangle factors that normalizes the symbol. As a consequence, the matrix element of the matter line at any value of the magnetic field is a single integral of $b$-$6j$ symbols against the Mellin kernel of the matter line,
\begin{equation}\label{eq:corr_Mb6j}
    M_b(s_1,s_2)=\frac{\overline{N_{s_1}}N_{s_2}}{2\pi i\,b}\int_{c_k+i\mathbb R}\frac{\dd k}{2\pi i}\;\rev{e^{2\pi b^2\nu k}}\,\frac{g^{\rm bb}_\Delta(k)}{D(\underline\alpha(k))}\,\Big\{\begin{matrix}\frac b2+i\frac{b}{2}(s_1-s_2)&\frac Q2-i\frac{b}{2}(s_1+s_2)&\frac Q2-ib\nu\\[2pt] \frac Q2+i \frac{b}{2}(s_1+s_2)&\frac b2-i\frac{b}{2}(s_1-s_2)&-bk\end{matrix}\Big\}_b .
\end{equation}
The labels of the $b$-$6j$ symbol have a transparent meaning. One edge carries the boundary dilaton, i.e.\ the magnetic field, as a principal-series label. The opposite edge is the matter line, with its Mellin momentum $k$, and it is integrated against the matter kernel. The remaining four edges carry the two energies through their sum and difference.

The full two-point function on the quantum disk, at any value of the magnetic field, is thus given by \eqref{eq:corr_spectral} with $M_{b}$ given in \eqref{eq:corr_Mb6j}.

In the JT limit $b\to0$, at fixed $s_i$, $\nu$ and $\Delta$, the double sines reduce to Gamma functions by \eqref{eq:app_Sbclassical} and \eqref{eq:corr_Mb6j} becomes a double Mellin--Barnes integral of Gamma functions. However, its most transparent form follows from the expansion \eqref{eq:desce}\rev{: the matrix element of each term between JT eigenstates was computed in closed form in~\cite{Griguolo:2026eoi}}, using an integral of two hypergeometric functions evaluated in~\cite{Groenevelt:2003}. Up to the normalization of the eigenfunctions, the JT limit of $M_b$ is therefore a sum of Wilson functions with increasing weight,
\begin{equation}\label{eq:corr_MJT}
    M_{\rm JT}(s_1,s_2)=\sum_{n\ge0}\frac{(\Delta)_n^2}{n!\,(2\Delta)_n}\;\Gamma(\Delta+n\pm is_1\pm is_2)\;\mathcal W\big(s_2,\nu;\tfrac12+is_1,\tfrac12-is_1,\Delta+n-i\nu,\Delta+n+i\nu\big),
\end{equation}
where $\mathcal W(\lambda,x;a,b,c,d)=\phi_\lambda(x;a,b,c,1-d)$ is the Wilson function~\cite{Groenevelt:2003}, \rev{the $SL(2,\mathbb R)$ analogue of a $6j$ symbol,} in the conventions of~\cite{Mertens:2017mtv,Griguolo:2026eoi}. \rev{This is the two-point function of finite-cutoff JT gravity of~\cite{Griguolo:2026eoi}.}

\subsection{Large loops in the strong field regime: Liouville gravity and DSSYK}
\label{subsec:corr_limits}
\label{subsec:qd_loops}

We now discuss the regime of strong magnetic field, which connects the quantum disk with Liouville gravity and DSSYK. The boundary of the disk is a closed classical orbit of the charged particle, a cyclotron circle around the horizon. In the JT limit its diameter is $L_*$ of \eqref{eq:mag_Lstar}, and at low energy compared with the cutoff, or equivalently at strong field and fixed energy, the orbits become large,
\begin{equation}
    L_*\simeq2\log(2\nu)-2\log s\qquad(s\ll\nu),
\end{equation}
and hug the conformal boundary. Instead, near the edge of the spectrum the orbits shrink towards $O$, deep in the bulk.

On the quantum disk the size of the orbit is given by the semiclassical analysis we will perform in Section~\ref{subsec:corr_semiclassical}, where the semiclassical picture is still described in terms of Larmor orbits of the magnetic particle, now in the Liouville frame of sinh dilaton gravity, i.e.\ the AdS$_2$ frame $g_-=e^{2\pi b^2\Phi}g$. \rev{There the boundary is a circle around the horizon, whose diameter is the length of the bridge at $t_L+t_R=0$, $\Lambda_*=\log(\rho_b^2/\rho_h^2-1)$ by \eqref{eq:sinh_ERB}. At strong field it sits at}
\begin{equation}\label{eq:corr_Lambdastar}
    \Lambda_*\simeq4\pi b^2\nu-2\log\sinh(2\pi b^2s) ,
\end{equation}
\rev{where we used the expressions \eqref{eq:mappings} of $\rho_b$ and $\rho_h$, with $\Phi_h\to s$ and $\Phi_b\to\nu$.} The orbit therefore moves away from $O$ linearly in the field, rather than logarithmically as in JT gravity, while its position relative to the scale $4\pi b^2\nu$ stays finite. This is the origin of the renormalization: the natural coordinate at strong field is the renormalized coordinate $\Lambda_{\rm ren}=\Lambda-4\pi b^2\nu$ of \eqref{eq:renorm}, and the little $q$-Jacobi variable of \eqref{eq:qd_identity} is precisely $X=q\,e^{\Lambda_{\rm ren}}$. At fixed renormalized coordinate and strong field, the magnetized quantum disk Laplacian \eqref{eq:qd_H} derived in Section \ref{sec:qdisk} becomes the Schrödinger equation of Liouville gravity \eqref{eq:sinh_liouville}.

The two-point diagram \eqref{eq:corr_spectral} follows the same pattern and at strong field it lives near the boundary of the quantum disk. At fixed renormalized coordinate the propagators \rev{\eqref{eq:corr_qprop}} become $q$-Bessel functions~\cite{Fan:2021bwt},
\begin{equation}
    \hat\chi_s(\Lambda_{\rm ren})=2\pi b\,e^{-i\pi b^2s^2}\int_{c+i\mathbb R}\frac{\dd t}{2\pi i}\,e^{t\Lambda_{\rm ren}-i\pi b^2t^2}S_b(bt\pm ibs),
\end{equation}
which solve the Liouville gravity equation \eqref{eq:sinh_liouville} with the Liouville gravity density. Moreover the full matter line propagator $G^{\rm bb}_\Delta(\rev{\Lambda})$ reduces to the pure exponential \eqref{eq:explimit}, since after renormalization all terms appearing in \eqref{eq:tower} are subdominant compared to it in the strong field regime\rev{. In terms of $\Lambda_{\rm ren}$ it reads $e^{-\Delta\Lambda}=e^{-4\pi b^2\nu\Delta}\,e^{-\Delta\Lambda_{\rm ren}}$, and the constant $e^{-4\pi b^2\nu\Delta}$ is the cutoff-dependent factor of the two boundary operators, which we absorb into their normalization}.  The radial integral again conserves the Mellin momentum and the remaining integral is the degenerate hyperbolic Barnes lemma \eqref{eq:corr_barnes2}. The result is
\begin{equation}\label{eq:corr_ML}
    M_L(s_1,s_2)=\int\dd\Lambda_{\rm ren}\;\overline{\hat\chi_{s_1}}\,e^{-\Delta\Lambda_{\rm ren}}\,\hat\chi_{s_2}=2\pi b\;\frac{S_b(b\Delta\pm ibs_1\pm ibs_2)}{S_b(2b\Delta)},
\end{equation}
\rev{which is the exact strong-field limit of $M_{b}$ in \eqref{eq:corr_Mb6j}.}

Moreover, at strong field and at leading order, the energy \rev{\eqref{eq:sinh_energy}} becomes \rev{$E_b\simeq e^{-\pi b^2\nu}\big(\cosh(2\pi b^2s)-1\big)/2\pi b^2$}\rev{, where the $s$-independent term cancels between the numerator and $Z_L$ in the normalized two-point function}. Defining the renormalized times $\beta_L=e^{-\pi b^2\nu}\beta$ and $\tau_L=e^{-\pi b^2\nu}\tau$,\footnote{This corresponds to relating the Tolman temperature to the Hawking temperature of the sinh dilaton gravity black hole, and performing the same rescaling for $\tau$.} the two-point function \eqref{eq:corr_spectral} becomes
\begin{equation}\label{eq:corr_Lfull}
\begin{aligned}
    \langle\mathcal O_\Delta(\tau)\mathcal O_\Delta(0)\rangle_\beta\simeq\frac{2\pi b}{Z_L(\beta_L)}\int_0^\infty\dd s_1\,\dd s_2\;&\rho_L(s_1)\,\rho_L(s_2)\;e^{-(\beta_L-\tau_L)\frac{\cosh(2\pi b^2s_1)}{2\pi b^2}-\tau_L\frac{\cosh(2\pi b^2s_2)}{2\pi b^2}}\\
    &\times\frac{S_b(b\Delta\pm ibs_1\pm ibs_2)}{S_b(2b\Delta)},
\end{aligned}
\end{equation}
with $\rho_L(s)=\sinh(2\pi b^2s)\sinh(2\pi s)$, up to a normalization of the operator. This is the two-point function of Liouville gravity~\cite{Mertens:2020hbs,Fan:2021bwt}, now obtained as the Witten diagram of Figure~\ref{fig:corr_witten} near the boundary of the quantum disk. This can thus be regarded as a new independent derivation of \eqref{eq:corr_Lfull}.

In the strong field regime, Liouville gravity has a natural continuation to sine dilaton gravity \cite{Blommaert:2024whf}.
Indeed continuing $b^2=i\lambda/2\pi$ with $\lambda>0$ the dilaton potential becomes $2\sin(\lambda\Phi)/\lambda$ and the shifts become real translations by $\lambda$. The equation of Liouville gravity then closes on the lattice $\Lambda_{\rm ren}=\lambda(n+\frac12)+i\pi$ and in terms of $\chi_n\equiv\chi(\lambda(n+\frac12)+i\pi)$ it becomes
\begin{equation}\label{eq:qd_dssyk}
    \chi_{n+1}+\big(1-q^{2n}\big)\chi_{n-1}=2\cos\theta\;\chi_n,\qquad q^2=e^{-\lambda},\quad\theta=\lambda s ,
\end{equation}
which is the transfer matrix of DSSYK in the chord basis \cite{Berkooz:2018jqr,Berkooz:2024lgq}, solved by continuous $q$-Hermite polynomials. The chord number is now identified with the radial level of the quantum disk and the lattice lies on the line $\mathrm{Im}\,\Lambda_{\rm ren}=\pi$, which is precisely the same shift by $i\pi$ as in the relabelling \eqref{eq:qd_swapvar}. Notice that the term $1-q^{2n}$ is the remnant, at strong field, of the field-dressed hopping of the centre term of the quantum disk Laplacian in \eqref{eq:qd_qsquaresB}. The $q$-Hermite polynomials can also be read off directly from the eigenfunction \eqref{eq:sinh_eigenfunction}, as we show in Appendix \ref{app:realq}. Continued to real $q$, the wavefunction \eqref{eq:sinh_eigenfunction} develops simple poles on the lattice of the quantum disk and its residues are Al-Salam--Chihara polynomials, which reduce to the continuous $q$-Hermite polynomials in the DSSYK limit (see Appendix \ref{app:realq}). \rev{This suggests that finite-cutoff sine dilaton gravity, which we do not develop here, is described by Al-Salam--Chihara polynomials on the lattice of the quantum disk at real $q$, such as \eqref{eq:rq_res}, defined by the recurrence \eqref{eq:rq_rec}. We return to this point, and to which parameters are realized, in Section~\ref{sec:discussion}.}

For the two-point diagram, on the lattice of \eqref{eq:qd_dssyk} the matter line becomes $q^{2\Delta n}$, the propagators become $q$-Hermite chord wavefunctions and the diagram becomes the \rev{two-point function of DSSYK, the real-$q$ form of \eqref{eq:corr_ML}.}\footnote{\rev{The matrix element of the matter line follows from the $q$-Mehler formula for the continuous $q$-Hermite polynomials~\cite{Koekoek:2010},}
\begin{equation}
    \sum_{n\ge0}\frac{H_n(\cos\theta_1|q^2)H_n(\cos\theta_2|q^2)}{(q^2;q^2)_n}\,q^{2\Delta n}=\frac{(q^{4\Delta};q^2)_\infty}{\prod_{\pm\pm}(q^{2\Delta} e^{i(\pm\theta_1\pm\theta_2)};q^2)_\infty},
\end{equation}
\rev{whose right-hand side is the matter matrix element of DSSYK~\cite{Berkooz:2018jqr}.}}

The physical picture that emerges is summarized in Figure \ref{fig:qd_web}. Finite-cutoff sinh dilaton gravity is a charged particle on the quantum hyperbolic disk for loops of all sizes. For large loops, at energies small compared with the cutoff, the particle stays close to the boundary of the disk and the dynamics reduces to Liouville gravity, to the Schwarzian theory in the JT limit and to DSSYK after continuation. For small loops, deep in the bulk, the dynamics is governed by the full little $q$-Jacobi problem, which is captured neither by Liouville gravity nor by DSSYK. The quantum disk is the description that holds at all energy scales.

\subsection{Localizing on the semiclassical length: the effective Larmor orbit}
\label{subsec:corr_semiclassical}

The exact two-point function \eqref{eq:corr_qWitten} has a transparent semiclassical limit. In the units of Section \ref{sec:sinh}, the effective Planck constant is $\varepsilon\equiv\hbar_{\rm eff}=2\pi b^2$ of~\eqref{eq:sinh_hbar}, and the semiclassical limit is
\begin{equation}
    \varepsilon\to0,\qquad \varphi_h\equiv\varepsilon s,\quad \varphi_b\equiv\varepsilon\nu\quad\text{fixed},
\end{equation}
i.e.\ at fixed horizon and boundary values of the dilaton and fixed $\Delta$, so that the matter field is a probe. In this limit the energy \eqref{eq:sinh_energy} and the density \eqref{eq:sinh_density} become
\begin{equation}
    E_b=\frac{e(\varphi_h)}{\varepsilon},\qquad e(\varphi)=\rev{\sqrt{2\cosh\varphi_b-2}-\sqrt{2\cosh\varphi_b-2\cosh\varphi}},\qquad \rho_b\simeq e^{\frac{2\pi}\varepsilon(\varphi_h-\varphi_b)} .
\end{equation}
The partition function $Z_b=\int\dd s\,\rho_b\,e^{-\beta E_b}$ is dominated by the stationary point of $2\pi(\varphi_h-\varphi_b)-\beta e(\varphi_h)$, i.e.
\begin{equation}\label{eq:corr_scbeta}
    \beta=\frac{2\pi\sqrt{2(\cosh\varphi_b-\cosh\varphi_h)}}{\sinh\varphi_h},\qquad \log Z_b\simeq\frac1\varepsilon\Big[2\pi(\varphi_h-\varphi_b)-\beta e(\varphi_h)\Big].
\end{equation}
The first relation is the inverse Tolman temperature of the black hole \eqref{eq:sinh_bh}, i.e.\ the period $4\pi/f'(\Phi_h)$ of its Euclidean time multiplied by the redshift factor $\sqrt{f(\Phi_b)}$ at the boundary. 

At fixed $\Delta$ the matter line becomes classical as well: the double sine functions in its kernel \eqref{eq:corr_gbb} become Gamma functions, and $G^{\rm bb}_\Delta$ becomes the AdS$_2$ propagator \eqref{eq:corr_GJT}. It is a smooth function of $\Lambda$, independent of $\varepsilon$, so it does not move the saddle point and the matter field is a probe.
In the two-point function \eqref{eq:corr_spectral} we first perform the two energy integrals of the boundary propagators by saddle point. As we show in Appendix \ref{app:semiclassical}, the remaining integral over the length then localizes on a single value:
\begin{equation}\label{eq:corr_scLE}
    \sinh\frac{L_E}2=\sqrt{\frac{\rho_b^2}{\rho_h^2}-1}\;\sin\frac{\pi \tau}{\beta},\qquad \frac{\rho_b}{\rho_h}=\frac{e^{\varphi_b}-\cosh\varphi_h}{\sinh\varphi_h},
\end{equation}
with $\beta$ given by the saddle \eqref{eq:corr_scbeta}. This is exactly the length of the Einstein--Rosen bridge \rev{\eqref{eq:sinh_ERB}} continued to Euclidean time, $\rho_{h} \left(t_{L}+t_{R}\right)\rightarrow i \left(\frac{2\pi \tau}{\beta}-\pi\right)$, since the Schwarzschild time of the
AdS$_2$ frame has period $2\pi /\rho_h$ while the boundary proper time has period $\beta$. The two-point function is then the matter line evaluated on this length,
\begin{equation}\label{eq:corr_scresult}
    \big\langle\mathcal O_\Delta(\tau)\mathcal O_\Delta(0)\big\rangle_\beta\simeq G^{\rm JT}_\Delta\big(L_{E}\big),
\end{equation}
where $G^{\rm JT}_\Delta$ is the AdS$_2$ propagator \eqref{eq:corr_GJT}.

We now investigate the meaning of this localization on the quantum disk. In the semiclassical limit the quantum disk becomes the classical hyperbolic disk and we have argued that the latter corresponds to the AdS$_2$ frame $g_-=e^{2\pi b^2\Phi}g$ of sinh dilaton gravity. As we saw in JT gravity in Section \ref{sec:ttbar}, the boundary is a circle around the Euclidean horizon $H$, which is the centre of the orbit. We now show that the semiclassical orbit is still a circle around the black hole horizon in the AdS$_2$ frame $g_-=e^{2\pi b^2\Phi}g$ and fix the semiclassical Euclidean length just in terms of a simple kinematic relation based on hyperbolic geometry and the gravitational boundary conditions, before comparing it with the result of the saddle point \eqref{eq:corr_scLE}.

\begin{figure}[t]
\centering
\includegraphics[width=0.8\textwidth]{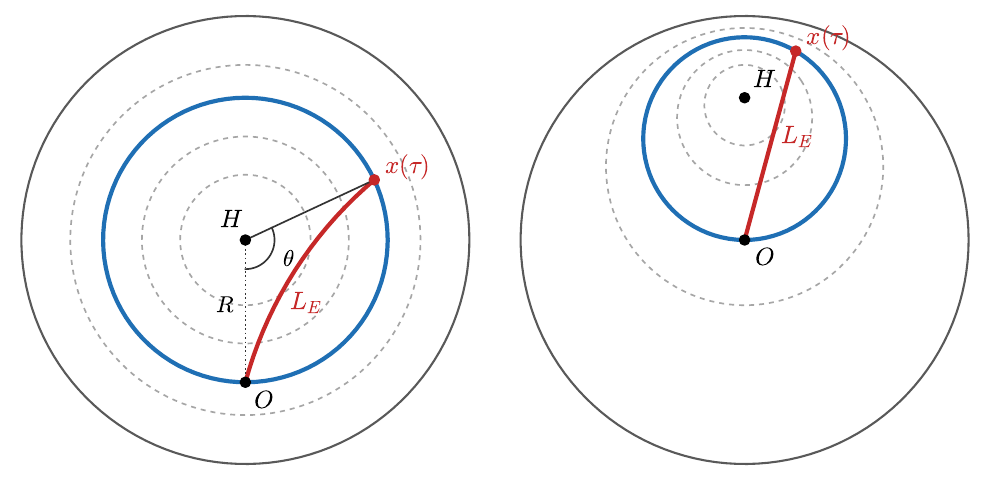}
\caption{The semiclassical Euclidean orbit on the Poincar\'e disk. \rev{Left:} the orbit (blue) is a circle of radius $R$ around the horizon $H$, which is the guiding centre; the dashed circles are the level sets of the dilaton. The particle starts at $O$ and, after a time $\tau$, is at the angle $\theta=2\pi\tau/\beta$. The Euclidean length $L_E$ (red) is the chord from $O$, and the diameter is $L_*=2R$. \rev{Right:} the same configuration with $O$ at the centre of the disk, as in Figure~\ref{fig:corr_witten}: $L_E$ is the radial position of the second insertion.}
\label{fig:corr_orbit}
\end{figure}

The kinematic configuration is displayed on the left side of Figure \ref{fig:corr_orbit}: the particle starts at $O$ on the boundary circle and, after a time $\tau$, has rotated by an angle $\theta$ around $H$. The Euclidean length is thus the chord from $O$ to its new position. It is the third side of the geodesic triangle with vertices $H$, $O$ and $x(\tau)$, whose other two sides are radii of length $R$ meeting at the angle $\theta$. The hyperbolic law of cosines gives
\begin{equation}\label{eq:corr_cosines}
    \cosh L_E=\cosh^2R-\sinh^2R\,\cos\theta ,
\end{equation}
and, using $\cosh L=1+2\sinh^2\frac L2$ and $\cosh^2R=1+\sinh^2R$,
\begin{equation}\label{eq:corr_chord}
    \sinh\frac{L_E}2=\sinh R\,\sin\frac\theta2 .
\end{equation}
\rev{This is the configuration of Figure~\ref{fig:corr_witten}b, seen from a different point:} if we move $O$ to the centre of the disk by an isometry, $L_E$ becomes the radial position of the second insertion. 

It remains to fix $\theta$ and $R$ in terms of the boundary data $\beta$ and $\Phi_b$. Both follow from the fact that the dilaton is constant on the boundary, so that the Weyl factor between the physical metric $g$ and the disk metric $g_-$ is constant there as well. First, the disk length of a boundary segment is a rigid rescaling of its proper length. Equal proper times therefore correspond to equal angles, the motion around $H$ is uniform and a full turn takes the time $\beta$, so that $\theta=2\pi\tau/\beta$. Second, the circumference $2\pi\sinh R$ of the orbit is the total length of the boundary on the disk, $\ell_-=e^{\pi b^2\Phi_b}\beta=e^{\varphi_b/2}\beta$, where $\beta$ is the boundary length in the original geometry of sinh dilaton gravity, i.e.\ its inverse Tolman temperature. Hence
\begin{equation}\label{eq:corr_scR}
    2\pi\sinh R=e^{\varphi_b/2}\beta,\qquad \sinh\frac{L_E}2=e^{\varphi_b/2}\,\frac\beta{2\pi}\,\sin\frac{\pi\tau}\beta ,
\end{equation}
with $\beta$ related to the horizon by the saddle \eqref{eq:corr_scbeta}. We can now compare with the semiclassical saddle \eqref{eq:corr_scLE} of the two-point function.
We find a perfect match between the two expressions thanks to the following manipulation:
\begin{equation}
    \frac{\rho_b^2-\rho_h^2}{\rho_h^2}=\frac{2e^{\varphi_b}(\cosh\varphi_b-\cosh\varphi_h)}{\sinh^2\varphi_h}=e^{\varphi_b}\Big(\frac\beta{2\pi}\Big)^2=\sinh^2R,
\end{equation}
where we finally used \eqref{eq:corr_scbeta}.

In JT gravity, the radius also fixes the momentum of the particle through the Larmor relation \rev{\eqref{eq:tt_larmor}}, and this momentum is the tension conjugate to the boundary length, as we saw in Section \ref{sec:ttbar}. In sinh dilaton gravity the type of orbit is the same, a circle around the horizon whose circumference is fixed by the boundary condition but, semiclassically, the momentum with which the particle runs along it is not the one given by the Larmor relation \eqref{eq:tt_larmor} with $B=\Phi_b$. The reason is that the boundary action on the disk contains, besides the flux, the terms produced by the Weyl rescaling and a non-vanishing bulk action, as we had already observed at the beginning of Section \ref{sec:qdisk}. We show in Appendix \ref{app:larmor} that the full on-shell action takes the form of a relativistic particle with a non-linear flux,
\begin{equation}\label{eq:corr_Inl}
    I=\mathcal B\,\ell_--G(A_-),\qquad \mathcal B=\rev{\frac{1-e^{-\varphi_b}}{\varepsilon}},
\end{equation}
where $A_-$ is the area enclosed by the orbit on the $g_{-}$ disk geometry and the counterterm gives $\mathcal B$\rev{, which reduces to $\Phi_b$ in the JT limit}. The tension conjugate to $\ell_-$ is the normal derivative of the dilaton in the disk metric, $P_-=\partial_{n_-}\Phi$, and we show in Appendix \ref{app:larmor} that it obeys a deformed Larmor relation,
\begin{equation}\label{eq:corr_deformedlarmor}
    P_-=B_{\rm eff}\tanh R,\qquad B_{\rm eff}=G'(A_-)=e^{-\varphi_b}\rho_b=\frac{1-e^{-\varphi_b}\cosh\varphi_h}{\varepsilon},
\end{equation}
in terms of an effective magnetic field $B_{\rm eff}$.
Up to the constant factor $e^{-\varphi_b}=\partial\Phi/\partial\rho$ at the boundary, this is the JT relation \rev{\eqref{eq:tt_larmor}} with the dilaton replaced by the radial coordinate of the AdS$_2$ frame, i.e.
\begin{equation}\label{eq:deformedlarmor}
\tanh R=\partial_{n_-}\rho/\rho_b .
\end{equation}
Comparing with \eqref{eq:tt_larmor} and \eqref{eq:tt_dn}, the on-shell geometry of sinh dilaton gravity is still that of a Larmor orbit, once the AdS$_2$ quantities are replaced by those of the effective AdS$_2$ $g_{-}$, i.e.\ $\left(\Phi, \Phi_{b}\right)\rightarrow \left(\rho,\rho_b\right)$. We find it remarkable that the semiclassical solution organizes in this way: it is a non-trivial manifestation, already at the classical level, of the description as a particle in a magnetic field, provided one uses the correct hyperbolic disk.
The effective field depends on the state and reduces to $\Phi_b$ only in the JT limit. Therefore, it should be regarded as a semiclassical reorganization of the result rather than as a coupling: the field that enters the $q$-difference operator is $\Phi_b$, as we showed in Section \ref{sec:qdisk}, and the dependence on $\varphi_h$ arises only on shell. 

\section{Non-perturbative structure away from the strong-field regime}
\label{sec:nonpert}

The observables computed in the previous sections contain corrections that vanish when the cutoff is removed, i.e.\ at strong magnetic field. They are most clearly visible in the spectral density. The density of finite-cutoff Liouville gravity \eqref{eq:sinh_density} can be expanded as
\begin{equation}\label{eq:np_series_b}
    \rho_b(s)=\frac{\sinh(2\pi b^2s)}{\pi b^2}\,\frac{\sinh2\pi s}{\cosh2\pi\nu+\cosh2\pi s}=\frac{2\sinh(2\pi b^2s)}{\pi b^2}\sum_{n\ge1}(-1)^{n-1}e^{-2\pi n\nu}\sinh(2\pi ns).
\end{equation}
At strong field only the first term survives, and after removing the factor $e^{-2\pi\nu}$ one recovers the density of Liouville gravity, $\sinh(2\pi b^2s)\sinh(2\pi s)$. The terms with $n\ge2$ are exponentially suppressed in the boundary dilaton, and are non-perturbative from the point of view of the large-cutoff expansion. The dependence on the cutoff sits entirely in the factor $\sinh2\pi s/(\cosh2\pi\nu+\cosh2\pi s)$, which is identical in JT gravity.
The goal of this section is therefore to study the structure of these corrections in the JT limit, where the spectral problem is the ordinary differential equation \eqref{eq:sinh_PT} and the full machinery of exact WKB is available. In~\cite{Griguolo:2026eoi} the $n$-th term of the expansion was associated with a Euclidean configuration in which the length returns $n-1$ times to zero. Here we derive the complete non-perturbative structure of the density from the Schr\"odinger problem, \rev{and show that it is generated by exactly three cycles, whose quantum periods receive no corrections in the WKB expansion. This explains why the on-shell action of the corresponding Euclidean saddles captures the full exact result.} In Section~\ref{subsec:np_gravity} we interpret the results gravitationally as instantons in which the boundary winds around the horizons. In Section~\ref{subsec:zz} we extend the analysis to sinh dilaton gravity, adopting the CFT perspective of the string theory formulation of the theory \cite{Mertens:2020hbs}. The technical details are given in Appendix~\ref{app:wkb}. 

\subsection{Exact WKB and the spectral density as a connection problem}
\label{subsec:np_problem}

\begin{figure}[t]
\centering
\includegraphics{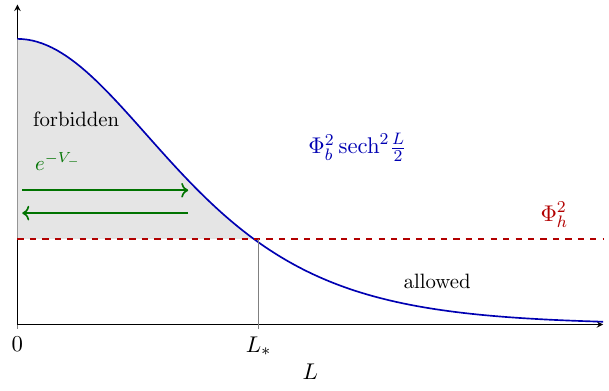}
\caption{The P\"oschl--Teller barrier for $\Phi_h<\Phi_b$. The origin, the centre of the hyperbolic plane, lies inside the classically forbidden region. A bounce is a round trip between the origin and the turning point $L_*$, with weight $e^{-V_-}$.}
\label{fig:wkb_barrier}
\end{figure}

To set up the semiclassical analysis we restore $\hbar$ in \eqref{eq:sinh_PT}, writing it as
\begin{equation}\label{eq:wkb_schr}
    -\hbar^2\psi''(L)+\Big[Q_0(L)+\hbar^2Q_2(L)\Big]\psi(L)=0,\qquad Q_0=\Phi_b^2\,\mathrm{sech}^2\frac L2-\Phi_h^2,\qquad Q_2=-\frac{1}{4\sinh^2L},
\end{equation}
with $\nu=\Phi_b/\hbar$ and $s=\Phi_h/\hbar$. The WKB expansion is an expansion in $\hbar$ at fixed $\Phi_b$ and $\Phi_h$, i.e.\ at large $\nu$ and $s$. Classically, $-Q_0$ is the square of the momentum. Since $\mathrm{sech}^2\frac L2$ decreases monotonically from one to zero, for horizons inside the boundary ($\Phi_h<\Phi_b$) there is a single turning point,
\begin{equation}\label{eq:wkb_turning}
    \cosh\frac{L_*}{2}=\frac{\Phi_b}{\Phi_h}=\frac\nu s ,
\end{equation}
which coincides with the diameter \eqref{eq:mag_Lstar} of the classical boundary circle. The region $0\le L<L_*$ is classically forbidden and the origin lies inside the barrier, see Figure \ref{fig:wkb_barrier}. However, as we saw in Section~\ref{sec:jt}, the origin is not a wall but the centre of the hyperbolic plane: writing $\psi=\sqrt{\sinh L}\,R$, the ordering term turns the problem into the radial Laplacian on the hyperbolic plane and a wave reaching $L=0$ passes through the centre rather than being reflected.

The physical solution is the one for which $R$ is regular at the centre,\footnote{One can easily check that near the origin \eqref{eq:wkb_schr} reduces to $-\psi''-\psi/4L^2+\kappa_0^2\psi=0$, with $\kappa_0^2=\nu^2-s^2$, whose solutions behave as $\sqrt L$ and $\sqrt L\log L$.} i.e.
\begin{equation}\label{eq:wkb_bc}
    \psi_{\rm reg}(L)\simeq\sqrt L\qquad(L\to0),
\end{equation}
while at large distance every solution oscillates, $\psi_{\rm reg}\simeq A_s\sin(sL+\delta_s)$. The normalization is decided at infinity, since the delta function in the overlap of two eigenfunctions comes entirely from the asymptotic region,
\begin{equation}
    \int_0^\infty\dd L\,A_s\sin(sL+\delta_s)\,A_{s'}\sin(s'L+\delta_{s'})=\frac{\pi}{2}A_s^2\,\delta(s-s')+(\text{regular}),
\end{equation}
while the value at the centre is fixed to one by \eqref{eq:wkb_bc}. Hence, with the normalization of~\cite{Griguolo:2026eoi},
\begin{equation}\label{eq:wkb_rho_def}
    \rho(s)=2\abs{R_s(0)}^2=\frac4\pi\,\frac{1}{A_s^2} .
\end{equation}
The density is the inverse square of the connection coefficient between the origin and infinity. The whole problem is therefore a connection problem: starting from the regular solution at the origin, we have to transport it through the barrier and out to infinity, and read off its amplitude there.

Exact WKB solves connection problems of this kind exactly, through the Borel resummation of the WKB series~\rev{\cite{Voros:1983,Silverstone:1985,Delabaere:1997,Delabaere:1999,Kawai:2005,Ito:2018eon,Sueishi:2020yux}}. Inserting $\psi=\exp\int^LS\,\dd L$ in \eqref{eq:wkb_schr}, the logarithmic derivative obeys the Riccati equation
\begin{equation}\label{eq:riccati}
    S^2+S'=\hbar^{-2}Q_0+Q_2 ,
\end{equation}
which can be solved order by order in $\hbar$, $S=\hbar^{-1}S_{-1}+S_0+\hbar S_1+\dots$, with $S_{-1}=\pm\sqrt{Q_0}$. The two solutions differ only in their odd part,
\begin{equation}\label{eq:2wkbsol}
    S_{\rm odd}=\frac12\big(S^{(+)}-S^{(-)}\big)=\frac{\sqrt{Q_0}}{\hbar}+\hbar S_1+\hbar^3S_3+\dots,
\end{equation}
since the even part is fixed by it.\footnote{It is fixed as $\frac12(S^{(+)}+S^{(-)})=-\frac12(\log S_{\rm odd})'$~\cite{Sueishi:2020yux}.} The two WKB solutions normalized at a point $a$ are then
\begin{equation}\label{eq:wkb_wkbsol}
    \psi^a_\pm(L)=\frac{1}{\sqrt{S_{\rm odd}}}\exp\Big(\pm\int_a^LS_{\rm odd}\,\dd L'\Big),
\end{equation}
and their Wronskian is exactly $-2$ to all orders. These are formal series in $\hbar$, which are divergent, and become actual solutions after Borel resummation. Three notions organize the exact WKB analysis.

The first is that of Stokes curves. The Borel transform of the series multiplying $e^{\pm\frac1\hbar\int_a^L\sqrt{Q_0}}$ is singular at the point where the two exponents differ and its Laplace integral becomes ambiguous when this singularity lies on the integration ray. This happens on the curves
\begin{equation}\label{eq:wkb_stokes_def}
    \mathrm{Im}\Big[\frac1\hbar\int_a^L\sqrt{Q_0}\,\dd L'\Big]=0
\end{equation}
emanating from the turning points and from the other special points of the equation. Away from them both WKB solutions are Borel summable; on them the dominant solution is not and its two lateral resummations differ by a multiple of the subdominant solution.

\rev{The second notion is that of Stokes jumps~\cite{Voros:1983,Sueishi:2020yux}.} Near a simple turning point $a$ the linearization of the potential reduces the equation to Airy's equation, characterized by three Stokes curves leaving $a$. On each of them one of the two solutions \eqref{eq:2wkbsol} is dominant. Crossing a Stokes curve, the Borel sums before and after the crossing are related by the Voros connection formula~\cite{Voros:1983},
\begin{equation}\label{eq:wkb_voros}
    \psi^a_{\rm dom}\big|_{\rm before}=\psi^a_{\rm dom}\big|_{\rm after}+i\,\psi^a_{\rm sub}\big|_{\rm after},\qquad \psi^a_{\rm sub}\big|_{\rm before}=\psi^a_{\rm sub}\big|_{\rm after},
\end{equation}
where $\psi^a_{\rm dom}$ and $\psi^a_{\rm sub}$ are the solutions that are dominant and subdominant on the curve.\footnote{Here the crossing is anticlockwise around $a$; for a clockwise crossing $i\to-i$.} We see from \eqref{eq:wkb_voros} that only the dominant solution jumps, by the Stokes constant $i$ times the subdominant one. We quickly review the derivation of the Stokes jumps \eqref{eq:wkb_voros} in Appendix \ref{app:wkb_stokes}.\footnote{In our problem we also have to consider the origin. At the origin, a regular singular point of Bessel type, the analogous rule has Stokes constant $2i$ and the regular solution is the median of the two lateral resummations. Both rules are derived from the Borel transforms of the Airy and Bessel functions in Appendix~\ref{app:wkb_stokes}.}

The third notion is that of Voros multipliers and quantum periods, which allow one to transport the WKB solution from one Stokes curve to the next. Indeed changing the normalization point of \eqref{eq:wkb_wkbsol} multiplies the WKB solutions by the exponentials of integrals of $S_{\rm odd}$,
\begin{equation}\label{eq:wkb_renorm}
    \psi^a_\pm=\exp\Big(\pm\int_a^bS_{\rm odd}\,\dd L\Big)\,\psi^b_\pm .
\end{equation}
When the integration path is a closed cycle $\gamma$, the integral defines a quantum period and its exponential a Voros multiplier,
\begin{equation}\label{eq:wkb_qperiod}
    V_\gamma=\oint_\gamma S_{\rm odd}\,\dd L=\frac1\hbar\oint_\gamma\sqrt{Q_0}\,\dd L+\hbar\oint_\gamma S_1\,\dd L+\dots,\qquad e^{-V_\gamma}.
\end{equation}
The connection coefficients between the origin and infinity are thus polynomials in the Stokes constants and in the Voros multipliers of the cycles traversed by the transport. In general the quantum periods are themselves divergent series, whose resummation produces further non-perturbative effects. We will see that in our problem this does not happen.

\begin{figure}[t]
\centering
\includegraphics{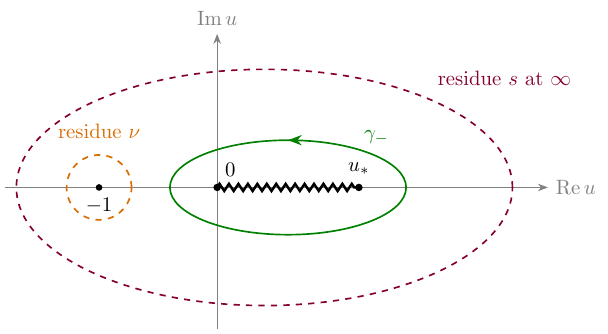}
\caption{The $u=\sinh^2(L/2)$ plane. The forbidden region is the cut between the origin $u=0$ and the turning point $u_*$. The bounce cycle $\gamma_-$ around it can be deformed into a large loop, with residue $s$ at infinity, minus a small loop around $u=-1$, with residue $\nu$, so that its period is $V_-=2\pi(\nu-s)$ exactly. The winding cycle is the loop at infinity, with period $V_\infty=4\pi s$.}
\label{fig:wkb_uplane}
\end{figure}

The equation \eqref{eq:wkb_schr} is periodic under $L\to L+2\pi i$, so that the turning point \rev{$T_0=L_*$} has images $T_k=L_*+2\pi ik$ and the origin \rev{$O=O_0$} has images $O_k=2\pi ik$. Only two independent cycles enter the connection problem and it is convenient to describe them in the variable $u=\sinh^2\frac L2$, in which the forbidden region is the segment between $u=0$ and the turning point $u_*$, and the only other singular points are $u=-1$ (i.e.\ $L=i\pi$) and $u=\infty$ (see Figure \ref{fig:wkb_uplane})\rev{, the same singular points of the radial equation \eqref{eq:qd_classical2} discussed in Section~\ref{sec:jt}}. The first cycle encircles the forbidden segment. Its quantum period is the action of the bounce, the round trip from the origin to the turning point and back. The second cycle connects two consecutive images of the turning point and its period is the action of a winding around the asymptotic region. Their leading-order values, which we compute in Appendix \ref{app:wkb_periods}, are
\begin{equation}\label{eq:wkb_periods}
    V_-=2\int_0^{L_*}S_{\rm odd}\,\dd L=2\pi(\nu-s),\qquad V_\infty=-2\int_{T_{-1}}^{T_0}S_{\rm odd}\,\dd L=4\pi s ,
\end{equation}
and their combination
\begin{equation}
    V_+=V_-+V_\infty=2\pi(\nu+s)
\end{equation}
is the action of a bounce combined with a winding. Deforming the bounce cycle in the $u$-plane, its period equals the residue at infinity minus the residue at $u=-1$, as shown in Figure \ref{fig:wkb_uplane}. As we show in Appendix~\ref{app:wkb_periods}, both residues coincide with the exact local exponents of the equation, so that the values \eqref{eq:wkb_periods} receive no corrections in $\hbar$. In other words, the quantum periods are exact. This relies crucially on the ordering term $Q_2$ in \eqref{eq:wkb_schr}, which played the role of embedding precisely the P\"oschl--Teller scattering problem into the disk Laplacian. We show in Appendix~\ref{app:wkb_periods} that without this choice of ordering, the one-loop exactness and the consequent transparent gravitational interpretation of Section~\ref{subsec:np_gravity} would be lost.

The map of Stokes curves, \rev{whose crossings with the path are shown schematically in Figure \ref{fig:wkb_path} of Appendix \ref{app:wkb_connection}}, tells us which connection rules have to be applied along the real axis, following the regular WKB wavefunction from the origin to infinity. For real $\hbar$ it is degenerate,\footnote{The forbidden segment is itself a Stokes segment joining the origin to the turning point.} so we give $\hbar$ a small phase to resolve the degeneracy. The forbidden segment is then crossed by no Stokes curve, so that the Bessel rule \rev{\eqref{eq:wkb_bessel}} at the origin can be applied on it. Passing below the turning point, the path crosses one of its Stokes curves, and on the allowed axis it crosses the curves rising from all the images $T_{-1},T_{-2},\dots$, each of which contributes one Voros multiplier of the winding. Following the regular solution along this path, as detailed in Appendix~\ref{app:wkb}, we find that on the allowed axis
\begin{equation}\label{eq:wkb_result_lateral}
    \sqrt{2\pi}\,\psi_{\rm reg}=e^{V_-/2}\,\frac{1+e^{-V_+}}{1-e^{-V_\infty}}\,\psi_+ + e^{V_-/2}\big(1+e^{-V_-}\big)\,\psi_- ,
\end{equation}
where \rev{$\psi_+=\psi_+^{T_0}$ and $\psi_-=i\,\psi_-^{T_0}$} are the outgoing and incoming WKB waves \rev{normalized} at the turning point\rev{, with the phase $i$ chosen so that at infinity both are proportional to the asymptotic waves $s^{-1/2}e^{\pm isL}$ with the same Wronskian}. The amplitude at infinity follows from the Wronskian and from the reality of the regular solution, and \eqref{eq:wkb_rho_def} gives
\begin{equation}\label{eq:wkb_main}
    \rho(s)=2s\,e^{-V_-}\,\frac{1-e^{-V_\infty}}{\big(1+e^{-V_-}\big)\big(1+e^{-V_+}\big)}=\frac{2s\,\sinh(2\pi s)}{\cosh(2\pi\nu)+\cosh(2\pi s)} .
\end{equation}

Exact WKB explains the structure of the density, its one-loop exactness and its sectors, which we discuss in Section~\ref{subsec:np_gravity}. It also makes concrete the relation with the Landau levels of the magnetic model of Section \ref{sec:jt}. In exact WKB discrete states follow from requiring that the wavefunction be normalizable, i.e.\ that the coefficient of the growing wave at infinity vanishes: this gives the exact quantization condition \cite{Sueishi:2020yux}. For a real magnetic field this is what happens: at $s=\pm i\kappa$ the two waves at infinity are $e^{\pm\kappa L}$ and removing the growing one through the zeros of $1+e^{-V_\mp}$ in \eqref{eq:wkb_result_lateral} leaves a bound state. This happens at $\kappa=B-n-\frac12$, which are precisely the Landau levels of Section~\ref{sec:jt}. On the gravity side, since $V_\mp=2\pi(\nu\mp s)$ exactly, the condition $e^{-V_\mp}=-1$ is a Bohr--Sommerfeld condition for the bounce action \eqref{eq:wkb_qperiod}, with the shift $\frac12$ coming from the two phases $i$ of the connection rules at its ends.\footnote{\rev{That is, $V_\mp=2\pi i\big(n+\frac12\big)$. The shift $\frac12$ is the Maslov correction: the turning point and the origin each contribute the reflection phase $i=e^{i\pi/2}$, see Section~\ref{subsec:np_gravity}.}}
In gravity the field is imaginary, $B=i\nu$, and the same condition is solved only at complex $s$,
\begin{equation}
    s=\pm\nu\pm i\big(k+\tfrac12\big),\qquad\Phi_h=\pm\Phi_b\pm i\hbar\big(k+\tfrac12\big),
\end{equation}
where both waves oscillate. The bound states are therefore absent from the gravitational spectrum, which is purely continuous, and survive only as resonances, i.e.\ poles of the density at complex horizons at the boundary.

\subsection{Instanton bounces and the gravitational interpretation}
\label{subsec:np_gravity}

We now interpret the sectors of \eqref{eq:wkb_main} in the gravitational path integral. Expanding the two bounce factors in geometric series and using $V_+=V_-+V_\infty$, the density becomes a sum over sectors labelled by an integer $n\ge1$,
\begin{equation}\label{eq:np_sectors}
    \rho(s)=2s\sum_{n\ge1}(-1)^{n-1}\Big[e^{-nV_-}-e^{-nV_+}\Big]=2s\sum_{n\ge1}(-1)^{n-1}e^{-2\pi n\nu}\Big[e^{2\pi ns}-e^{-2\pi ns}\Big].
\end{equation}
Equivalently,
\begin{equation}\label{eq:np_kemble}
    \rho(s)=2s\Big[\mathcal T(V_-)-\mathcal T(V_+)\Big],\qquad \mathcal T(V)=\frac{1}{1+e^{V}} .
\end{equation}
The function $\mathcal T$ is Kemble's transmission coefficient~\cite{Kemble:1935}, the probability of transmission through a potential barrier with tunnelling exponent $V$, which improves on the naive $e^{-V}$ by resumming the multiple reflections inside the barrier. Here the barrier is the forbidden region between the origin and the turning point, and \eqref{eq:np_kemble} holds exactly due to the exact WKB analysis we performed above.

Each sector in \eqref{eq:np_sectors} consists of its classical action and the one-loop prefactor $2s$, with no higher-loop corrections. This is because the quantum periods \eqref{eq:wkb_periods} are exact and the only other numbers entering the connection problem, the phases of the local connection rules, are exact as well. This extends the one-loop exactness of the Schwarzian density~\cite{Stanford:2017thb} to finite cutoff, including all non-perturbative sectors. Note that this is a statement about the density at fixed energy; the canonical partition function acquires non-trivial fluctuation factors from the energy $E(s)=\nu-\sqrt{\nu^2-s^2}$, not from the density.

To understand the two terms in \eqref{eq:np_kemble}, consider the Euclidean black hole of JT gravity in the Schwarzschild gauge, which is the $b\to0$ limit of the sinh dilaton gravity black hole \eqref{eq:sinh_bh},
\begin{equation}\label{eq:np_schwarzschild}
    \dd s^2=\big(\Phi^2-\Phi_h^2\big)\,\dd\tau^2+\frac{\dd\Phi^2}{\Phi^2-\Phi_h^2},\qquad \Phi_h\le\Phi\le\Phi_b ,
\end{equation}
where the dilaton plays the role of the radial coordinate. At the outer horizon, $\Phi=\Phi_h$, the geometry closes off smoothly provided $\tau\sim\tau+2\pi/\Phi_h$, and the Euclidean section is the ordinary disk; the proper length of the boundary at $\Phi=\Phi_b$, i.e.\ the inverse Tolman temperature, is then $\beta=2\pi\sqrt{\Phi_b^2-\Phi_h^2}/\Phi_h$. At the inner horizon, $\Phi=-\Phi_h$, the dilaton changes sign and a Euclidean section closing off there is a complex saddle obtained from the previous one by $\Phi_h\to-\Phi_h$, i.e.\ $s\to-s$. Consider now a geometry in which the boundary winds $n$ times around the outer horizon, i.e.\ with
\begin{equation}\label{eq:np_nwound}
    \tau\sim\tau+\frac{2\pi n}{\Phi_h},\qquad \beta=\frac{2\pi n}{\Phi_h}\sqrt{\Phi_b^2-\Phi_h^2}.
\end{equation}
The total angle around the horizon is $2\pi n$, i.e.\ there is a conical excess $2\pi(n-1)$. In the Euclidean picture of Figure \ref{fig:mag_disk}, this is precisely what one bounce from $O$ to $L_*$ and back does $n$ times: each bounce is one revolution of the boundary point around the horizon. The weight of such a geometry at fixed energy has two factors. \rev{The first is the weight of the boundary curve. The canonical quantization computes the heat kernel of the charged particle, which weights a curve by the Aharonov--Bohm flux $e^{\Phi_bA/\hbar}$ of the signed area it encloses, rather than by the gravitational boundary term $e^{\Phi_b\oint K/\hbar}$. By the Gauss--Bonnet argument of Section \ref{sec:jt}, $\oint K=2\pi n+A$ for a curve that winds $n$ times, so the two differ by $e^{-2\pi n\Phi_b/\hbar}$.\footnote{\rev{In~\cite{Griguolo:2026eoi} the same sectors were described as necklaces of $n$ disks pinched at $n-1$ points, where each pinch contributes a corner term $2\pi\Phi_b$ to the action. The two pictures agree: the necklace and the $n$-wound disk have the same boundary arcs, hence the same flux and the same $\oint K$ along them, and by Gauss--Bonnet the extra turning $2\pi(n-1)$ is compensated either by the conical excess at the horizon or by a corner angle $-2\pi$ at each pinch.}} It is this factor that suppresses the wound geometries;} the turning number playing the role of an effective Euler characteristic and the boundary dilaton that of the ground-state entropy\rev{. The second factor is the horizon:} each turn around the horizon contributes the Bekenstein--Hawking entropy $S_{\rm BH}=2\pi\Phi_h/\hbar$.\footnote{The entropy $2\pi n\Phi_h/\hbar$ of the $n$-wound geometry is indeed the expectation value of the bulk defect operator \eqref{eq:np_defect} with $\gamma=2\pi n$ on the disk topology.} Together,
\begin{equation}
    e^{\,n(S_{\rm BH}-2\pi\Phi_b/\hbar)}=e^{-2\pi n(\nu-s)}=e^{-nV_-},
\end{equation}
which is the $n$-th term of $2s\,\mathcal T(V_-)$. At the inner horizon the entropy changes sign and the weight becomes $e^{-nV_+}$; the winding around the asymptotic region of Section~\ref{subsec:np_problem} is precisely what exchanges the two horizons. As a check, already observed in~\cite{Griguolo:2025kpi}, the $n$-th sector of the partition function,
\begin{equation}
    \int\dd s\;2s\,e^{2\pi ns-2\pi n\nu-\beta E(s)},\qquad E(s)=\nu-\sqrt{\nu^2-s^2},
\end{equation}
has a saddle point at $s_n=2\pi n\nu/\sqrt{4\pi^2n^2+\beta^2}$, i.e.\ at the horizon value fixed by \eqref{eq:np_nwound}, and its exponent evaluates to
\begin{equation}
    -I_n=\nu\Big(\sqrt{4\pi^2n^2+\beta^2}-\beta\Big)-2\pi n\nu ,
\end{equation}
\rev{i.e.\ the gravitational on-shell action of the $n$-wound geometry, multiplied by the factor $e^{-2\pi n\nu}$ discussed above.} The non-perturbative sectors of the finite-cutoff density are therefore instantons in which the boundary winds several times around the outer or the inner horizon. \rev{They become important near the edge of the spectrum, where the horizon approaches the boundary.}

The alternating sign $(-1)^{n-1}$ in \eqref{eq:np_sectors} means that each additional bounce multiplies the weight by $-1$. This weight has a clear interpretation at the level of the connection problem \rev{of Section~\ref{subsec:np_problem}}, where it is given by the product of two phases $i$, one from the Airy rule at the turning point and one from the Bessel rule at the origin, \eqref{eq:wkb_airy_lateral} and \eqref{eq:wkb_bessel}. They are the reflection phases of semiclassical quantization: a wave reflected at a smooth turning point is shifted by $\frac\pi2$ with respect to the incident one,\footnote{The standing wave near a turning point is $\cos(x-\frac\pi4)=e^{-i\pi/4}\big(e^{ix}+\rev{e^{i\pi/2}}e^{-ix}\big)/2$, with $x$ the classical phase measured from the turning point. The familiar $\frac\pi4$ is the phase of the standing wave, while the reflected wave is shifted by twice as much.} and the centre of the disk reflects the wave in the same way.\footnote{\rev{Near the origin the regular solution is $\sqrt L\,I_0(\sqrt{\nu^2-s^2}\,L)$, from which the Bessel rule \eqref{eq:wkb_bessel} of Appendix~\ref{app:wkb_stokes} follows. Continued to an allowed region, $\sqrt{\nu^2-s^2}=ik$, it becomes the regular radial wave in two dimensions, $J_0(kL)\propto\cos(kL-\frac\pi4)$ at large $kL$, which has the same reflection phase as a smooth turning point.}} A bounce is reflected once at each end, so its phase is $i\cdot i=-1$, as for a one-dimensional orbit between two smooth turning points, whose Bohr--Sommerfeld condition is shifted by $\frac12$. \rev{A hard wall at the origin would instead reflect with phase $\pi$, so that a bounce would carry the phase $-i$ and the shift would be $\frac34$. This is precisely what happens with the flat ordering and a Dirichlet condition at $L=0$: the density obtained in this case in Appendix A of~\cite{Griguolo:2026eoi} contains $\Gamma\big(\frac34\pm i\sqrt{\nu^2-\frac1{16}}\pm is\big)$ in place of the $\Gamma\big(\frac12\pm i\nu\pm is\big)$ of \eqref{eq:wkb_main}, and the shift of $\nu$ is the quantum correction to the period found in Appendix~\ref{app:wkb_periods}.} The sign of the sectors thus probes the fact that the origin is the regular centre of the disk. In the gravitational path integral, the same sign arises from the fluctuations of the boundary curve around the $n$-wound saddle: at finite cutoff, the quadratic action has $n-1$ complex negative modes, whose number does not depend on the cutoff, and each of them contributes a factor $-1$ once its contour is rotated. We prove this statement exactly in Appendix \ref{app:negmodes}.

The sectors \eqref{eq:np_sectors} also have a natural interpretation in terms of defects. In JT gravity, a conical defect of opening angle $\gamma$ is created by inserting the bulk operator~\cite{Mertens:2019tcm,Blommaert:2025avl}
\begin{equation}\label{eq:np_defect}
    V_\gamma=\int\dd^2x\sqrt g\;e^{-(2\pi-\gamma)\Phi/\hbar},
\end{equation}
and the disk with one such defect has density of states proportional to $\cosh(\gamma s)$~\cite{Mertens:2019tcm}. Comparing with \eqref{eq:np_sectors}, the $n$-th sector is a defect amplitude with $\gamma=2\pi n$,
\begin{equation}\label{eq:np_defectsector}
    2s\big(e^{2\pi ns}-e^{-2\pi ns}\big)=4\,\partial_\gamma\cosh(\gamma s)\Big|_{\gamma=2\pi n}.
\end{equation}
As we explain in Section~\ref{subsec:zz}, the derivative is the JT limit of a finite difference between two defects at degenerate angles, which for $n=1$ is the identity operator of the Liouville description. The excess is located at the horizon, and it is precisely the conical excess $2\pi(n-1)$ of the $n$-wound geometry \eqref{eq:np_nwound}. The finite-cutoff density is therefore a sum over defect amplitudes, weighted by the boundary action,
\begin{equation}\label{eq:np_defectsum}
    \rho(s)=\sum_{n\ge1}(-1)^{n-1}\,e^{-2\pi n\nu}\;4\,\partial_\gamma\cosh(\gamma s)\Big|_{\gamma=2\pi n}.
\end{equation}
The same structure holds in sinh dilaton gravity, where the derivative with respect to the angle is replaced by a finite difference. By \eqref{eq:np_series_b},
\begin{equation}\label{eq:np_defectsum_b}
    \rho_b(s)=\sum_{n\ge1}(-1)^{n-1}\,e^{-2\pi n\nu}\;\frac{\cosh(\gamma_+s)-\cosh(\gamma_-s)}{\pi b^2},\qquad \gamma_\pm=2\pi\big(n\pm b^2\big),
\end{equation}
which reduces to \eqref{eq:np_defectsum} in the JT limit. The $n$-th sector is now the difference of two defects, with angles split by $\pm2\pi b^2$ around the classical excess $2\pi n$. This structure is the starting point of the CFT interpretation of the next subsection.

Let us conclude the discussion by analyzing the large-cutoff limit, or strong magnetic field regime, in the language of bounces and exact WKB developed so far. When the cutoff is large, at fixed horizon the bounce action grows without bound, the barrier between the origin and the turning point becoming infinitely thick. Therefore the wave reflected at the origin, which is the only place where the origin enters the WKB computation, never returns. Only the singly wound sector survives,
\begin{equation}
    \rho(s)\;\longrightarrow\;2s\,e^{-V_-}\big(1-e^{-V_\infty}\big)=4s\,e^{-2\pi\nu}\sinh(2\pi s),
\end{equation}
which, after removing the \rev{turning-number factor $e^{-2\pi\nu}$ of the smooth disk}, is the Schwarzian density\rev{~\cite{Bagrets:2016cdf,Stanford:2017thb,Mertens:2017mtv}}. \rev{In the canonical normalization} an additional revolution of the boundary costs \rev{a factor $e^{-2\pi\Phi_b/\hbar}$}, so that at large cutoff only the smooth disk and its inner-horizon image survive. \rev{The two terms $e^{\pm2\pi s}$ are thus localized on the outer and the inner horizon. The same localization on the outer and the inner horizon, at large cutoff, was found in~\cite{Blommaert:2024whf} for general dilaton potentials, by integrating by parts the exact density obtained from the gas of defects~\cite{Maxfield:2020ale,Witten:2020wvy,Turiaci:2020fjj} and picking up the pole at the horizon on each branch of the inverse prepotential.}

\subsection{A CFT interpretation: degenerate bulk vertex operators}
\label{subsec:zz}
We have already seen in \eqref{eq:qd_liouvilleaction} that sinh dilaton gravity can be formulated as a pair of Liouville theories, a spacelike and a timelike one. If we couple them to the $bc$ ghosts, all the gravitational amplitudes should be understood as amplitudes of this worldsheet theory \cite{Mertens:2020hbs,Collier:2023cyw}. What follows should not be regarded as a full CFT derivation of the results but rather as an identification on the CFT side of the key ingredients that have emerged in the previous sections, together with their interpretation.

\begin{figure}[t]
\centering
\includegraphics[width=0.8\textwidth]{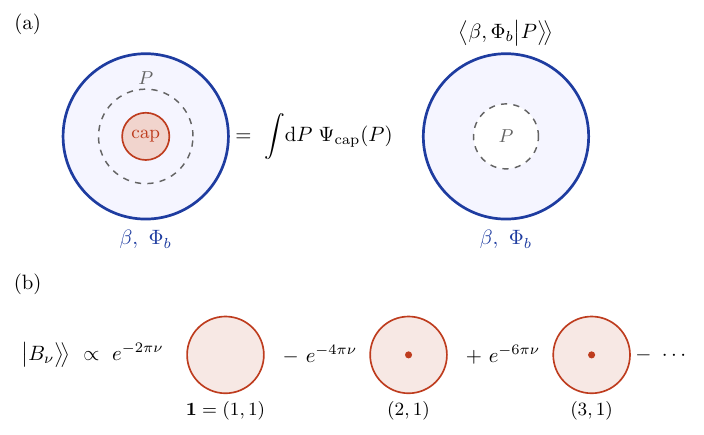}
\caption{\rev{(a) In the closed channel the disk is cut along a circle carrying a Liouville primary $P$: the outer part is the finite-cutoff trumpet \eqref{eq:zz_fctrumpet}, bounded by the Dirichlet boundary at fixed $\beta$ and $\Phi_b$, and the inner part is the cap, which contributes its wavefunction $\Psi_{\rm cap}(P)$. (b) The cap state \eqref{eq:zz_Bnu}: the $n$-th sector is a cone of total angle $2\pi n$ at the horizon, created by the degenerate insertion $(n,1)$, weighted by $e^{-2\pi n\nu}$ and with alternating sign. The first sector, the identity, is the smooth disk. Each degenerate state is the difference of two non-degenerate ones, i.e.\ of two defects with angles $2\pi(n\pm b^2)$.}}
\label{fig:zz}
\end{figure}

The amplitude we are interested in is the disk, which on the worldsheet can be viewed as an annulus with one end given by the gravitational boundary and the other by the point where the geometry closes off, which we call the cap. Both ends are described by boundary states of the two Liouville theories, which can be expanded in Ishibashi states,\footnote{Ishibashi states \cite{Ishibashi:1988kg} are boundary states built from the Virasoro representation of a single bulk primary. They diagonalize the propagation along a cylinder.} with momenta $P$ and $\tilde P$. Following~\cite{Collier:2024mlg}, in the closed channel the annulus is a cylinder of length $t$ and the amplitude becomes
\begin{equation}\label{eq:zz_cylinder}
    Z=\int_0^\infty\dd t\int\dd P\,\dd\tilde P\;\Psi_1(P,\tilde P)\,\Psi_2(P,\tilde P)\;\chi_P(\tilde q)\,\tilde\chi_{\tilde P}(\tilde q)\,\eta(\tilde q)^2,\qquad\tilde q=e^{-2\pi t},
\end{equation}
where $\Psi_1$ and $\Psi_2$ are the wavefunctions of the gravitational boundary and of the cap, and $\chi_P=\tilde q^{P^2}/\eta$ and $\tilde\chi_{\tilde P}=\tilde q^{-\tilde P^2}/\eta$ are the non-degenerate characters of the spacelike and timelike theories, whose oscillators are cancelled by the ghost contribution $\eta^2$. The integral over the modulus,
\begin{equation}
    \int_0^\infty\dd t\;e^{-2\pi t(P^2-\tilde P^2)}=\frac{1}{2\pi\big(P^2-\tilde P^2\big)},
\label{eq:zz_pole}
\end{equation}
produces a pole and the integral over the timelike momentum picks it up, setting $\tilde P=P$~\cite{Collier:2024mlg}.\footnote{The poles at $\tilde P=\pm P$ lie on the real contour, so the integral over $t$ is defined by analytic continuation.} This is the Wheeler--DeWitt constraint~\cite{Blommaert:2025avl}. \rev{The constraint identifies the two momenta, so that the amplitude becomes an overlap in a single Liouville theory. We write it in the spacelike theory, where the boundary carries the FZZT condition;\footnote{\rev{It measures $g_+$ rather than the quantum-disk frame $g_-$. This affects neither the cap, since the two frames differ by a Weyl factor regular at the horizon, nor the boundary, where the two lengths differ by a constant factor, see \eqref{eq:zz_lpm}.}}} equivalently, the timelike field can be gauge fixed as a clock~\cite{Blommaert:2025rgw}. The amplitude reduces to an overlap in the spacelike theory,
\begin{equation}\label{eq:zz_overlap}
    Z=\big\langle\mathrm{bdy}\big|\mathrm{cap}\big\rangle=\int\dd P\;\Psi_{\rm cap}(P)\,\big\langle\mathrm{bdy}\big|P\big\rangle\!\big\rangle .
\end{equation}
Each term is the overlap of the gravitational boundary with a single primary $P$, i.e.\ a trumpet, weighted by the wavefunction $\Psi_{\rm cap}(P)$ of the cap. In the open channel, the same amplitude is an integral over the label $s$ of the boundary, which is the energy label of the density of states, and the two channels are related by the modular kernel $\cos(4\pi bPs)$ below. At finite cutoff both ends are modified, and we discuss them in turn. \rev{The construction is summarized in Figure~\ref{fig:zz}.}

For a boundary of fixed length $\ell$, as in Liouville gravity, the overlap with a non-degenerate primary of momentum $\alpha=Q/2+iP$ is the trumpet~\cite{Mertens:2020hbs,Blommaert:2025rgw}
\begin{equation}\label{eq:zz_trumpet}
    Z_{\rm tr}(\ell;P)=\big\langle\ell\big|P\big\rangle\!\big\rangle=\int_0^\infty\dd s\;\cos(4\pi bPs)\,e^{-\ell\mu_B(s)},
\end{equation}
where $\cos(4\pi bPs)$ is the wavefunction of the primary $P$ in the presence of the FZZT brane with label $s$~\cite{Fateev:2000ik,Teschner:2000md}, whose boundary cosmological constant is
\begin{equation}\label{eq:zz_muB}
    \mu_B(s)=\frac{\cosh2\pi b^2s}{2\pi b^2}
\end{equation}
in the units of \eqref{eq:sinh_energy}.\footnote{The boundary cosmological constant \eqref{eq:zz_muB} corresponds to the ADM energy of sinh dilaton gravity, the conjugate to the inverse Hawking temperature $\ell\sim\beta_H$.} At finite cutoff, the Dirichlet boundary conditions have to be translated into Liouville variables. The finite-cutoff boundary fixes the dilaton $\Phi_b$ and its own proper length, i.e.\ the inverse Tolman temperature $\beta$. Each Liouville field measures its own metric $g_\pm=e^{\mp2\pi b^2\Phi}g$, and since the dilaton is constant along the boundary, the boundary lengths $\ell_+=\oint e^{b\phi}$ and $\ell_-=\oint e^{b\tilde\phi}$ seen by the two fields are rigid rescalings of the proper length,
\begin{equation}\label{eq:zz_lpm}
    \ell_\pm=e^{\mp\pi b^2\Phi_b}\,\beta, \qquad  \frac{\ell_-}{\ell_+}=e^{2\pi b^2\Phi_b}.
\end{equation}
On the black hole \eqref{eq:sinh_bh} as $\Phi_b\to\infty$, the length $\ell_+$ tends to $\beta_H/(2\pi b^2)$ where $\beta_H$ is the inverse Hawking temperature and $\ell_-$ blows up, while at finite cutoff neither length blows up. Assuming that the reduction \eqref{eq:zz_cylinder} can be carried out in the same way, this leaves the spacelike theory with a boundary of fixed length $\ell_+$, which is now related to the Tolman temperature instead of the Hawking temperature. The variable conjugate to $\beta$ is the quasi-local energy $E_b(s)$,\footnote{As one can verify from the first law $\beta=2\pi(\dd E_b/\dd\Phi_h)^{-1}$.} so that the variable conjugate to $\ell_+$ is
\begin{equation}\label{eq:zz_calE}
    \mathcal E(s)=e^{\pi b^2\nu}E_b(s),\qquad e^{-\beta E_b(s)}=e^{-\ell_+\mathcal E(s)},\qquad \mathcal E(s)\xrightarrow{\;\nu\to\infty\;}\mu_B(s)\rev{-\frac{1}{2\pi b^2}},
\end{equation}
in the rescaled units of \eqref{eq:sinh_energy}. At large cutoff it reduces to the boundary cosmological constant of the FZZT brane \eqref{eq:zz_muB}\rev{, up to the $s$-independent shift $1/2\pi b^2$ produced by the counterterm of \eqref{eq:sinh_action}, which only multiplies the fixed-length amplitude by $e^{\ell_+/2\pi b^2}$}.\footnote{\rev{By \eqref{eq:zz_lpm}, the counterterm contributes $\beta F(\Phi_b)=(\ell_--\ell_+)/2\pi b^2$ to the action. The term in $\ell_-$ cancels the divergent part of $\beta\,\partial_n\Phi$, while the term in $\ell_+$ is the shift.}} 

The transform to the fixed-length basis then proceeds as in Liouville gravity~\cite{Mertens:2020hbs,Blommaert:2025rgw}. One has to consider that the FZZT amplitude is unmarked: it is invariant under translations along the boundary circle and in terms of the fixed-length amplitude it reads $Z_{\rm FZZT}=\int\dd\ell\,\ell^{-1}e^{-\mathcal E\ell}Z(\ell)$, where $\ell^{-1}$ removes the volume of these translations. On the other hand the fixed-length amplitude is marked, as is manifest in the canonical formulation where the disk is the transition amplitude $\bra{L=0}e^{-\beta\hat H}\ket{L=0}$ between slices at a definite boundary point. Marking a point corresponds to performing a derivative with respect to $\mathcal{E}$ which cancels the factor $\ell^{-1}$. Schematically,
\begin{equation}\label{eq:zz_marked}
    Z_\bullet(\mathcal E)=-\partial_{\mathcal E}Z_{\rm FZZT}(\mathcal E)=\int_0^\infty\dd\ell\;e^{-\mathcal E\ell}\,Z(\ell)=\int\dd s\;\frac{\rho(s)}{\mathcal E+\mathcal E(s)} .
\end{equation}
The marked amplitude is therefore the resolvent of the spectrum, with simple poles that accumulate into a cut at $\mathcal E=-\mathcal E(s)$, and the fixed-length amplitude is the inverse Laplace transform of $Z_\bullet(\mathcal E)$, whose contour can be wrapped around the cut. The discontinuity yields $2\pi i\rho(s)/\mathcal E'(s)$,\footnote{Each pole contributes a delta function $\delta(\mathcal E+\mathcal E(s))$ to the discontinuity.} and when the integral along the cut is written in terms of $s$, the measure $\dd\mathcal E=\mathcal E'(s)\dd s$ supplies exactly the same Jacobian:
\begin{equation}\label{eq:zz_fc}
    Z(\beta)=\int\frac{\dd\mathcal E}{2\pi i}\;e^{\ell_+\mathcal E}\;\mathrm{Disc}\,Z_\bullet(\mathcal E)=\int\dd s\;\rho(s)\,e^{-\ell_+\mathcal E(s)}=\int\dd s\;\rho(s)\,e^{-\beta E_b(s)} .
\end{equation}
Therefore the density does not depend on which function of $s$ is chosen as the conjugate variable. The only difference with infinite cutoff is the location of the cut: it now runs over the spectrum of the $T\bar T$-deformed operator, $\mathcal E(s)$ instead of $\mu_B(s)$. The physical spectrum lies on the perturbative branch of $\mathcal E(s)$, the one continuously connected to $\mu_B$ at large cutoff. The cut of \eqref{eq:zz_marked} is the corresponding segment and closing the contour through the branch point reproduces the spectrum truncated to $s<\nu$, i.e.\ the projected prescription of~\cite{Griguolo:2026eoi,SinhBox}.

The same reasoning yields the finite-cutoff trumpet. The transform to the basis of fixed Tolman temperature produces the $T\bar T$-deformed Boltzmann factor,
\begin{equation}\label{eq:zz_fctrumpet}
    Z_{\rm tr}(\beta;P)=\big\langle\beta,\Phi_b\big|P\big\rangle\!\big\rangle=\int_{\mathcal{C}}\dd s\;\cos(4\pi bPs)\,e^{-\beta E_b(s)} ,
\end{equation}
which reduces to \eqref{eq:zz_trumpet} at large cutoff. The contour $\mathcal C$ depends on how the spectrum is treated beyond the branch point of $E_b(s)$ at $s=\nu$ (Section~\ref{sec:ttbar}): it is the interval $0<s<\nu$ if only the perturbative branch of the energy is kept, as we do in this paper, while it also runs on the non-perturbative branch if the theory is extended beyond that point~\cite{Griguolo:2025kpi}.\footnote{In the JT limit, if both the perturbative and the non-perturbative branches are included, \eqref{eq:zz_fctrumpet} reproduces the finite-cutoff trumpet computed in~\cite{Griguolo:2025kpi} from the path integral.}

We now turn to the cap. A bulk operator of the gravitational theory must be a marginal operator of the combined worldsheet theory, and the simplest such operators are the dressed vertex operators $e^{2\alpha\phi}e^{2(b-\alpha)\tilde\phi}$, in which the timelike Liouville factor dresses a spacelike Liouville primary of momentum $\alpha$ so that the product has total dimension one~\cite{Mertens:2020hbs}. Using \eqref{eq:zz_map}, the exponents combine so that in gravitational variables one has
\begin{equation}\label{eq:zz_dressing}
    e^{2\alpha\phi}\,e^{2(b-\alpha)\tilde\phi}=\sqrt g\;e^{2\pi b(b-2\alpha)\Phi},
\end{equation}
which is precisely the defect operator \eqref{eq:np_defect}, in units $\hbar=1$, with opening angle
\begin{equation}\label{eq:zz_angle}
    \gamma(\alpha)=2\pi+2\pi b\,(b-2\alpha)=-4\pi ibP,\qquad \alpha=\frac Q2+iP.
\end{equation}
Bulk vertex operators of Liouville gravity are therefore conical defects of dilaton gravity~\cite{Maxfield:2020ale}. The degenerate momenta of Liouville theory and their continuations~\cite{Mertens:2020hbs},
\begin{equation}
    \alpha_{m,\pm n}=\frac{1-m}{2b}+\frac{(1\mp n)b}{2},\qquad P_{m,\pm n}=\frac i2\Big(\frac mb\pm nb\Big),\qquad m,n\in\mathbb Z_{>0},
\end{equation}
correspond by \eqref{eq:zz_angle} to the angles
\begin{equation}\label{eq:zz_degangle}
    \gamma_{m,\pm n}=2\pi\big(m\pm nb^2\big)
\end{equation}
of the dressed vertex operators. 

The smooth disk is the simplest cap. As emphasized in Appendix~A of~\cite{Blommaert:2025rgw}, it is the overlap of the boundary state with the identity Ishibashi state: nothing is inserted where the geometry closes off. The identity is a degenerate representation: its null vector at level one is removed, so that its character is the difference of two non-degenerate characters at the continued momenta $P_{1,\pm1}$. Correspondingly,
\begin{equation}\label{eq:zz_identity}
    \big|\mathbf 1\big\rangle\!\big\rangle=\big|P_{1,1}\big\rangle\!\big\rangle-\big|P_{1,-1}\big\rangle\!\big\rangle,\qquad \Psi_{\mathbf 1}(P)=\delta\big(P-P_{1,1}\big)-\delta\big(P-P_{1,-1}\big),
\end{equation}
understood after continuation of the $P$ contour. The same holds for every degenerate representation: removing the null vector at level $mn$, the Ishibashi state of the representation $(m,n)$ is $|(m,n)\rangle\!\rangle=|P_{m,n}\rangle\!\rangle-|P_{m,-n}\rangle\!\rangle$, and its wavefunction is localized at the two momenta $P_{m,\pm n}$.

At finite cutoff the cap is largely fixed by the \rev{weight of the wound boundaries}. When $\nu$ is finite, the \rev{factor} $e^{-2\pi n\nu}$ \rev{that weights} a geometry whose boundary winds $n$ times around the horizon \rev{(Section~\ref{subsec:np_gravity})} can no longer be neglected, and, as discussed in Section~\ref{subsec:np_gravity}, such a geometry has a conical excess $2\pi(n-1)$ at the horizon. The Liouville fields measure the metrics $g_\pm$, which differ from $g$ by a Weyl factor that is regular at the horizon, so they see the same cone of total angle $2\pi n$. By \eqref{eq:zz_angle}, this is created classically by a heavy insertion with $\alpha\simeq(1-n)/2b$, i.e.\ with the label $m=n$ multiplying $1/b$. The label multiplying $b$ is not fixed classically, encoding quantum fluctuations around the defect background, and we take it equal to the one of the identity, so that the sector $n=1$ is the smooth disk. The $n$-th sector is then the degenerate representation $(n,1)$, weighted by \rev{the turning-number factor $e^{-2\pi n\nu}$}, and with an alternating sign. In the JT limit this sign has several independent explanations: it is the product of the two reflection phases $i\cdot i$ of each bounce in the exact WKB analysis (Section~\ref{subsec:np_gravity}), and it is produced by the negative modes of the fluctuations of the $n$-wound boundary at finite cutoff (Appendix~\ref{app:negmodes}). We assume that it persists at finite $b$, so that the cap state and wavefunction are given by:
\begin{equation}\label{eq:zz_Bnu}
\begin{gathered}
    \ket{B_\nu}=\frac{1}{\pi b^2}\sum_{n\ge1}(-1)^{n-1}\,e^{-2\pi n\nu}\,\big|(n,1)\big\rangle\!\big\rangle,\\ \Psi_{B_\nu}(P)=\frac{1}{\pi b^2}\sum_{n\ge1}(-1)^{n-1}e^{-2\pi n\nu}\Big[\delta\big(P-P_{n,1}\big)-\delta\big(P-P_{n,-1}\big)\Big],
\end{gathered}
\end{equation}
i.e.\ a weighted sum over degenerate insertions.

Putting everything together, the finite-cutoff partition function is the overlap \eqref{eq:zz_overlap} between the state of the gravitational boundary, whose overlap with $|P\rangle\!\rangle$ is the finite-cutoff trumpet \eqref{eq:zz_fctrumpet},\footnote{The residue at the pole \eqref{eq:zz_pole} produces a factor $1/P$, up to normalization. It is cancelled by the marking: the derivative with respect to $s$ acting on the FZZT wavefunction $\cos(4\pi bPs)$ brings down a factor $4\pi bP$, as in eqs.~(5.25)--(5.32) of \cite{Blommaert:2025rgw}. The discontinuity in $s$ then turns the resulting $\sin(4\pi bPs)$ back into $\cos(4\pi bPs)$, times a function of $P$ that combines with the normalization of the Ishibashi states.} and the wavefunction \eqref{eq:zz_Bnu} of the cap:
\begin{equation}\label{eq:zz_closedchannel}
    Z(\beta)=\int\dd P\;\Psi_{B_\nu}(P)\,Z_{\rm tr}(\beta;P)=\frac{1}{\pi b^2}\sum_{n\ge1}(-1)^{n-1}e^{-2\pi n\nu}\Big[Z_{\rm tr}\big(\beta;P_{n,1}\big)-Z_{\rm tr}\big(\beta;P_{n,-1}\big)\Big],
\end{equation}
a signed sum of finite-cutoff trumpets at the degenerate momenta. The $n=1$ term is the smooth disk, while higher terms modify the smooth no-boundary state. Inserting the trumpet \eqref{eq:zz_fctrumpet} and exchanging the integrals, the partition function takes the form $Z(\beta)=\int_{\mathcal C}\dd s\,\rho_b(s)\,e^{-\beta E_b(s)}$, with the density given by the transform of the wavefunction of the cap to the open channel,
\begin{equation}\label{eq:zz_density}
    \rho_b(s)=\int\dd P\;\Psi_{B_\nu}(P)\cos(4\pi bPs)=\frac{1}{\pi b^2}\sum_{n\ge1}(-1)^{n-1}e^{-2\pi n\nu}\Big[\cosh\big(\gamma_{n,1}s\big)-\cosh\big(\gamma_{n,-1}s\big)\Big],
\end{equation}
where we used $4\pi bP_{n,\pm1}=i\gamma_{n,\pm1}$. Each term is a pair of defects at the degenerate angles \eqref{eq:zz_degangle}, which is the defect sum \eqref{eq:np_defectsum_b}. In the JT limit the two angles collide and the difference becomes the derivative in \eqref{eq:np_defectsum}. 

Using $\cosh(\gamma_{n,1}s)-\cosh(\gamma_{n,-1}s)=2\sinh(2\pi b^2s)\sinh(2\pi ns)$ and the generating function of the Chebyshev polynomials $U_{n-1}(\cosh2\pi s)=\sinh(2\pi ns)/\sinh(2\pi s)$, the sum gives
\begin{equation}
    \rho_b(s)=\frac{1}{\pi b^2}\,\frac{\sinh(2\pi b^2s)\,\sinh(2\pi s)}{\cosh(2\pi\nu)+\cosh(2\pi s)} ,
\end{equation}
which is the finite-cutoff density \eqref{eq:sinh_density}. A derivation of the choice of the label multiplying $b$, and of the alternating sign directly from the worldsheet theory with Dirichlet boundary conditions at finite cutoff, remains an interesting open problem.

\section{Conclusion and outlook}
\label{sec:discussion}

In this paper we have shown that finite-cutoff sinh dilaton gravity, i.e.\ Liouville gravity with a Dirichlet boundary at finite cutoff, describes the rotationally symmetric sector of a charged particle on the quantum hyperbolic disk, in a magnetic field set by the boundary value of the dilaton. The quantum disk is the quantization of the AdS$_2$ geometry measured by one of the two Liouville fields, and its non-commutativity is the canonical commutator of the gravitational phase space. On this space, correlators are dressed Witten diagrams, which we evaluated exactly for the two-point function. \rev{The identification extends to the Hilbert spaces: the Hilbert space of finite-cutoff sinh dilaton gravity was constructed in~\cite{SinhBox}, and its flat measure in $\Lambda$ is the invariant measure of the quantum disk after the conjugation in \eqref{eq:qd_vaksman}. The physical region $L>0$ lies outside the unit disk, $\abs{z}>1$, and it is connected to the interior, where the Fock representation of \eqref{eq:qd_relation} is defined, by the analytic continuation of Appendix \ref{app:realq}.} We conclude by discussing \rev{three} directions that we find particularly interesting.

\paragraph{The $q$-CFT on the boundary of the quantum disk.} The quantum disk carries an action of $U_q(\mathfrak{su}(1,1))$, and its boundary circle is the natural home of a one-dimensional conformal theory with quantum-group symmetry: the $q$-CFT of $q$-holography~\cite{Almheiri:2024ayc,Belaey:2025}. \rev{As in JT gravity, where the Schwarzian theory arises both as the infrared limit of the SYK model~\cite{Maldacena:2016hyu,Kitaev:2017awl} and as a limit of two-dimensional Liouville conformal field theory~\cite{Mertens:2017mtv,Mertens:2018fds}, one expects this boundary theory to be a suitable dimensional reduction of a two-dimensional conformal field theory, here with quantum-group symmetry.} \rev{Two observations in this paper support this picture. First, consider the primary part of the matter insertion, the $b$-deformation of $(1+e^{\Lambda})^{-\Delta}=\cosh(\frac L2)^{-2\Delta}$, obtained by replacing the Beta function in its Mellin kernel by the $b$-Beta function. By the $b$-binomial theorem \eqref{eq:corr_binom} it reads
\begin{equation}\label{eq:disc_primary}
    G^{(b)}_\Delta(\Lambda)=\int\frac{\dd k}{2\pi i}\,e^{k\Lambda}\,2\pi b\,\frac{S_b(b\Delta+bk)\,S_b(-bk)}{S_b(b\Delta)}=e^{-\Delta\Lambda/2}\,\Big|S_b\Big(\frac Q2-\frac{b\Delta}2+\frac{i\Lambda}{2\pi b}\Big)\Big|^2 ,
\end{equation}
and it has exactly the same structure as the solution of the $q$-Ward identities of the $q$-CFT two-point function~\cite{Belaey:2025}. Moreover, the full matter line \eqref{eq:corr_gbb} is built from these primary insertions. By the degenerate hyperbolic Barnes lemma \eqref{eq:corr_barnes2},
\begin{equation}\label{eq:disc_decomp}
    G^{\rm bb}_\Delta(\Lambda)=\frac{S_b(2b\Delta)}{S_b(b\Delta)^2}\,e^{-i\pi b^2\Delta^2}\int\frac{\dd\sigma}{i}\,e^{-i\pi Q\sigma}\,\frac{S_b(b\Delta+\sigma)^2\,S_b(-\sigma)}{S_b(2b\Delta+\sigma)}\;G^{(b)}_{\Delta+\sigma/b}\big(\Lambda-i\pi b(b\Delta+\sigma)\big).
\end{equation}
Closing the contour to the right, the residue decomposition of \eqref{eq:disc_decomp} writes the matter line \eqref{eq:corr_GFb} as a sum of the primary insertions \eqref{eq:disc_primary}. The poles of $S_b(-\sigma)$ give primaries of weights $\Delta+m+n/b^2$, the leading one with unit coefficient at large $\Lambda$, while the zeros of $S_b(2b\Delta+\sigma)$ give a reflected family of weights $Q/b-\Delta+m+n/b^2$; together they reproduce the two towers of \eqref{eq:tower}. Each primary is evaluated at a point shifted by $-i\pi b^2$ times its weight, a shift that vanishes in the JT limit, where the coefficients reduce to those of \eqref{eq:desce}, i.e.\ to the series coefficients of ${}_2F_1\big(\Delta,\Delta;2\Delta;\mathrm{sech}^2\frac L2\big)$. A solution of the $q$-Ward identities was found to emerge close to the boundary of sinh dilaton gravity in~\cite{Belaey:2025}, where it was related to the effective AdS$_2$ geometry. Here the quantum disk follows from the quantization of the phase space at finite cutoff and from the matching of the Hamiltonians, and the decomposition above shows that, at any distance from the boundary, the matter line is a sum of insertions with the structure of this solution. Second, the two-point function \eqref{eq:corr_Mb6j} is an integral of $b$-$6j$ symbols, which are the natural crossing kernels} of a theory with this symmetry. A precise holographic statement would be that the bulk-to-boundary limit of the matter line \eqref{eq:corr_gbb} reproduces the $q$-CFT two-point function, and that the \rev{Witten diagrams with more insertions} reproduce its higher-point functions.

\rev{\paragraph{The eigenfunctions as conformal partial waves.} The $q$-CFT picture suggests a further link, between the gravitational wavefunctions and the conformal partial waves of the boundary theory. In the JT limit it rests on a known fact of $SL(2,\mathbb R)$ harmonic analysis: one-dimensional conformal partial waves with unequal external weights are matrix elements of $SL(2,\mathbb R)$ between boost eigenstates~\cite{Dolan:2003hv,Isachenkov:2016gim,Schomerus:2016epl,Kitaev:2017hnr}. The finite-cutoff wavefunctions are matrix elements of exactly this type, with both boost charges fixed by the boundary dilaton~\cite{Griguolo:2026eoi,SinhBox}. Accordingly, after the change of variable $\zeta=-1/u=1-\abs{z}^2$, the radial equation \eqref{eq:qd_classical2} coincides with the Casimir equation of a four-point partial wave~\cite{Dolan:2003hv,Maldacena:2016hyu}. In this reading the cross ratio is the radial variable of the disk, the energy is the Casimir of the exchanged weight $h=\frac12+is$, and the magnetic field is the difference of the external weights. The wavefunction regular at $O$ is the combination of block and shadow block, i.e.\ the partial wave.

At finite $b$ one may expect the same structure, with $SL(2,\mathbb R)$ replaced by its quantum deformation. The eigenfunctions \eqref{eq:sinh_chiFb} are $b$-hypergeometric functions. In the harmonic analysis of the modular double, Clebsch--Gordan coefficients and $b$-$6j$ symbols are built from these functions~\cite{Ponsot:1999uf,Ponsot:2001}, and the two-point function of Section~\ref{sec:corr} is indeed an integral of $b$-$6j$ symbols. On the boundary side, $q$-deformed conformal blocks arise in the deformed Virasoro algebra as $q$-hypergeometric functions~\cite{Shiraishi:1995rp,Awata:2009ur}, and paired with their modular duals they give double-sine integrals of the same type as $F_b$~\cite{Nieri:2013yra}. This suggests that the eigenfunctions of finite-cutoff sinh dilaton gravity are the conformal partial waves of the $q$-CFT, with exchanged weight $\frac12+is$, external weights fixed by the boundary dilaton and cross ratio given by the radial variable of the quantum disk.}

\rev{\paragraph{Line bundles on the quantum disk and sine dilaton gravity.} A magnetic field can also be placed on the quantum disk through its homogeneous line bundles~\cite{Shklyarov:1998,Shklyarov:1998b}. The invariant Laplacian on sections of the line bundle of weight $\mu$ was considered in Section 5.3 of~\cite{Berkooz:2025classS}, as a $q$-deformation of the particle on the hyperbolic disk of~\cite{Kitaev:2018wpr}, and its partition function was shown to reproduce the Schur half-index of the $n_F=4$ theory once $q^\mu$ is continued to a phase. On rotationally symmetric sections its eigenvalue equation is a three-term recursion on the Fock lattice $y=q^{2n}$ of the disk, which terminates at the centre $n=0$ and is solved by Al-Salam--Chihara polynomials with parameters $q^{1\pm\mu}$. Written in the disk variable $y=-1/(qu)$ of \eqref{eq:qd_swapvar}, the sinh operator \eqref{eq:qd_H} becomes a recursion on the same lattice, and the two read
\begin{equation}\label{eq:disc_BKS}
\begin{aligned}
    &\mathcal H_{\rm sinh}\,\phi_n=\tfrac12\big(1-q^{2n+2}\big)\phi_{n+1}+\tfrac12\big(1-q^{2n-4B}\big)\phi_{n-1}+q^{2n+1-2B}\,\phi_n,\\
    &\frac{(1-q^2)^2}{q}\,C_\mu\,\phi_n=q^{\mu}\big(1-q^{2n+2}\big)\phi_{n+1}+q^{2-\mu}\big(1-q^{2n}\big)\phi_{n-1}+\big(q^{2n+2}(q^\mu+q^{-\mu})-1-q^2\big)\phi_n,
\end{aligned}
\end{equation}
with $C_\mu$ the Casimir of~\cite{Berkooz:2025classS} on sections of weight $\mu$. Reading the gauge invariants of Appendix \ref{app:exponents} off \eqref{eq:disc_BKS}, the two operators carry the same field and differ in where it is regular. In the second one the zeros of the hopping product coincide and the recursion terminates at the centre, so the field carries no flux there, while the exponents at $x=0$, i.e.\ at $O$, differ by $\mu$, the Aharonov--Bohm flux seen from $O$. In the sinh operator the zeros of the hopping product are split by $q^{-4B}$, the holonomy $2B$ around the centre, while the exponents at $O$ coincide. For $\mu=2B$ these are the two descriptions of the field discussed in Section \ref{subsec:qd_neutral}, regular around the centre of the disk or around $O$, and correspondingly the Al-Salam--Chihara parameters $(q^{1-2B},q^{1+2B})$ of~\cite{Berkooz:2025classS} differ from ours, $(q^{1-2B},q^{1-2B})$ in Appendix \ref{app:realq}, by $q^{-4B}$ in one entry. In fact, the operator of~\cite{Berkooz:2025classS} follows from the construction of Section \ref{subsec:qd_field} with the roles of the two points exchanged. Starting from the Laplacian on waves of angular momentum $m$ in the disk variable, one keeps the diagonal term, which fixes the exponents at $O$, and removes the flux at the centre by requiring that the zeros of the hopping product coincide.\footnote{\rev{The backward hopping $1-y$ of \eqref{eq:qd_boxm} already vanishes at the centre, and the zeros of the hopping product coincide only if the constant term of the forward hopping is replaced, $q^{-2}\to q^{2m-2}$. The resulting operator coincides with the second line of \eqref{eq:disc_BKS} for $\mu=m$, up to normalization, a radial gauge and a constant shift of the eigenvalue. Unlike the condition at $O$, this condition is linear and has a unique solution. Its classical counterpart is $\mathcal L_{m=2B}-B^2/(1+u)$, in which the barrier of \eqref{eq:qd_Lm} is removed at $L=i\pi$ rather than at $O$, and which is the image of \eqref{eq:qd_LBLm} under $u\to-1-u$.}} That the same construction reproduces the operator obtained in~\cite{Berkooz:2025classS} from the line-bundle Laplacian is a non-trivial check of the way the field is coupled in Section~\ref{subsec:qd_field}.

For real $q$, where the quantum disk is originally defined, the Fock lattice is anchored at the centre (Appendix~\ref{app:realq}), so that the field should be regular there rather than at $O$. This suggests that the construction of~\cite{Berkooz:2025classS} is the natural candidate for the bulk description of sine dilaton gravity at finite cutoff, a regime for which deformations of DSSYK have also been proposed from the boundary perspective~\cite{Aguilar-Gutierrez:2026dssyk,Aguilar-Gutierrez:2026cmc,Aguilar-Gutierrez:2026askey}. The discreteness of the lattice would then be the discreteness of the length of the Einstein--Rosen bridge in sine dilaton gravity~\cite{Blommaert:2024ymv}. It would be interesting to establish this identification by a canonical quantization of sine dilaton gravity at finite cutoff along the lines of Section \ref{sec:sinh}.}

\section*{Acknowledgments}
We thank Andreas Belaey, Andreas Blommaert, Luca Griguolo, Thomas Mertens, Lorenzo Russo, Domenico Seminara and Alex Tarana for useful discussions. JP acknowledges
financial support from the European Research Council (grant BHHQG-101040024). Funded
by the European Union. Views and opinions expressed are however those of the author(s)
only and do not necessarily reflect those of the European Union or the European Research
Council. Neither the European Union nor the granting authority can be held responsible for
them.

\appendix

\section{Double sine functions and integral identities}
\label{app:doublesine}

In this appendix we collect the properties of the double sine function used in the main text. We define it in terms of the Barnes double Gamma function as
\begin{equation}
    S_b(x)=\frac{\Gamma_b(x)}{\Gamma_b(Q-x)},\qquad Q=b+b^{-1},
\end{equation}
so that it satisfies the shift and reflection relations
\begin{equation}
    S_b(x+b^{\pm1})=2\sin(\pi b^{\pm1}x)\,S_b(x),\qquad S_b(x)\,S_b(Q-x)=1 .
\end{equation}
For large imaginary argument it behaves as a Gaussian, $S_b(x)\sim e^{\pm\frac{i\pi}2x(Q-x)}$ for $\mathrm{Im}\,x\to\pm\infty$, and it is convenient to introduce the function
\begin{equation}
    G(x)=e^{i\pi x(x-Q)/2}S_b(x).
\end{equation}
In the limit $b\to0$ the double sine reduces to the Gamma function,
\begin{equation}\label{eq:app_Sbclassical}
    S_b(bx)\simeq(2\pi b^2)^{x-\frac12}\,\frac{\Gamma(x)}{\sqrt{2\pi}} ,
\end{equation}
and every identity below reduces to its classical counterpart.

The first identity is the $b$-binomial theorem~\cite{Faddeev:2000}, which is the $b$-analogue of the Mellin representation of $(1+x)^{-\alpha}$,
\begin{equation}\label{eq:corr_binom}
    \int_{i\mathbb R}\frac{\dd\tau}{i}\,e^{2\pi i\tau\beta}\,\frac{G(\alpha+\tau)}{G(Q+\tau)}=\frac{G(\alpha)G(\beta)}{G(\alpha+\beta)},
\end{equation}
where the contour separates the poles at $-\alpha$ and at the origin. The second is the hyperbolic Barnes lemma, valid when the parameters satisfy the balancing condition $\sum_j(\alpha_j+\beta_j)=Q$,
\begin{equation}\label{eq:corr_barnes3}
    \int_{i\mathbb R}\frac{\dd\tau}{i}\prod_{j=1}^3S_b(\alpha_j+\tau)S_b(\beta_j-\tau)=\prod_{j,k=1}^3S_b(\alpha_j+\beta_k) .
\end{equation}
Its degenerate version, with only two pairs of double sines and an exponential, is the one needed for the Liouville gravity two-point function. Denoting by $\Sigma$ the sum of the four parameters, it reads
\begin{equation}\label{eq:corr_barnes2}
    \int_{i\mathbb R}\frac{\dd\tau}{i}\,e^{\mp i\pi\Sigma\tau}\prod_{j=1}^2S_b(\alpha_j+\tau)S_b(\beta_j-\tau)=e^{\pm i\pi(\alpha_1\alpha_2-\beta_1\beta_2)}\,\frac{\prod_{j,k=1}^2S_b(\alpha_j+\beta_k)}{S_b(\Sigma)} .
\end{equation}
Finally, the $b$-$6j$ symbol of the modular double~\cite{Ponsot:2001,Teschner:2012} admits a representation as a balanced integral of eight double sines,
\begin{equation}\label{eq:corr_TV}
    \Big\{\begin{matrix}\alpha_1&\alpha_2&\alpha_s\\ \alpha_3&\alpha_4&\alpha_t\end{matrix}\Big\}_b=D(\underline\alpha)\int_{\mathcal C}\dd u\;\prod_{k=1}^4S_b(u-U_k)\,S_b(V_k-u),
\end{equation}
where the parameters are the sums of the labels around the faces and vertices of the tetrahedron,
\begin{equation}
    U=(\alpha_{12s},\alpha_{s34},\alpha_{23t},\alpha_{1t4}),\qquad V=(\alpha_{1234},\alpha_{st13},\alpha_{st24},2Q),
\end{equation}
with $\alpha_{ijk}=\alpha_i+\alpha_j+\alpha_k$ and similarly for four indices. They satisfy the balancing condition $\sum_k(V_k-U_k)=2Q$. The prefactor $D$ is a product of four triangle factors, and the contour separates the two families of poles.

\section{The Laplacian on the quantum disk}
\label{app:qdlap}

In this appendix we collect the computations on the Laplacian of the quantum disk used in Section~\ref{sec:qdisk}: its action on radial functions, its action on waves of definite angular momentum, and the way in which the exponents of the classical equation, and in particular the magnetic field, are encoded in the resulting $q$-difference operators.

\subsection{On radial functions}
\label{app:qdlap_radial}

We first derive the form \eqref{eq:qd_SSV} of the Laplacian \eqref{eq:qd_AP} on radial functions. To this end we compare the action of $\Box$ on a basis of radial functions with that of the $q$-difference operator on the right-hand side of \eqref{eq:qd_SSV}. We use the notation
\begin{equation}
    [k]_t=\frac{1-t^k}{1-t},\qquad y=1-zz^* .
\end{equation}

Rotation-invariant polynomials on the quantum disk are spanned by the normal-ordered monomials $z^kz^{*k}$ and the first step is to express them in terms of the radial variable. Since $zz^*=1-y$ and the relation \eqref{eq:qd_relation} implies $zy=q^{-2}yz$, one finds
\begin{equation}
    z^kz^{*k}=z^{k-1}(1-y)z^{*\,k-1}=\big(1-q^{-2(k-1)}y\big)\,z^{k-1}z^{*\,k-1},
\end{equation}
and repeating $k$ times one obtains the $q$-Pochhammer symbol
\begin{equation}\label{eq:app_Pk}
    z^kz^{*k}=P_k(y)\equiv(y;q^{-2})_k=(1-y)(1-q^{-2}y)\cdots(1-q^{-2(k-1)}y).
\end{equation}

Next we compute the action of the $q$-derivatives on a monomial. The derivatives are defined by their commutation relations \eqref{eq:qd_qder} with the coordinates, together with the condition that they annihilate the constant function, so that to differentiate a monomial we move the derivatives to the right until they hit the identity. One can easily show that each derivative passing through its own coordinate produces a $q$-number,
\begin{equation}
    \partial_{z^*}z^{*b}=[b]_{q^2}\,z^{*\,b-1}+q^{2b}z^{*b}\,\partial_{z^*},\qquad \partial_zz^a=[a]_{q^{-2}}\,z^{a-1}+q^{-2a}z^a\,\partial_z ,
\end{equation}
while passing through the other coordinate only produces a power of $q$. Acting first with $\partial_{z^*}$ and then with $\partial_z$, and multiplying by the prefactor of \eqref{eq:qd_AP}, we find
\begin{equation}\label{eq:qd_boxmono}
    \Box\big(z^az^{*b}\big)=q^{-1-2a}\,[a]_{q^{-2}}[b]_{q^2}\;(1-zz^*)^2\,z^{a-1}z^{*\,b-1},
\end{equation}
which reduces to $(1-\abs{z}^2)^2\partial\bar\partial$ as $q\to1$. On the radial basis \eqref{eq:app_Pk}, this gives
\begin{equation}\label{eq:app_APPk}
    \Box P_k=\frac{q\,(1-q^{-2k})^2}{(1-q^2)^2}\,y^2\,P_{k-1}.
\end{equation}
We now compute the right-hand side of \eqref{eq:qd_SSV} on the same basis. Rewritten in terms of $y=1/x$, it acts as
\begin{equation}\label{eq:qd_SSVy}
    q^3(q^{-1}-q)^2\,\Box_0f=(q^{-2}-y)\big[f(q^2y)-f(y)\big]-(1-y)\big[f(y)-f(q^{-2}y)\big].
\end{equation}
Setting $f = P_k$ and using the product form of the $q$-Pochhammer symbols, the two brackets combine into a single difference and we obtain
\begin{equation}
    q^3(q^{-1}-q)^2\,\Box_0P_k=q^2(1-q^{-2k})^2\,y^2\,P_{k-1}.
\end{equation}
Comparing with \eqref{eq:app_APPk} we find that this coincides with $\Box P_k$ for every element of the basis, which proves \eqref{eq:qd_SSV}. Only algebraic manipulations were used, so the result holds for $\abs{q}=1$ as well.

Finally, we derive the identity \eqref{eq:qd_vaksman}. Written in terms of $x$, the radial Laplacian is
\begin{equation}
    q^3(q^{-1}-q)^2\,\Box_0f=\Big(q^{-2}-\frac1x\Big)\big[f(q^{-2}x)-f(x)\big]+\Big(1-\frac1x\Big)\big[f(q^2x)-f(x)\big].
\end{equation}
We perform the change of variable $x=-qu$ and conjugate by $u^{1/2}$, writing $f=u^{-1/2}g$. Since $f(q^{\mp2}x)=q^{\pm1}u^{-1/2}g(q^{\mp2}u)$ we find
\begin{equation}
    q^3(q^{-1}-q)^2\,u^{\frac12}\Box_0u^{-\frac12}g=\frac1q\Big(1+\frac qu\Big)g(q^{-2}u)+\frac1q\Big(1+\frac{q^{-1}}u\Big)g(q^2u)-\Big(q^{-2}+1+\frac{2}{qu}\Big)g .
\end{equation}
Multiplying by $q/2$ and adding $(q+q^{-1})/2$ reproduces exactly the sinh operator \eqref{eq:qd_H} at $B=0$.

\subsection{On waves of angular momentum $m$}
\label{app:mwave}

We now derive the action \eqref{eq:qd_boxm} of the quantum disk Laplacian on wavefunctions of definite angular momentum and its form \eqref{eq:qd_Hm} after the relabelling. The relation \eqref{eq:qd_relation} is invariant under the rotations $z\to e^{i\alpha}z$, $z^*\to e^{-i\alpha}z^*$, under which a normal-ordered monomial $z^az^{*b}$ has charge $a-b$. The wavefunctions of angular momentum $m\ge0$ are therefore of the form $z^mf(y)$, with $z^m$ written on the left. As a basis for the radial functions we use once again the $q$-Pochhammer symbols $P_k(y)$ of \eqref{eq:app_Pk}, so that $z^mP_k(y)=z^{m+k}z^{*k}$ is a single normal-ordered monomial.

We already computed the action on a generic monomial in \eqref{eq:qd_boxmono}. For $a=m+k$ and $b=k$ the result is proportional to $(1-zz^*)^2z^mP_{k-1}(y)$ and we thus bring $z^m$ to the left using $yz=q^2zy$. Altogether we get
\begin{equation}\label{eq:qd_boxpk}
    \Box\big(z^mP_k\big)=z^m\;q^{-1-2(m+k)}\,q^{4m}\,[m+k]_{q^{-2}}\,[k]_{q^2}\;y^2\,P_{k-1}(y).
\end{equation}
The Laplacian thus maps waves of angular momentum $m$ into waves of the same angular momentum, as it should since it commutes with rotations, and on the radial profile it lowers the degree $k$ by one as in the neutral case.

To find the operator that reproduces this action we look for an operator with the same shape as the radial Laplacian \eqref{eq:qd_SSVy}, i.e.\ a forward and a backward $q$-difference, each multiplied by a coefficient linear in $y$. Our ansatz is of the form
\begin{equation}\label{eq:form}
    \mathcal D_mf(y)=(\alpha-\beta y)\big[f(q^2y)-f(y)\big]-(1-y)\big[f(y)-f(q^{-2}y)\big],
\end{equation}
and fix $\alpha$ and $\beta$ so as to reproduce \eqref{eq:qd_boxpk}.
Setting $f=P_k$ in \eqref{eq:form} and exploiting the product form of the $q$-Pochhammer symbols, one finds
\begin{equation}
    \mathcal D_m\,P_k=-(1-q^{-2k})\,y\,P_{k-1}(y)\,\Big[(q^2\alpha-1)+q^2\,(q^{-2k}-\beta)\,y\Big].
\end{equation}
In order to match with \eqref{eq:qd_boxpk}, one is led to set
\begin{equation}
    \alpha=q^{-2},\qquad \beta=q^{2m},
\end{equation}
for which
\begin{equation}
    \mathcal D_m\,P_k=q^{2+2m}(1-q^{-2k})(1-q^{-2m-2k})\,y^2\,P_{k-1}.
\end{equation}
Writing the $q$-numbers in \eqref{eq:qd_boxpk} in the same form, the constants agree for every $k$ provided the radial operator is normalized as $q^{-1}(1-q^2)^{-2}\mathcal D_m$. Since the $P_k$ form a basis, this proves \eqref{eq:qd_boxm} for every radial profile. At $m=0$ it reduces to $\Box_0$, by \eqref{eq:qd_SSVy}. 

It remains to perform the relabelling. Setting $y=-1/(qu)$, a dilation of $y$ by $q^{\pm2}$ becomes a dilation of $u$ by $q^{\mp2}$ and the two coefficients become
\begin{equation}
    q^{-2}-q^{2m}y=q^{-2}\Big(1+\frac{q^{2m+1}}u\Big),\qquad 1-y=q^{-1}\Big(q+\frac1u\Big).
\end{equation}
We normalize the operator as in the neutral identity \eqref{eq:qd_vaksman}, i.e.\ we set $\mathcal H_m=\frac{q^4}2(q^{-1}-q)^2\,\Box_m+\frac{q+q^{-1}}2$, without the conjugation by $u^{1/2}$, which only changes the radial gauge, fixed below. In terms of $u$ this gives
\begin{equation}\label{eq:qd_Hmraw}
    \mathcal H_m\chi=\frac12\Big(q^{-1}+\frac{q^{2m}}u\Big)\chi(q^{-2}u)+\frac12\Big(q+\frac1u\Big)\chi(q^2u)-\frac{1+q^{2m}}{2u}\,\chi(u).
\end{equation}
The hopping product of this operator, i.e.\ the product $C_m(u)=A_-(u)\,A_+(q^{-2}u)$ of its backward and shifted forward hopping coefficients, defined in \eqref{eq:qd_hopping} below, is
\begin{equation}
    C_m(u)=\frac14\Big(1+\frac qu\Big)\Big(1+\frac{q^{2m+1}}u\Big).
\end{equation}
As we discuss in Appendix~\ref{app:exponents}, the hopping product and the diagonal term are invariant under rescalings of the wavefunction, which only decide how the two zeros of $C_m$ are shared between the forward and the backward hopping. We are therefore led to choose the radial gauge for which they are shared as in the sinh operator, with the zero at $u=-q$ in the backward hopping, which gives the form \eqref{eq:qd_Hm}. 

\subsection{Exponents of the $q$-deformed operator and their physical interpretation}
\label{app:exponents}

In this subsection we explain the property of $q$-difference operators used in Section \ref{subsec:qd_field}. The general theory of linear $q$-difference equations, including the notion of characteristic exponents at the fixed points $0$ and $\infty$ of the dilations, goes back to~\cite{Adams:1928,Sauloy:2003} and the relevant class of equations is that of Heine's basic hypergeometric equation~\cite{Gasper:2004}. The property we need is that the exponents at $O$ are fixed by the diagonal term, while the gauge potential at $L=i\pi$ is encoded in the hopping coefficients. The interpretation in terms of angular momentum, flux and holonomy on the quantum disk is, to our knowledge, new. Requiring no flux at $O$ and exponent difference $2B$ at $L=i\pi$ leaves only the sinh operator \eqref{eq:qd_H} of the main text, up to gauge transformations.

\paragraph{Gauge invariants.} Consider a three-term operator
\begin{equation}\label{eq:qd_heine}
    \mathcal A\chi(u)=A_+(u)\,\chi(q^2u)+A_-(u)\,\chi(q^{-2}u)+A_0(u)\,\chi(u),
\end{equation}
with hopping coefficients $A_\pm$ and diagonal term $A_0$. A rescaling of the wavefunction, $\chi\to g\chi$, multiplies the hopping coefficients by $g(q^{\pm2}u)/g(u)$, but leaves invariant the diagonal term and the hopping product
\begin{equation}\label{eq:qd_hopping}
    C(u)=A_-(u)\,A_+(q^{-2}u).
\end{equation}
As in the JT limit treated in Section \ref{sec:jt}, these rescalings are radial gauge transformations and $A_0$ and $C$ are the gauge invariants: two operators related by a rescaling describe the same physics. A rescaling only decides how the zeros of $C$ are shared between the forward and the backward hopping.

An example is the relation between the operator \eqref{eq:qd_H} used in the main text and the symmetrically ordered Hamiltonian introduced in \cite{SinhBox}. The latter acts on $\psi=\Phi_\nu\chi$ in \eqref{eq:sinh_Phinu}, and the equation obeyed by $\psi$ is~\cite{SinhBox}:
\begin{equation}\label{eq:app_symmetric}
    \mathcal H_{\rm sym}\psi=\sqrt{C(q^2u)}\;\psi(q^2u)+\sqrt{C(u)}\;\psi(q^{-2}u)-\frac{q^{-2B}}{u}\,\psi(u),\qquad C(u)=\frac14\Big(1+\frac qu\Big)\Big(1+\frac{q^{1-4B}}u\Big).
\end{equation}
Here the hopping product is shared equally between the two directions, which makes the operator symmetric with respect to the flat measure in $\Lambda$. The two forms have the same diagonal term and the same hopping product: they are the same operator in two radial gauges, related by the Faddeev phase in \eqref{eq:sinh_Phinu}, which is the $q$-analogue of the radial phase of \eqref{eq:qd_chiR}.

For operators whose coefficients are linear in $1/u$, the class of Heine's basic hypergeometric equation, we can use this freedom, together with an affine redefinition of the operator, to fix a convenient gauge. We set the constant parts of the hopping coefficients to $\frac12$ and of the diagonal term to zero, and assign one zero of the hopping product to each direction:
\begin{equation}\label{eq:app_heinegauge}
    \begin{gathered}
    A_-(u)=\frac12\Big(1+\frac{c_1}u\Big),\qquad A_+(u)=\frac12\Big(1+\frac{q^{-2}c_2}u\Big),\qquad A_0(u)=-\frac au,\\
    C(u)=\frac14\Big(1+\frac{c_1}u\Big)\Big(1+\frac{c_2}u\Big).
    \end{gathered}
\end{equation}
The operator is then specified by three numbers: the locations $u=-c_1$, $u=-c_2$ of the zeros of the hopping product and the coefficient $a$ of the diagonal term. 

\paragraph{Exponents at $O$: the diagonal term.} The dilations of $u$ have two fixed points, $u=0$ and $u=\infty$, and only there do solutions behave as powers. Near $u=0$ a dilation acts on $\chi\simeq u^\rho$ by multiplication and the terms proportional to $1/u$ cancel if
\begin{equation}
    \frac{q^{-2}c_2}{2}\,X+\frac{c_1}{2X}-a=0,\qquad X=q^{2\rho}.
\end{equation}
The two roots of this indicial equation are the exponents at $O$, and they are fixed by the diagonal term $a$. Their meaning follows from the classical operator \eqref{eq:qd_Lm}: near $u=0$ the centrifugal term $-m^2/4u$ dominates the potential, and $u^\rho$ solves the equation for $\rho=\pm m/2$. The exponent difference at $O$ is therefore the angular momentum around $O$ shifted by any flux of the field through $O$, i.e.\ the effective angular momentum around $O$.

At $u=\infty$ only the constant parts survive, and the exponents $X=q^{\pm2is}$ encode the energy.

\paragraph{The gauge potential at $L=i\pi$: the hopping product.} The point $L=i\pi$, i.e.\ $u=-1$, is not a fixed point of the dilations and there is no local power-law expansion around it. Its information is instead stored in the two zeros of the hopping product and their splitting reduces to the exponent difference at the third classical singular point $u=-1$. 

As $q\to1$ both $c_i$ tend to one, so the zeros merge at $u=-1$. We thus write $c_i=e^{\varepsilon\gamma_i}$, with $\varepsilon=2\pi ib^2$, where $\gamma_i$ parametrizes the splitting of the two zeroes, and we now show that $\gamma_2-\gamma_1$ is the exponent difference at $u=-1$.

Since $C$ is gauge invariant, we can use the symmetric form \eqref{eq:app_symmetric}, and we define the hopping amplitude $f=2\sqrt C$. Writing $u=e^\Lambda$ and expanding in $\varepsilon$ we find
\begin{equation}\label{eq:qd_hopexp}
    \mathcal A\psi=\big(f+A_0\big)\psi+\frac{\varepsilon^2}2\,\partial_\Lambda\big(f\,\partial_\Lambda\psi\big)+\frac{\varepsilon^2}{8}\,(\partial_\Lambda^2f)\,\psi+O(\varepsilon^4).
\end{equation}
The hopping amplitude $f$ is the coefficient of the kinetic term, but through the first term it is also a potential. A singularity of $f$ at $u=-1$ is therefore a singularity of the potential there.

To find it, write $1+c_i/u=1+1/u+\delta_i$, with $\delta_i=(c_i-1)/u\simeq\varepsilon\gamma_i/u$, and introduce the mean and half-difference of the two shifts
\begin{equation}
\bar\delta=\frac12(\delta_1+\delta_2) ,\qquad d=\frac12(\delta_1-\delta_2).
\end{equation}
Then the hopping amplitude $f$ can be rewritten as:
\begin{equation}
    f=\sqrt{\Big(1+\frac1u+\bar\delta\Big)^2-d^2}=1+\frac1u+\bar\delta-\frac{d^2}{2(1+\frac1u)}+\dots ,
\end{equation}
where we expanded in both $d,\bar{\delta}$.
If the two zeros coincide, i.e.\ $d=0$, $f$ is a polynomial in $1/u$ with no singularity at $u=-1$. If they are split, the last term has a pole at $u=-1$ and in the classical limit it reads
\begin{equation}
    -\frac{d^2}{2(1+\frac1u)}\simeq-\frac{\varepsilon^2}{8}\,\frac{(\gamma_2-\gamma_1)^2}{u\,(1+u)} .
\end{equation}
This term is of order $\varepsilon^2$, like the kinetic term. The remaining terms of order $\varepsilon^2$ in \eqref{eq:qd_hopexp} reorganize into the neutral operator of \eqref{eq:qd_classical2}, together with terms that only have poles at $u=0$ and affect only the exponents at $O$. The pole just found is exactly the centrifugal barrier of \eqref{eq:qd_Lm}, with angular momentum $m=\gamma_2-\gamma_1$. The exponent difference at $L=i\pi$ is therefore $m$ for the wave \eqref{eq:qd_Hm}, and $2B$ for the sinh operator, so the continuation performed in \eqref{eq:hoppingterms} correctly reconstructs the expected holonomy of the field around $L=i\pi$.

\paragraph{The little $q$-Jacobi operator.}\label{app:qjacobi} To corroborate this identification of the parameters, we now show that the sinh operator \eqref{eq:qd_H} is the little $q$-Jacobi operator with parameters $(\alpha,\beta)=(0,2B)$. The little $q$-Jacobi functions are the $q$-analogues of the Jacobi functions and arise as matrix elements of representations of $U_q(\mathfrak{su}(1,1))$~\cite{Kakehi:1995,Koelink:2000}. Little $q$-Jacobi functions are expressed as Heine hypergeometric series~\cite{Gasper:2004}: 
\begin{equation}\label{eq:Heineseries}
    {}_2\varphi_1\Big(\genfrac{}{}{0pt}{}{a_1,\ a_2}{c};p,\,x\Big)=\sum_{m\ge0}\frac{(a_1,a_2;p)_m}{(p,c;p)_m}\,x^m .
\end{equation}
Setting, for parameters $(\alpha,\beta)$,
\begin{equation}
    a=q^{\alpha+\beta+1},\qquad b_J=q^{\alpha-\beta+1} ,
\end{equation}
one finds that the functions ${}_2\varphi_1(a\sigma,a/\sigma;ab_J;q^2;-b_JX/a)$ are eigenfunctions of the little $q$-Jacobi operator $\mathcal{L}_{(\alpha,\beta)}$:
\begin{equation}
    \mathcal L_{(\alpha,\beta)}f=\frac{1}{2a}\Big[a^2\Big(1+\frac1X\Big)\big(f(q^2X)-f(X)\big)+\Big(1+\frac{aq^2}{b_JX}\Big)\big(f(q^{-2}X)-f(X)\big)\Big]+\frac{a+a^{-1}}{2}\,f ,
\end{equation}
with eigenvalue $\frac12(\sigma+\sigma^{-1})$.
Conjugating by the power of $X$ that is multiplied by $1/a$ under $X\to q^2X$, i.e.\ $f=a^{-\log_{q^2}X}g$, we can take the above operator to a form where both hopping coefficients start with $\frac12$, i.e.
\begin{equation}
    a^{\log_{q^2}X}\,\mathcal L_{(\alpha,\beta)}\,a^{-\log_{q^2}X} g=\tfrac12\Big(1+\frac1X\Big)g(q^2X)+\tfrac12\Big(1+\frac{aq^2}{b_JX}\Big)g(q^{-2}X)-\frac{a+q^2/b_J}{2X}\,g .
\end{equation}
Comparing term by term with \eqref{eq:qd_H}, the forward hopping fixes the rescaling $X=q^{1+4B}u$, and the backward hopping and the diagonal term give
\begin{equation}
    \frac{a}{b_J}=q^{4B},\qquad a+\frac{q^2}{b_J}=2q^{1+2B},
\end{equation}
which together imply $(a-q^{1+2B})^2=0$. The solution is therefore unique, $a=q^{1+2B}$ and $b_J=q^{1-2B}$, i.e.\ $(\alpha,\beta)=(0,2B)$, and we recover \eqref{eq:qd_identity},
\begin{equation}
    \mathcal{H}_{\mathrm{sinh}}=X^{\frac12+B}\,\mathcal L_{(0,2B)}\,X^{-\frac12-B},\qquad X=q^{1+4B}u .
\end{equation}
With $\sigma=q^{-2is}$ the eigenfunctions are the Heine series \eqref{eq:qd_heinesol}. As $q\to1$ the Heine series tends to ${}_2F_1$ and the little $q$-Jacobi operator tends to the Jacobi operator~\cite{Koornwinder:1984}. The deformation therefore preserves the meaning of both parameters: $\alpha=0$ is the absence of flux at $O$ and $\beta=2B$ is the holonomy of the magnetic field around $L=i\pi$, exactly as found from the gauge invariants above.

\section{Analytic continuation to real $q$ and the swap of anchors}
\label{app:realq}

In Section \ref{sec:qdisk} the sinh operator was identified with the quantum-disk Laplacian after the relabelling $x=-qu$, which places the endpoint $O$ of the Einstein--Rosen bridge at the point $L_{\rm disk}=i\pi$. In this appendix we start from our eigenfunction \eqref{eq:sinh_eigenfunction} and continue it to real $q$, where the quantum disk is originally defined. We show two things. First, around $O$ it is a Heine series in the variable $x$ up to an infinite tower of non-perturbative corrections. Second, beyond the centre of the disk it develops simple poles on the Fock lattice anchored at the centre and the residues are Al-Salam--Chihara polynomials. The second step can be regarded as a generalization at any value of the field $B$ of the argument in \cite{Blommaert:2024whf}, where a sum over periodic images reduced the strong-field wavefunctions to $q$-Hermite polynomials.

We set $b^2=i\lambda/2\pi$ with $\lambda>0$, so that $q=e^{-\lambda/2}$, and $E=\cosh(2\pi b^2s)=\cos\theta$ with $\theta=\lambda s$. We take the boundary dilaton on the imaginary contour, $\nu=-i\kappa$, so that $B=\kappa$ is real and the operator \eqref{eq:qd_H} has real coefficients. This is the natural contour deformation that should be relevant for sine dilaton gravity, where the holographic screen is located at complex infinity in the imaginary direction~\cite{Blommaert:2024ymv,Belaey:2025}. We write the eigenfunction \eqref{eq:sinh_eigenfunction} as
\begin{equation}\label{eq:rq_mellin}
    \chi_s(\Lambda)=N_s\int_{c+i\mathbb R}\frac{\dd t}{2\pi i}\,u^t f(t),\qquad f(t)=e^{-2\pi b^2\nu t}\,\frac{S_b(bt\pm ibs)\,S_b(\frac b2+ib\nu-bt)}{S_b(\frac b2-ib\nu+bt)},\qquad u=e^\Lambda .
\end{equation}
We denote the dual base as $\tilde q=e^{-2\pi i/b^2}=e^{-4\pi^2/\lambda}$ and we express the double sine function in terms of the Faddeev quantum dilogarithm, which in turn can be expressed as a ratio of $q$-Pochhammer symbols~\cite{Faddeev:1993rs,Faddeev:2000}:
\begin{equation}\label{eq:rq_prod}
    S_b(x)=C\,e^{\frac{i\pi}2x(x-Q)}\,\frac{(q^2e^{-2\pi ibx};q^2)_\infty}{(e^{-2\pi ix/b};\tilde q)_\infty},\qquad (a;p)_\infty=\prod_{k\ge0}(1-ap^k),
\end{equation}
with $C$ a constant. We use \eqref{eq:rq_prod} to write each double sine function in \eqref{eq:rq_mellin} in this form. The Gaussians cancel up to a linear exponential and one finds
\begin{equation}\label{eq:rq_f}
    f(t)=K\,(-q)^t\,G(z)\,\omega(t),\qquad z=q^{2t},
\end{equation}
with $K$ a constant and
\begin{equation}\label{eq:rq_G}
    \begin{gathered}
    G(z)=\frac{(q^{1-2B}z,\,q^{1-2B}z;q^2)_\infty}{(ze^{i\theta},\,ze^{-i\theta};q^2)_\infty}=\sum_{n\ge0}\frac{Q_n(E;q^{1-2B},q^{1-2B}\,|\,q^2)}{(q^2;q^2)_n}\,z^n,\\
    \omega(t)=\frac{(\tilde q\,e^{2\pi i t\pm2\pi s};\tilde q)_\infty}{(-e^{2\pi i(t-B)},\,-\tilde q\,e^{2\pi i(t-B)};\tilde q)_\infty}.
    \end{gathered}
\end{equation}
The second expression for $G$ is the generating function of the Al-Salam--Chihara polynomials $Q_n$~\cite{Koekoek:2010}. The factorization \eqref{eq:rq_f} separates the two periodicities of the problem:
\begin{itemize}
    \item $G$ depends on $t$ only through $z$ so it is invariant under $t\to t+2\pi i/\lambda$;
    \item $\omega$ depends on $t$ only through $e^{2\pi it}$ so it is invariant under $t\to t+1$. It is a dual copy of $G$, with numerator and denominator exchanged.
\end{itemize}
The difference equation \eqref{eq:qd_H}, written in Mellin space, is a first-order recursion under $t\to t+1$ which fixes $(-q)^tG(z)$. It leaves a periodic factor free and $\omega$ is the choice made by our boundary condition.

Let us see how. As $u\to0$ the integral \eqref{eq:rq_mellin} is controlled by the poles of $f$ to the right of the contour. $G$ has no poles there, since its poles sit at $t=\pm is-k$ with $k\ge0$. The poles on the right therefore come from $\omega$, at $t=B+\frac12+m$. The same poles repeat on the left and there they must not survive, since the poles of $f$ on the left control the behaviour of the wavefunction as $u\to\infty$, where the difference equation only allows the plane waves $u^{\pm is}=e^{\pm is\Lambda}$ (up to powers of $1/u$): these are produced precisely by the poles of $G$ at $t=\pm is-k$.\footnote{A left copy of the poles at $t\in B+\frac12+\mathbb Z$ would instead produce powers $u^{B+\frac12-k}$, which the equation does not allow at infinity.} The left copies are absorbed by the double zeros of $G$ at $t=B+\frac12-k$ with $k\ge1$ and this works precisely if the poles of $\omega$ are at most double. The order of these poles is therefore not fixed by the equation and on the right the two options differ: a simple pole produces a pure power, $u^{\frac12+B+m}$, while a double pole produces $u^{\frac12+B+m}(a+b\log u)$.
These are the two solutions associated with the exponent $\frac12+B$ at $O$, which is a double root of the indicial equation.\footnote{Inserting $u^t$ into \eqref{eq:qd_H} as $u\to0$, the terms proportional to $1/u$ dominate and give $\big(q^{t-\frac12-B}-q^{\frac12+B-t}\big)^2=0$.} 

The solution regular at $O$ is the one without branch cut singularity, so it corresponds to simple poles. Indeed $\omega$ in \eqref{eq:rq_G} has simple poles at $t=B+\frac12+m$ coming from the first factor of its denominator. Picking them up, the residue of $\omega$ is the same at every $m$ and by using
\begin{equation}
    G(q^{1+2B+2m})=\frac{(q^2;q^2)^2_\infty}{(q^{1+2B}e^{\pm i\theta};q^2)_\infty}\,\frac{(q^{1+2B}e^{\pm i\theta};q^2)_m}{(q^2;q^2)_m^2},
\end{equation}
their contribution is reorganized as:
\begin{equation}\label{eq:rq_heine}
    \chi_s(\Lambda)\supset\mathcal C_s\,(qu)^{\frac12+B}\;{}_2\varphi_1\Big(\genfrac{}{}{0pt}{}{q^{1+2B}e^{i\theta},\ q^{1+2B}e^{-i\theta}}{q^2};q^2,\,x\Big),\qquad x=-qu,
\end{equation}
where we exploited the definition \eqref{eq:Heineseries} of Heine's basic hypergeometric series~\cite{Gasper:2004}.\footnote{Equation \eqref{eq:rq_heine} is the $q$-analogue of the JT wavefunction $\sinh\frac L2\cosh^{2i\nu}\frac L2\;{}_2F_1(\frac12+i\nu\pm is;1;-\sinh^2\frac L2)$ written in the relabelled variable $x$.}

Let us now have a look at the remaining poles of $\omega$, located at $e^{2\pi i(t-B)}=-\tilde q^{-k}$ with $k\ge1$, i.e.\ at $t=B+\frac12+m-2\pi ik/\lambda$, which are double poles. They are the continuation to real $q$ of the dual series of the double sine functions, which at $\abs{q}=1$ gives subleading terms $u^{\frac12+B+m+k/b^2}(a+b\log u)$. At real $q$ they add to the Heine series only exponentially small corrections.\footnote{At real $q$ $u^{k/b^2}$ becomes the phase $e^{-2\pi ik\Lambda/\lambda}$. The terms with $m=0$ then add to $\chi_s$ a small admixture of the logarithmic solution, which is of order $e^{-2\pi^2/\lambda}$ on the real $\Lambda$ axis.}

The Heine series converges for $\abs{x}<1$. In the classical limit, the ${}_2F_1(a,b;c;x)$ has a branch cut for $x\ge1$ and the eigenfunction cannot be continued past the centre of the disk, $x=1$. At real $q$ the cut is replaced by a sequence of simple poles which merge into the cut as $q\to1$. In terms of $\Lambda$, the positive $x$ axis is the line $\operatorname{Im}\Lambda=-\pi$ where $u<0$, and the poles sit at
\begin{equation}\label{eq:rq_lattice}
    \Lambda_n=\lambda\big(n+\tfrac12\big)-i\pi,\qquad x_n=q^{-2n},\qquad n\ge0 .
\end{equation}
The first one, $x_0=1$, is precisely the centre of the disk.

To see how these poles appear we use the same argument of~\cite{Blommaert:2024whf}, which exploits the two periodicities of \eqref{eq:rq_f}. Along the vertical contour $t=c+iy$, one has that $G$ is periodic in $y$ with period $2\pi/\lambda$, while $\omega$ tends to one as $y\to+\infty$, since every $q$-Pochhammer symbol in \eqref{eq:rq_G} tends to one.
At the points \eqref{eq:rq_lattice} the factor $u^t(-q)^t=e^{\lambda nt}=z^{-n}$ is invariant under $t\to t+2\pi i/\lambda$ and therefore the whole integrand is invariant up to the factor $\omega$. However, far up the contour $\omega$ can be replaced by one up to corrections of order $e^{-2\pi y}$, as we have just observed.

We thus split the upper part of the contour into periods, $t=t_0+2\pi ik/\lambda$ with $k\ge0$, and write $\Lambda=\Lambda_n+\delta$ where $\delta$ parametrizes a small displacement from each pole. Each period then contributes the same integral, multiplied by $e^{2\pi ik\delta/\lambda}$, and the sum over $k$ is geometric:
\begin{equation}\label{eq:poles}
    \chi_s(\Lambda_n+\delta)=(\text{regular at }\delta=0)+\frac{N_sK}{1-e^{2\pi i\delta/\lambda}}\int_{\rm period}\frac{\dd t}{2\pi i}\,e^{\delta t}\,z^{-n}G(z).
\end{equation}
The corrections from $\omega-1$ decay exponentially in $k$, so they sum to the regular part.\footnote{The images going down the contour do not produce poles on this line. There $e^{2\pi it}\to\infty$ and $\omega$ is not asymptotically constant. Under one downward period, $e^{2\pi it}\to\tilde q^{-1}e^{2\pi it}$ and $\omega(t-2\pi i/\lambda)/\omega(t)\to\tilde q\,e^{4\pi iB}$. On the line $\operatorname{Im}\Lambda=-\pi$ the rest of the integrand is unchanged at the lattice points, so the downward sum converges and just contributes to the regular part.} As $\delta \rightarrow 0$ with $\mathrm{Im}(\delta)$ positive, the expression \eqref{eq:poles} thus develops a simple pole as we anticipated. The residue is the remaining integral at $\delta=0$. Over one period $z=e^{-\lambda t}$ winds once around the origin, so by Cauchy's theorem the integral extracts the $n$-th Taylor coefficient of $G$:
\begin{equation}\label{eq:rq_res}
    \operatorname*{Res}_{\Lambda=\Lambda_n}\chi_s=\mathcal N_s\,\frac{Q_n(E;q^{1-2B},q^{1-2B}\,|\,q^2)}{(q^2;q^2)_n}\quad(n\ge0),\qquad \operatorname*{Res}_{\Lambda=\Lambda_n}\chi_s=0\quad(n<0),
\end{equation}
with $\mathcal N_s$ independent of $n$. The residues vanish for $n<0$ because $G$ has no negative powers of $z$. Equivalently the points associated with $n<0$ are regular because they lie inside the circle of convergence of \eqref{eq:rq_heine}. 

\begin{figure}[t]
\centering
\includegraphics[width=0.78\textwidth]{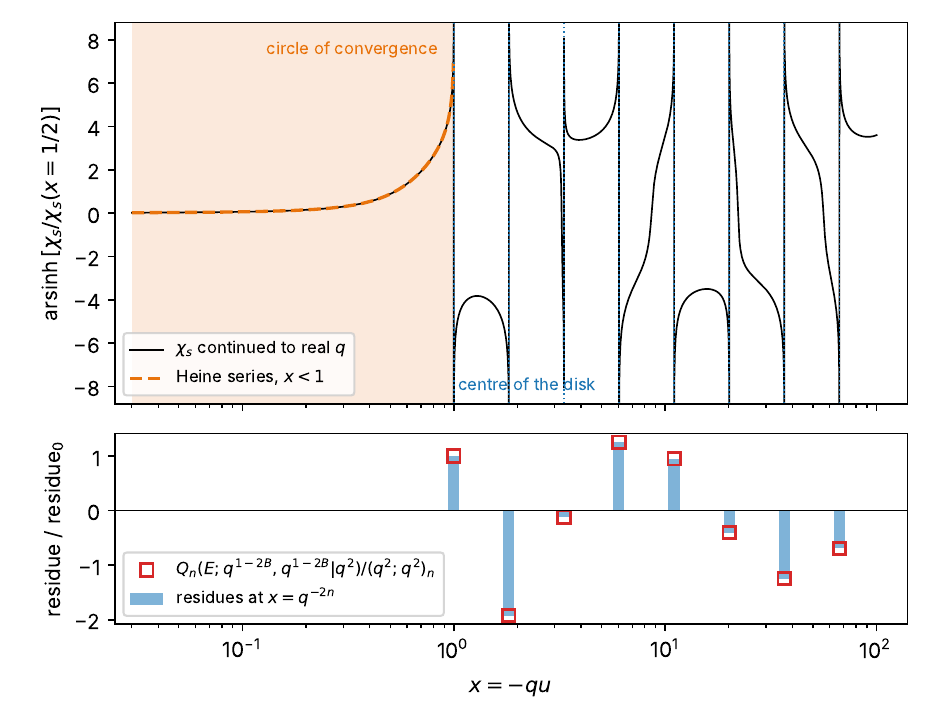}
\caption{Top: the eigenfunction \eqref{eq:sinh_eigenfunction} continued to $b^2=i\lambda/2\pi$, on the line $\operatorname{Im}\Lambda=-\pi$ where $x=-qu>0$ (solid, plotted as $\operatorname{arsinh}$ to accommodate the poles), compared with the Heine series \eqref{eq:rq_heine} inside its circle of convergence (dashed). Beyond the centre of the disk, $x=1$, simple poles appear at $x=q^{-2n}$ (dotted). Bottom: their residues (bars, normalized to $n=0$) compared with the Al-Salam--Chihara polynomials \eqref{eq:rq_res} (squares). Here $\lambda=0.6$, $q^{2B}=e^{-0.18}$ and $\theta=1.1$. \rev{The eigenfunction is obtained by evaluating numerically the Mellin integral \eqref{eq:rq_mellin}, with the integrand in the product form \eqref{eq:rq_f}, at $\mathrm{Im}\,\Lambda=-\pi+10^{-7}$. The Heine series is summed up to $2\cdot10^4$ terms. The residues are extracted as $\delta\,\chi_s(\Lambda_n+\delta)$ at $\delta=10^{-6}i$.}}
\label{fig:rq}
\end{figure}

The same type of structure and the emergence of these poles can also be appreciated directly at the level of the difference equation. On the lattice \eqref{eq:rq_lattice}, i.e.\ $u_n=-q^{-1-2n}$, the difference equation~\eqref{eq:qd_H} becomes the following recurrence:
\begin{equation}\label{eq:rq_rec}
    2E\,\chi_n=\big(1-q^{2n+2}\big)\chi_{n+1}+2q^{1-2B+2n}\chi_n+\big(1-q^{2n-4B}\big)\chi_{n-1},\qquad \chi_n\equiv\chi_s(\Lambda_n).
\end{equation}
The coefficient of $\chi_{n+1}$ vanishes at $n=-1$: this is the mechanism that forces the poles. Indeed at $n=-1$ and $n=-2$ we are inside the circle of convergence and $\chi_{-1}$, $\chi_{-2}$ are finite and generically non-zero. Approaching the lattice, $\Lambda=\Lambda_{-1}+\delta$, the equation at $n=-1$ then requires
\begin{equation}
    \big(1-e^{-\delta}\big)\,\chi_s(\Lambda_0+\delta)\ \longrightarrow\ 2\big(E-q^{-1-2B}\big)\chi_{-1}-\big(1-q^{-2-4B}\big)\chi_{-2},
\end{equation}
i.e.\ a finite limit. Since the coefficient vanishes linearly in $\delta$, $\chi_s$ must have a simple pole at $\Lambda_0$, i.e.\ at the centre $x=1$, so that $0\cdot\infty$ is finite. Its residue is fixed by the values inside the circle of convergence. The equations at $n\ge0$ then propagate the pole outwards: at each step the coefficient of $\chi_{n+1}$ is non-zero, so a pole at $x_n$ forces one at $x_{n+1}$. The residues obey \eqref{eq:rq_rec}, starting at $n=0$, which is precisely the Al-Salam--Chihara recurrence for \eqref{eq:rq_res}. We have evaluated the analytically continued wavefunction numerically and in Figure \ref{fig:rq} we display the eigenfunction on the line $\mathrm{Im}\,\Lambda=-\pi$, inside and outside the circle of convergence, together with the residues. This confirms the analytic results above.

In the disk variable $y=1/x$ the poles sit at $y=q^{2n}$, $n\ge0$. This is the Fock lattice of the quantum disk at real $q$, anchored at the centre $y=1$ and extending towards the boundary. The eigenfunction we started from is defined by regularity at $O$ and around $O$ it is a power series in the relabelled variable $x=-qu$. Continued to real $q$, it nevertheless carries the discrete structure of the disk anchored at the centre. In this sense the relabelling does not discard any structure: the description anchored at $O$ contains the one anchored at the centre. 

Finally, the residues reproduce the chord wavefunctions of DSSYK. Consider the limit in which the boundary dilaton is sent to infinity along the imaginary direction, such that $q^{-2B}\to0$. In this limit the two $q$-Pochhammer symbols in the numerator of $G$, which carry all the dependence on $B$, tend to one. What remains is
\begin{equation}
    G(z)\ \longrightarrow\ \frac{1}{(ze^{i\theta},ze^{-i\theta};q^2)_\infty}=\sum_{n\ge0}\frac{H_n(\cos\theta\,|\,q^2)}{(q^2;q^2)_n}\,z^n,
\end{equation}
the generating function of the continuous $q$-Hermite polynomials, so the residues \eqref{eq:rq_res} become the chord wavefunctions.

\section{Semiclassical evaluation of the two-point function}
\label{app:semiclassical}

In this appendix we derive the localization \eqref{eq:corr_scLE} of the length in the semiclassical limit. 

We use the units and the limit of Section \ref{subsec:corr_semiclassical}, with $\varepsilon=2\pi b^2=\hbar_{\rm eff}$, in which the matter line is a smooth function of $\Lambda$ that does not affect the saddle point. We need the eigenfunctions in the same limit and, in order to do so, we perform a WKB ansatz $\chi\sim e^{\mathcal W/\varepsilon}$ for them. The difference equation \eqref{eq:sinh_difference} involves the shifted eigenfunctions $\chi_s(\Lambda\pm i\varepsilon)=e^{\pm i\varepsilon\partial_\Lambda}\chi(\Lambda)$. On the WKB ansatz $\chi\sim e^{\mathcal W/\varepsilon}$ it acts as a phase,
\begin{equation}
    \chi(\Lambda\pm i\varepsilon)\simeq e^{\pm i\mathcal W'(\Lambda)}\chi(\Lambda).
\end{equation}
The difference equation then becomes an algebraic equation for $\mathcal W'$. It is more convenient to work with the symmetric form of the sinh dilaton operator~\cite{SinhBox}, i.e.\ the one obtained after removing the unitary factor $\Phi_\nu$ in \eqref{eq:sinh_Phinu}, which only rearranges the coefficients. It reads
\begin{equation}\label{eq:sc_HJ}
    \begin{gathered}
    \cosh\varphi_h=\mathcal A(\Lambda)\cos\mathcal W'(\Lambda)-\mathcal B(\Lambda),\\
    \mathcal A=\sqrt{(1+e^{-\Lambda})(1+e^{2\varphi_b-\Lambda})},\qquad \mathcal B=e^{\varphi_b-\Lambda}.
    \end{gathered}
\end{equation}
This is the classical Hamiltonian \eqref{eq:sinh_cosh_hamiltonian}, written in the variables $(\Lambda,p)$ of \eqref{eq:sinh_LambdaPi}, where it reads $\cosh\varphi_h=\mathcal A\cosh p-\mathcal B$, with the Euclidean momentum $p=-i\mathcal W'$. Where $p$ is real $\mathcal W'$ is imaginary and the eigenfunction oscillates: this is the region $L>L_*$ explored by the Lorentzian motion. Where $\mathcal W'$ is real, the eigenfunction grows or decays: this is the region $L<L_*$, between $O$ and the turning point $\mathcal W'=0$, where the Euclidean motion takes place.

We will now find a useful parametrization for the Euclidean region. Solving \eqref{eq:sc_HJ} for $\mathcal W'$ involves the combination $\mathcal A^2-(\cosh\varphi_h+\mathcal B)^2$. Expanding both terms of the latter one finds
\begin{equation}\label{eq:combination}
    \mathcal A^2-(\cosh\varphi_h+\mathcal B)^2=\sinh^2\!\varphi_h\,\big(e^{\Lambda_*-\Lambda}-1\big),
\end{equation}
where we defined the turning point
\begin{equation}\label{eq:sc_turning}
    e^{\Lambda_*}=\frac{2e^{\varphi_b}(\cosh\varphi_b-\cosh\varphi_h)}{\sinh^2\varphi_h} ,
\end{equation}
at which $\mathcal A(\Lambda_{*})=\cosh\varphi_h+\mathcal B(\Lambda_{*})$.
Notice that \eqref{eq:sc_turning} at strong field reduces to \eqref{eq:corr_Lambdastar} of the main text. The combination \eqref{eq:combination} is positive between $O$ and the turning point. It is natural to parametrize this region by an angle $\alpha$,
\begin{equation}\label{eq:sc_alpha}
    e^{\Lambda}=e^{\Lambda_*}\sin^2\alpha ,
\end{equation}
which runs from $\alpha=0$ at $O$ (i.e.\ at $\Lambda \rightarrow -\infty$) to $\alpha=\frac\pi2$ at the turning point. 

We will see later that the saddle point involves the derivative of $\mathcal W$ with respect to the energy at fixed $\Lambda$. Writing \eqref{eq:sc_HJ} as $\cos\mathcal W'=(\cosh\varphi_h+\mathcal B)/\mathcal A$ and differentiating, one finds\footnote{Indeed $-\sin\mathcal W'\,\partial_{\varphi_h}\mathcal W'=\sinh\varphi_h/\mathcal A$ and then one uses $\mathcal A\sin\mathcal W'=\sqrt{\mathcal A^2-(\cosh\varphi_h+\mathcal B)^2}$.}
\begin{equation}\label{eq:saddle1}
    \partial_{\varphi_h}\mathcal W'=-\frac{1}{\sqrt{e^{\Lambda_*-\Lambda}-1}} .
\end{equation}
The eigenfunction is normalized at $O$ in an energy-independent way, $\chi_s\to e^{\Lambda/2}$, so $\partial_{\varphi_h}\mathcal W$ is obtained by integrating from $O$. With $u=e^\Lambda$ and then using \eqref{eq:sc_alpha}, the integral of \eqref{eq:saddle1} can be done in closed form:
\begin{equation}
    \int_{-\infty}^\Lambda\frac{\dd\Lambda'}{\sqrt{e^{\Lambda_*-\Lambda'}-1}}=\int_0^u\frac{\dd u'}{\sqrt{u'}\sqrt{e^{\Lambda_*}-u'}}=\int_0^\alpha2\,\dd\alpha'=2\alpha.
\end{equation}
Therefore we found
\begin{equation}\label{eq:sc_dW}
    \partial_{\varphi_h}\mathcal W\big|_\Lambda=-2\alpha .
\end{equation}
A closed orbit from $O$ goes out to the turning point and comes back, so each $\Lambda$ is visited twice, at $\alpha$ and at $\pi-\alpha$, which give the same value in \eqref{eq:sc_alpha}. Relation \eqref{eq:sc_dW} continues to $\alpha\in[\frac\pi2,\pi]$, and the Euclidean momentum $\partial_\Lambda\mathcal W$ changes sign at the turning point.

We can now evaluate the two-point function. Each propagator in \eqref{eq:corr_qWitten} is an energy integral of $\rho_b\,e^{-\tau E_b}\psi_s$, so
\begin{equation}
    K(\Lambda;\tau)\simeq\int\dd\varphi\;\exp\frac1\varepsilon\Big[2\pi(\varphi-\varphi_b)-\tau\,e(\varphi)+\mathcal W(\Lambda;\varphi)\Big],
\end{equation}
where the three terms are the entropy, the Euclidean evolution and the WKB form of the eigenfunction. With energies $\varphi_1$ and $\varphi_2$ for the propagators of duration $\tau$ and $\beta-\tau$, and $\alpha_{1,2}$ the corresponding angles \eqref{eq:sc_alpha} at the same $\Lambda$, the exponent of the two-point function is
\begin{equation}
    \frac1\varepsilon\Big[2\pi(\varphi_1-\varphi_b)-\tau e(\varphi_1)+\mathcal W(\varphi_1,\alpha_1)+2\pi(\varphi_2-\varphi_b)-(\beta-\tau)e(\varphi_2)+\mathcal W(\varphi_2,\alpha_2)\Big].
\end{equation}
Using \eqref{eq:sc_dW}, stationarity with respect to $\varphi_1$, $\varphi_2$ and $\Lambda$ gives
\begin{equation}
    \begin{gathered}
    2\pi-\tau e'(\varphi_1)-2\alpha_1=0,\qquad 2\pi-(\beta-\tau)e'(\varphi_2)-2\alpha_2=0,\\
    \partial_\Lambda\mathcal W(\varphi_1,\alpha_1)+\partial_\Lambda\mathcal W(\varphi_2,\alpha_2)=0 .
    \end{gathered}
\end{equation}
The last equation requires opposite Euclidean momenta at the second insertion, so that the two arcs join smoothly. Since the momentum changes sign at the turning point, this is achieved by the same energy on both arcs, with one arc reaching $\Lambda$ before the turning point and the other after it, i.e.
\begin{equation}
    \varphi_1=\varphi_2\equiv\varphi,\qquad \alpha_1+\alpha_2=\pi .
\end{equation}
Adding the first two equations then gives $4\pi-\beta e'(\varphi)-2\pi=0$, i.e.\ the thermal condition \eqref{eq:corr_scbeta}: the orbit closes after the time $\beta$. Plugging this condition into the first two equations then separately fixes the angles,
\begin{equation}
    \alpha_2=\frac{\pi\tau}\beta,\qquad \alpha_1=\pi-\frac{\pi\tau}\beta .
\end{equation}
Inserting $\alpha=\pi\tau/\beta$ in \eqref{eq:sc_alpha} and using \eqref{eq:corr_scbeta}, which on the saddle gives $e^{\Lambda_*}=e^{\varphi_b}(\beta/2\pi)^2$, we obtain (recall $e^{\Lambda}=\sinh^2 (L/2)$)
\begin{equation}
\sinh\frac{L_E}2=e^{\varphi_b/2}\frac\beta{2\pi}\sin\frac{\pi\tau}\beta,
\end{equation}
the second form in \eqref{eq:corr_scR}. As shown in Section~\ref{subsec:corr_semiclassical}, this equals \eqref{eq:sinh_ERB} after the continuation to Euclidean time.

Finally, we have checked that the exponent at the saddle, i.e.\ the sum of the entropies, of the Euclidean evolution and of the WKB exponents accumulated along the closed orbit, equals $\log Z_b$ in \eqref{eq:corr_scbeta} and therefore cancels against the normalization.

\subsection{The deformed Larmor relation from the on-shell action of sinh dilaton gravity}
\label{app:larmor}

In this subsection we derive the deformed Larmor relations \eqref{eq:corr_Inl} and \eqref{eq:corr_deformedlarmor} from the on-shell action of sinh dilaton gravity. We use $\varepsilon=2\pi b^2$ and $\varphi=\varepsilon\Phi$ as in the main text. The Euclidean action is
\begin{equation}
    I=-\frac12\int\dd^2x\sqrt g\,\big(\Phi R+W(\Phi)\big)-\oint\dd\tau\sqrt h\,\big(\Phi K-F(\Phi)\big),
\end{equation}
with the counterterm \rev{$F(\Phi_b)=\sqrt{2\cosh\varphi_b-2}/\varepsilon=2\sinh(\varphi_b/2)/\varepsilon$ of \eqref{eq:sinh_action}}, so that the Brown--York energy is $E=F-\partial_n\Phi$, as in \eqref{eq:sinh_quasilocal}.
The static solutions are black holes \eqref{eq:sinh_bh}, which we report here for the reader's convenience. One has $\dd s^2=f(r)\,\dd\tau^2+\dd r^2/f(r)$, with $\Phi=r$ and
\begin{equation}
    f(r)=\frac{2\big(\cosh\varepsilon r-\cosh\varphi_h\big)}{\varepsilon^2} ,
\end{equation}
which obeys $f'=W$ and $R=-f''$. Smoothness at the horizon $r=\Phi_h$ fixes the period of $\tau$ (i.e.\ the Hawking inverse temperature) to $4\pi/f'(\Phi_h)$ and the boundary at $r=\Phi_b$ has proper length $\beta=4\pi\sqrt{f(\Phi_b)}/f'(\Phi_h)$. The normal derivative of the dilaton at the boundary is $\partial_n\Phi=\sqrt{f(\Phi_b)}$. Evaluating the bulk term, the Gibbons--Hawking term and the counterterm on these solutions,\footnote{The bulk integrand is $\Phi R+W=-rf''+f'$, the total derivative of $-rf'+2f$. Integrating from the horizon, where $f$ vanishes, to the boundary gives $I_{\rm bulk}=\frac{2\pi}{f'(\Phi_h)}\Phi_bf'(\Phi_b)-\beta\,\partial_n\Phi-2\pi\Phi_h$. With $K=f'/(2\sqrt f)$ the Gibbons--Hawking term is $-\frac{2\pi}{f'(\Phi_h)}\Phi_bf'(\Phi_b)$, which cancels the first bulk term.} we find
\begin{equation}\label{eq:onshellsinh}
    I=\beta F-\beta\,\partial_n\Phi-2\pi\Phi_h=\beta E-S,\qquad S=2\pi\Phi_h .
\end{equation}
The goal is now to write this action in the variables of the disk $g_{-}$, i.e.\ the Weyl-rescaled geometry. The disk length of the boundary is $\ell_-=e^{\varphi_b/2}\beta$, so the on-shell action can be reorganized as a term proportional to the length $\ell_{-}$ and what would be the analog of a flux term $G$, i.e.
\begin{equation}\label{eq:app_IG}
    I=\mathcal B\,\ell_--G,\qquad G\equiv\beta\,\partial_n\Phi+2\pi\Phi_h ,
\end{equation}
with $\mathcal B=e^{-\varphi_b/2}F\rev{=(1-e^{-\varphi_b})/\varepsilon}$ as in \eqref{eq:corr_Inl}.
Indeed in JT gravity $G$ is the ordinary magnetic flux, $G=\Phi_b(A+2\pi)$, and \eqref{eq:app_IG} is the relativistic particle \eqref{eq:tt_action} on shell.\footnote{With $f=r^2-\Phi_h^2$ one has $\beta=2\pi\sinh R$, $\partial_n\Phi=\Phi_h\sinh R$ and $\Phi_h=\Phi_b/\cosh R$, so that $G=2\pi\Phi_h(\sinh^2R+1)=2\pi\Phi_b\cosh R$, while the area enclosed by the circle is $A=2\pi(\cosh R-1)$.} In sinh dilaton gravity $G$ is not linear in the area, but we can still interpret it as a flux term by computing its derivative with respect to the area $A_{-}$, with the slope defining a notion of effective field $B_{\mathrm{eff}}$. At fixed $\varphi_b$ the solutions form a one-parameter family labelled by $\varphi_h$ and both $G$ and the area $A_-$ enclosed by the orbit on the disk are functions of $\varphi_h$ that are known:\footnote{We used $f'(\Phi_h)=2\sinh\varphi_h/\varepsilon$, $f(\Phi_b)=2(\cosh\varphi_b-\cosh\varphi_h)/\varepsilon^2$ and the radius of the orbit in the effective disk, $\cosh R=\rho_b/\rho_h=(e^{\varphi_b}-\cosh\varphi_h)/\sinh\varphi_h$, which was found in \eqref{eq:corr_scR}.}
\begin{equation}
    G=\frac{2\pi}\varepsilon\Big[\frac{2(\cosh\varphi_b-\cosh\varphi_h)}{\sinh\varphi_h}+\varphi_h\Big],\qquad
    A_-=2\pi(\cosh R-1)=2\pi\Big[\frac{e^{\varphi_b}-\cosh\varphi_h}{\sinh\varphi_h}-1\Big].
\end{equation}
Differentiating both with respect to $\varphi_h$, the ratio simplifies to
\begin{equation}
    B_{\rm eff}\equiv\frac{\dd G}{\dd A_-}=\frac{\dd G}{\dd \varphi_h}\frac{\dd \varphi_h}{\dd A_{-}}=\frac{1-e^{-\varphi_b}\cosh\varphi_h}{\varepsilon}=e^{-\varphi_b}\rho_b ,
\end{equation}
where we used \eqref{eq:mappings}.
Along the family of circles, the area and the length are related by $\dd A_-/\dd\ell_-=\tanh R$. We thus differentiate \eqref{eq:app_IG} to obtain:
\begin{equation}
\frac{\dd I}{\dd \ell_{-}}=\mathcal B-\frac{\dd G}{\dd A_{-}}\frac{\dd A_{-}}{\dd \ell_{-}}=\mathcal B-B_{\rm eff}\tanh R.
\end{equation}
On the other hand, the first law $\dd S=\beta\,\dd E$ implies $\dd I/\dd\beta=E$, so that
\begin{equation}
\frac{\dd I}{\dd \ell_{-}}=e^{-\varphi_b/2}E=\mathcal B-P_- ,\qquad P_-=e^{-\varphi_b/2}\partial_n\Phi=\partial_{n_-}\Phi ,
\end{equation}
where we used \eqref{eq:onshellsinh} and $\partial_{n_{-}}$ is the normal derivative in the disk metric. Comparing the two expressions above gives the deformed Larmor relation
\begin{equation}
P_-=B_{\rm eff}\tanh R \qquad \Longrightarrow \qquad \tanh R=\frac{\partial_{n_-}\rho}{\rho_b} ,
\end{equation}
where in the second step we used $\partial_{n_-}\rho=e^{\varphi_b}\partial_{n_-}\Phi$. The equation above is indeed equation \eqref{eq:deformedlarmor} of the main text and explains once again why the magnetic particle interpretation, on the gravity side, has a natural manifestation in the effective geometry $g_{-}$ of sinh dilaton gravity.

\section{Applying exact WKB to the P\"oschl--Teller problem}
\label{app:wkb}

In this appendix we derive the density \eqref{eq:wkb_main} from the Schr\"odinger equation \eqref{eq:wkb_schr} with the methods of exact WKB, following~\cite{Voros:1983,Delabaere:1997,Sueishi:2020yux}. The derivation uses only local connection rules and the periods of the WKB differential.

\subsection{The Stokes phenomenon and the local connection rules}
\label{app:wkb_stokes}

The WKB solutions \eqref{eq:wkb_wkbsol} are formal series. Writing one of them as an exponential times a power series,
\begin{equation}\label{eq:wkbseries}
    \psi_\pm=e^{\pm\frac1\hbar F(L)}\sum_{n\ge0}c^{(\pm)}_n(L)\,\hbar^{n+\frac12},\qquad F(L)=\int_a^L\sqrt{Q_0}\,\dd L',
\end{equation}
the coefficients typically grow factorially, $c_n\sim n!/A^n$, and the series has zero radius of convergence. Borel summation divides out the factorial and restores it through a Laplace integral,
\begin{equation}\label{eq:wkb_borel}
    B(t)=\sum_n\frac{c_n}{n!}\,t^n,\qquad \mathcal S\Big[\sum_nc_n\hbar^n\Big]=\frac1\hbar\int_0^\infty\dd t\,e^{-t/\hbar}\,B(t),
\end{equation}
where for complex $\hbar$ the integration ray runs along $\arg t=\arg\hbar$. In general the Borel transform has singularities, whose position for WKB series of the form \eqref{eq:wkbseries} is located at $\zeta_*=\pm2F(L)$, i.e.\ at the difference between the two exponents of $\psi_{\pm}$. The Laplace integral of $\psi_\mp$ fails when the singularity lies on the integration ray, i.e.\ when $\mp2F/\hbar$ is real and positive. This defines the Stokes curves \eqref{eq:wkb_stokes_def}. On a Stokes curve the Laplace integral of the dominant solution must avoid the singularity from above or from below and the two lateral sums differ by a multiple of the subdominant solution, as we will see explicitly in \eqref{eq:laterals}. For real $\hbar$ the Stokes curves lie on the integration path and therefore one usually gives $\hbar$ a small phase
\begin{equation}
    \delta=\arg(1/\hbar)=0^\pm,
\end{equation}
which rotates the Stokes curves slightly away from the real axis.\footnote{One should perform the computation and then check that the result does not depend on the sign of $\delta$.}

\begin{figure}[t]
\centering
\includegraphics{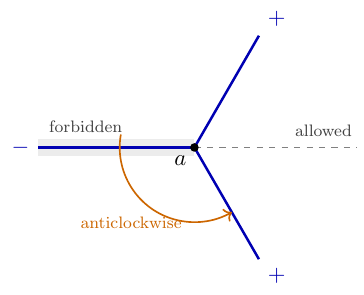}
\caption{The three Stokes curves leaving a simple turning point $a$, with the forbidden region on the left. On the curve inside the barrier $\psi_-^a$ dominates; on the other two $\psi_+^a$ dominates. Going from the forbidden segment to the allowed axis below $a$ crosses the lower curve anticlockwise.}
\label{fig:wkb_local}
\end{figure}

The Stokes phenomenon is local. We therefore expect the connection rules to depend only on local data and the Stokes graph of the full problem to be assembled from the elementary Stokes graphs of the simplest local problems, the Airy problem at each simple turning point \cite{Sueishi:2020yux}. Near a turning point $a$ with the forbidden region on its left, we linearize the potential as $Q_0\simeq g\,(a-L)$ with $g>0$ and the equation becomes Airy's equation $y''=zy$ with $z\propto(a-L)/\hbar^{2/3}$. The definition \eqref{eq:wkb_stokes_def} of the Stokes curves becomes $\mathrm{Im}\,(a-L)^{3/2}=0$ for real $\hbar$. Writing $L-a=re^{i\varphi}$ the Stokes curves are thus determined by the three directions
\begin{equation}
    \varphi=\pi,\qquad\varphi=\pm\frac\pi3 ,
\end{equation}
as shown in Figure \ref{fig:wkb_local}. With $\zeta=\frac23z^{3/2}$, the two WKB solutions of Airy's equation are
\begin{equation}\label{eq:asy}
    \psi_\pm^{\rm Airy}=\frac{e^{\mp\zeta}}{z^{1/4}}\sum_{k\ge0}(\mp1)^k\frac{u_k}{\zeta^k},\qquad u_k=\frac{\Gamma(3k+\frac12)}{54^k\,k!\,\Gamma(k+\frac12)}\sim\frac{(k-1)!}{2^k},
\end{equation}
where $1/\zeta$ plays the role of $\hbar$.\footnote{On the curve inside the barrier the growing solution $\psi_-$ dominates, while on the other two $\psi_+$ dominates.} Starting from the asymptotic expansions \eqref{eq:asy}, the Borel transforms \eqref{eq:wkb_borel} can be computed in closed form and yield
\begin{equation}\label{eq:wkb_airy_borel}
    B_\pm(t)=\sum_{k\ge0}\frac{(\mp1)^ku_k}{k!}\,t^k={}_2F_1\Big(\frac16,\frac56;1;\mp\frac t2\Big),\qquad \psi_\pm^{\rm Airy}=\frac{e^{\mp\zeta}}{z^{1/4}}\,\zeta\!\int_0^\infty\!\dd t\,e^{-\zeta t}B_\pm(t).
\end{equation}
The decaying series has its singularity at $t=-2$, away from the integration ray, so it is Borel summable and indeed the Borel sum of $B_{+}$ on the right-hand side of \eqref{eq:wkb_airy_borel} yields directly $2\sqrt\pi\,\mathrm{Ai}(z)$. On the other hand, the growing series $B_-$ has its singularity at $t=2$, the logarithmic branch point of the hypergeometric function, which lies on the integration ray. One can show that its discontinuity across the cut is $i\,B_+(t-2)$, i.e.\ the Borel transform of the decaying series, shifted to the singularity.\footnote{For $c=a+b$, the function ${}_2F_1(a,b;c;x)$ has a logarithmic branch point at $x=1$, with coefficient $-\frac{\Gamma(a+b)}{\Gamma(a)\Gamma(b)}\,{}_2F_1(a,b;1;1-x)\log(1-x)$. Hence, for $t>2$,
\begin{equation*}
    B_-(t+i0)-B_-(t-i0)=\frac{2\pi i}{\Gamma(\frac16)\Gamma(\frac56)}\,{}_2F_1\Big(\frac16,\frac56;1;1-\frac t2\Big)=i\,B_+(t-2),
\end{equation*}
where we used $\Gamma(\frac16)\Gamma(\frac56)=\pi/\sin\frac\pi6=2\pi$.} More precisely, for $\arg\zeta=0^\pm$ the ray of the Laplace transform in \eqref{eq:wkb_airy_borel} passes below or above the singularity and the difference between the two lateral sums is given by:
\begin{equation}\label{eq:laterals}
    \mathcal S_{0^-}[\psi_-]-\mathcal S_{0^+}[\psi_-]=\frac{e^{\zeta}}{z^{1/4}}\,\zeta\!\int_2^\infty\!\dd t\,e^{-\zeta t}\,i\,B_+(t-2)=i\,\frac{e^{-\zeta}}{z^{1/4}}\,\zeta\!\int_0^\infty\!\dd t\,e^{-\zeta t}B_+(t)=i\;\mathcal S[\psi_+].
\end{equation}
This is the resurgent structure of the Airy series, and the Stokes constant $i$ can be read off from \eqref{eq:laterals}. Comparing with the Airy functions, one can then identify
\begin{equation}\label{eq:wkb_airy_lateral}
    \mathcal S_{0^\pm}[\psi_-^a]=\sqrt\pi\,\big(\mathrm{Bi}\mp i\,\mathrm{Ai}\big),\qquad \mathcal S[\psi_+^a]=2\sqrt\pi\,\mathrm{Ai},
\end{equation}
so that $\mathrm{Bi}$ is the median of the two lateral sums. The computation above proves the Voros rule \eqref{eq:wkb_voros} for an anticlockwise crossing of the Stokes curve inside the barrier; the rule on the other two curves follows from the symmetry $z\to ze^{\pm2\pi i/3}$ of Airy's equation.

In our specific problem \eqref{eq:wkb_schr}, the origin $O$ requires its own rule. It is not a turning point since $Q_0(0)>0$, but it is a regular singular point of the full equation because of the ordering term. Near $L=0$ the equation \eqref{eq:wkb_schr} reduces to a Bessel equation
\begin{equation}
    -\psi''-\frac{\psi}{4L^2}+\left(\nu^2-s^2\right)\psi=0,
\end{equation}
and the regular solution \eqref{eq:wkb_bc} is given by $\psi_{\rm reg}=\sqrt L\,I_0(\sqrt{\nu^2-s^2}L)$. For large argument $I_0$ is dominated by the growing exponential but it also contains an exponentially small decaying piece whose coefficient depends on the side of the real axis~\cite{DLMF},
\begin{equation}
    I_0(z)\simeq\frac{e^{z}\pm i\,e^{-z}}{\sqrt{2\pi z}}\qquad(\arg z=0^\pm).
\end{equation}
Since the WKB solutions \eqref{eq:wkb_wkbsol} normalized at the origin, with $Q_0=\Phi_b^2-\Phi_h^2$ as $L \rightarrow 0$ from \eqref{eq:wkb_schr}, behave as $\psi^O_\pm\simeq \frac{e^{\pm \left(\nu^2-s^2\right)^{1/2}L}}{\left(\nu^2-s^2\right)^{1/4}}$, this gives the Bessel connection rule at the origin
\begin{equation}\label{eq:wkb_bessel}
    \psi_{\rm reg}=\frac{1}{\sqrt{2\pi}}\Big(\psi_+^O\pm i\,\psi_-^O\Big)\qquad(\delta=0^\pm).
\end{equation}
The Airy and Bessel rules are the two building blocks of all the monodromy coefficients used in the connection problem below.

\subsection{Exact quantum periods}
\label{app:wkb_periods}

Two quantum periods enter the connection problem. The bounce period is twice the integral of $S_{\rm odd}$ from the origin to the turning point. At leading order,
\begin{equation}
    V_-=\frac2\hbar\int_0^{L_*}\sqrt{Q_0}\,\dd L .
\end{equation}
We evaluate it by the contour argument shown in Figure \ref{fig:wkb_uplane}. In the variable $u=\sinh^2\frac L2$ the integrand is
\begin{equation}
    \sqrt{Q_0}\,\dd L=\frac{\sqrt{\Phi_b^2-\Phi_h^2(1+u)}}{(1+u)\sqrt u}\,\dd u ,
\end{equation}
which is single-valued outside the forbidden segment $[0,u_*]$, with $1+u_*=\Phi_b^2/\Phi_h^2$. Twice the integral from $0$ to $u_*$ is therefore the integral along a closed contour around the segment. This contour can be deformed outward into a large circle, minus a small circle around the only other singularity, the pole at $u=-1$. The residue at $u=-1$ is $\pm i\Phi_b$, and the residue at infinity is $\pm i\Phi_h$, so that
\begin{equation}
    \hbar V_-=2\pi(\Phi_b-\Phi_h),\qquad V_-=2\pi(\nu-s).
\end{equation}
The two residues are the classical exponents of the equation at $u=-1$ and at the boundary, and they are the only data needed for the exactness argument below.

The winding period follows from the periodicity of the equation under $L\to L+2\pi i$. The integral between $T_{-1}$ and $T_0$ can be deformed into two horizontal pieces, which cancel by periodicity, and a vertical piece of length $2\pi$ at large real part, where $Q_0\to-\Phi_h^2$. Therefore, with the branch $\sqrt{Q_0}\to+i\Phi_h$, one has that $S_{\rm odd}\to is$. Hence
\begin{equation}\label{eq:wkb_P}
    \int_{T_{-1}}^{T_0}S_{\rm odd}\,\dd L=is\cdot2\pi i=-2\pi s,\qquad V_\infty=4\pi s,
\end{equation}
and more generally $\int_{T_{-k}}^{T_0}S_{\rm odd}=-\frac k2V_\infty$, using the same type of contour deformation.

So far we have used only the leading term of $S_{\rm odd}$. The full quantum period is a series in $\hbar$,
\begin{equation}\label{eq:wkb_period_series}
\begin{gathered}
    V_-=\frac{2}{\hbar}\int_0^{L_*}\!\sqrt{Q_0}\,\dd L+2\hbar\int_0^{L_*}\!S_1\,\dd L+2\hbar^3\int_0^{L_*}\!S_3\,\dd L+\dots,\\
    S_0=-\frac{Q_0'}{4Q_0},\qquad S_1=\frac{Q_2-S_0^2-S_0'}{2\sqrt{Q_0}},
\end{gathered}
\end{equation}
where $S_n$ can be found recursively from the WKB Riccati equation \eqref{eq:riccati}.
In general \eqref{eq:wkb_period_series} is itself divergent, and its resummation produces additional non-perturbative effects. We now show that for our ordering all the corrections vanish, so that the series has a single term and needs no resummation.

The contour argument used above applies to the full $S_{\rm odd}$ and not only to its leading term. Each coefficient of $S_{\rm odd}$ is single-valued outside the forbidden segment, so the period is the integral along a closed contour around the segment, which can be deformed onto the only other singular points of the equation, $u=-1$ ($L=i\pi$) and $u=\infty$ (see Figure \ref{fig:wkb_uplane}). The period is therefore fixed by the residues of $S_{\rm odd}\,\dd L$ at these two points and it is exact if and only if these residues are given exactly by their classical values.

The residues of $S_{\rm odd}$ are determined by the local exponents of the equation. At a regular singular point $x_0$ of $\psi''=V\psi$, the potential has a double pole, $V\simeq\mathcal C/(x-x_0)^2$. The solutions behave as $(x-x_0)^{\varrho}$, with exponents fixed exactly by the indicial equation,
\begin{equation}
    \varrho(\varrho-1)=\mathcal C,\qquad\varrho_\pm=\frac12\pm\sqrt{\frac14+\mathcal C}.
\end{equation}
If $S_{\rm odd}$ has the residue $r$ at $x_0$, the two WKB solutions \eqref{eq:wkb_wkbsol} behave as $(x-x_0)^{\frac12\pm r}$,\footnote{If $S_{\rm odd}\simeq r/(x-x_0)$, then $\int^xS_{\rm odd}\simeq r\log(x-x_0)$, so that $\exp(\pm\int^xS_{\rm odd})\simeq(x-x_0)^{\pm r}$, while the prefactor in \eqref{eq:wkb_wkbsol} gives $S_{\rm odd}^{-1/2}\propto(x-x_0)^{1/2}$.} and since they solve the equation, they must reproduce the exact exponents:
\begin{equation}\label{eq:wkb_residue_rule}
    r=\sqrt{\tfrac14+\mathcal C} .
\end{equation}
Let us assume that the potential has a classical part and a quantum correction, $V=Q_0/\hbar^2+Q_2$, so that $\mathcal C=\mathcal C_0/\hbar^2+\mathcal C_2$. The leading term $\sqrt{Q_0}/\hbar$ of $S_{\rm odd}$ has the classical residue $r_{\rm cl}=\sqrt{\mathcal C_0}/\hbar$, while the exact residue is
\begin{equation}
    r=\sqrt{\frac{\mathcal C_0}{\hbar^2}+\mathcal C_2+\frac14}=r_{\rm cl}\Big[1+\Big(\mathcal C_2+\frac14\Big)\frac{\hbar^2}{2\mathcal C_0}+O(\hbar^4)\Big].
\end{equation}
The higher terms $S_1,S_3,\dots$ of $S_{\rm odd}$ therefore have no residue at $x_0$ if and only if $\mathcal C_2=-\frac14$, i.e.\ if the quantum correction to the potential exactly compensates the term $\frac14$ in the indicial equation.

This is exactly what happens in our equation. Near $L=i\pi+\epsilon$ we have $\mathrm{sech}^2\frac L2\simeq-4/\epsilon^2$ and $\sinh L\simeq-\epsilon$, so that from \eqref{eq:wkb_schr} one has:
\begin{equation}
    \frac{Q_0}{\hbar^2}+Q_2\simeq-\frac{4\nu^2}{\epsilon^2}-\frac{1}{4\epsilon^2},\qquad\mathcal C_0/\hbar^2=-4\nu^2,\quad\mathcal C_2=-\frac14 .
\end{equation}
The exponents at $u=-1$ thus receive no quantum correction and the residue of $S_{\rm odd}$ there is exactly the classical one, $r=2i\nu$. At infinity the potential tends to a constant and the solutions are exactly $e^{\pm isL}$, so that $S_{\rm odd}\to is$ exactly; this is also why \eqref{eq:wkb_P} is exact. Both quantum periods are therefore given exactly by their classical values.

For comparison, with the flat ordering $Q_2=0$ one has $\mathcal C_2=0$ at $L=i\pi$, so that the residue $r=\sqrt{\frac14-4\nu^2}$ receives corrections at every order in $\hbar$ and so does the period. As a consistency check, by computing the first correction $\hbar\oint S_1\,\dd L$ along a closed contour around $[-L_*,L_*]$ one can verify that it vanishes with the ordering of \eqref{eq:wkb_schr} and it does not vanish with the flat ordering, where it agrees with the expansion of $r$. The exactness of the periods is the reason why the density has no perturbative corrections in $\hbar$.

\subsection{The connection problem}
\label{app:wkb_connection}

We now transport the regular solution from the origin to infinity, working at $\delta=0^+$. The first task is to determine which Stokes curves the path crosses. For $\delta=0$ the Stokes curves from $O$ and $T_0$ lie on the forbidden segment itself. For small $\delta$ they move off the real axis and to find the direction we look for the points $L$ of each curve near the segment, with $\mathrm{Im}\,L$ small.\footnote{The integration path stays on the real axis.} 

From the definition \eqref{eq:wkb_stokes_def}, the Stokes curve from the origin satisfies $\mathrm{Im}[e^{i\delta}F_O(L)]=0$, with $F_O(L)=\int_0^L\sqrt{Q_0}$. On the segment $\sqrt{Q_0}$ is real and positive, so that $F_O(L)\simeq F_O(\mathrm{Re}\,L)+i\sqrt{Q_0(\mathrm{Re}\,L)}\;\mathrm{Im}\,L$ and to first order
\begin{equation}
\delta\,F_O(\mathrm{Re}\,L)+\sqrt{Q_0(\mathrm{Re}\,L)}\;\mathrm{Im}\,L=0\quad\Longrightarrow\quad \mathrm{Im}\,L=-\delta\,\frac{F_O(\mathrm{Re}\,L)}{\sqrt{Q_0(\mathrm{Re}\,L)}}<0 ,
\end{equation}
i.e.\ the curve from the origin bends below the real axis. One can perform the same computation for the curve from the turning point, with $F_{T_0}<0$ on the segment, showing that it bends above. For $\delta=0^+$ the forbidden segment therefore lies between the two split curves and is crossed by neither, so that both the Bessel rule at the origin and the Airy rule at the turning point can be applied on it. 

\begin{figure}[t]
\centering
\includegraphics{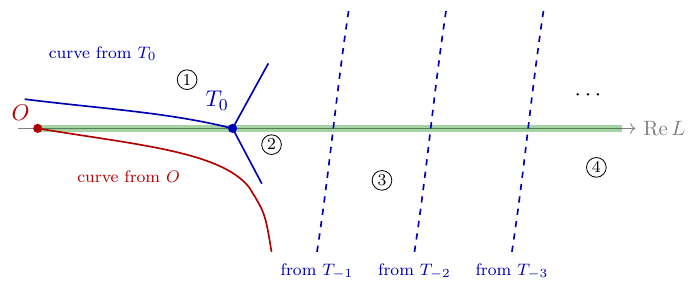}
\caption{Schematic of the crossings met by the path (green) for $\delta=0^+$. (1) On the forbidden segment nothing is crossed, and the Bessel rule holds. (2) Passing below $T_0$, the path crosses the lower Stokes curve of $T_0$ anticlockwise. (3) On the allowed axis it crosses the curves rising from $T_{-1},T_{-2},\dots$ clockwise, at phases $\Xi_k=2\pi ks\tan\delta$. (4) Beyond them there are no more crossings.}
\label{fig:wkb_path}
\end{figure}

On the allowed axis, we denote by $\Xi(L)$ the classical phase $\int_{L_*}^L\sqrt{Q_0}=i\hbar\,\Xi(L)$. The Stokes curve from the image $T_{-k}$ is defined by the exponent measured from $T_{-k}$ itself. To find where it meets the real axis, we evaluate this exponent at a point $L$ of the allowed axis, transporting it from $T_{-k}$ first to the turning point $T_0$, which by \eqref{eq:wkb_P} gives the winding period, and then along the real axis. This process gives the classical phase:
\begin{equation}
    \frac1\hbar F_{T_{-k}}(L)=\frac1\hbar\int_{T_{-k}}^{T_0}\sqrt{Q_0}+\frac1\hbar\int_{T_0}^{L}\sqrt{Q_0}=-2\pi ks+i\,\Xi(L).
\end{equation}
Therefore the Stokes condition $\mathrm{Im}[e^{i\delta}(-2\pi ks+i\Xi)]=0$ yields
\begin{equation}
    \Xi_k=2\pi ks\tan\delta>0,\qquad \mathrm{Re}\big[e^{i\delta}(-2\pi ks+i\Xi_k)\big]=-\frac{2\pi ks}{\cos\delta}<0 .
\end{equation}
The Stokes curve from $T_{-k}$ therefore crosses the allowed axis at the point where the classical phase equals $\Xi_k$ and there the solution $\psi_-^{T_{-k}}$ dominates since the real part is negative. Since $T_{-k}$ lies below the axis and its curve comes up, moving to the right across it is a clockwise crossing. With a similar argument one can show the images above the axis give $\Xi<0$ and therefore do not cross the allowed axis. The path thus meets the sequence of crossings shown in Figure \ref{fig:wkb_path}.

We now follow the regular solution through the four steps depicted in Figure \ref{fig:wkb_path}, keeping all constants. On the forbidden segment, the Bessel rule \eqref{eq:wkb_bessel} gives $\sqrt{2\pi}\,\psi_{\rm reg}=\psi_+^O+i\,\psi_-^O$. Re-expressing it in terms of solutions normalized at the turning point with \eqref{eq:wkb_renorm}, i.e.\ $\psi_\pm^O=e^{\pm V_-/2}\psi_\pm^{T_0}$, we obtain
\begin{equation}\label{eq:wkb_step1}
    \sqrt{2\pi}\,\psi_{\rm reg}=e^{V_-/2}\,\psi_+^{T_0}+i\,e^{-V_-/2}\,\psi_-^{T_0}.
\end{equation}
Then to reach the allowed axis we pass just below $T_0$ and cross its lower Stokes curve, on which $\psi^{T_0}_+$ dominates, anticlockwise. By the local Voros rule \eqref{eq:wkb_voros} associated to the Airy turning point, one has
\begin{equation}
    \psi_+^{T_0}\to\psi_+^{T_0}+i\,\psi_-^{T_0},\qquad\psi_-^{T_0}\to\psi_-^{T_0},
\end{equation}
and \eqref{eq:wkb_step1} becomes
\begin{equation}\label{eq:wkb_step2}
    \sqrt{2\pi}\,\psi_{\rm reg}=e^{V_-/2}\,\psi_+^{T_0}+i\,e^{V_-/2}\big(1+e^{-V_-}\big)\,\psi_-^{T_0}.
\end{equation}
Moving further to the right, we cross the curves from $T_{-1},T_{-2},\dots$, on which the incoming wave dominates. At the $k$-th crossing the clockwise Voros rule, in the normalization of $T_{-k}$, is $\psi_-^{T_{-k}}\to\psi_-^{T_{-k}}-i\,\psi_+^{T_{-k}}$. Transporting the solution as in \eqref{eq:wkb_renorm} with the winding period \eqref{eq:wkb_P}, one has that $\psi_\pm^{T_{-k}}=e^{\mp2\pi ks}\psi_\pm^{T_0}$ so that in the normalization of $T_0$ the local Voros rule for the $k$-th crossing is:
\begin{equation}
    \psi_-^{T_0}\to\psi_-^{T_0}-i\,e^{-kV_\infty}\,\psi_+^{T_0},\qquad \psi_+^{T_0}\to\psi_+^{T_0} .
\end{equation}
Each crossing feeds a fraction of the incoming wave into the outgoing one, with the Voros multiplier of $k$ windings. After all the crossings the incoming wave of \eqref{eq:wkb_step2} has produced the outgoing wave $-i\sum_{k\ge1}e^{-kV_\infty}\psi_+^{T_0}=-i\,\frac{e^{-V_\infty}}{1-e^{-V_\infty}}\,\psi_+^{T_0}$, and the overall coefficient of $\psi_+^{T_0}$ thus becomes
\begin{equation}
    e^{V_-/2}\Big[1+\big(1+e^{-V_-}\big)\frac{e^{-V_\infty}}{1-e^{-V_\infty}}\Big]=e^{V_-/2}\,\frac{1+e^{-V_+}}{1-e^{-V_\infty}},
\end{equation}
where we used $V_-+V_\infty=V_+$. \rev{The coefficient of $\psi_-^{T_0}$ instead remains the one in \eqref{eq:wkb_step2}.} This proves formula \eqref{eq:wkb_result_lateral} of the main text.

It remains to read off the amplitude at infinity. There the potential tends to a constant, so $\psi_+^{T_0}$ and $i\psi_-^{T_0}$ are proportional to the exact asymptotic waves $s^{-1/2}e^{\pm isL}$. These have the same Wronskian, $-2i$, and since the Wronskian of two solutions does not depend on $L$, the product of the two proportionality constants is one. Moreover, $\psi_{\rm reg}$ is real so its components on the two complex conjugate waves have the same modulus. Writing $\psi_{\rm reg}\simeq A_s\sin(sL+\delta_s)$ at infinity, we thus find\footnote{Call $a$ and $b$ the coefficients of $\psi_+^{T_0}$ and $i\psi_-^{T_0}$ in \eqref{eq:wkb_result_lateral}, and write $\psi_+^{T_0}=N_+s^{-1/2}e^{isL}$ and $i\psi_-^{T_0}=N_-s^{-1/2}e^{-isL}$ at infinity, so that $\sqrt{2\pi}\,\psi_{\rm reg}=aN_+s^{-1/2}e^{isL}+bN_-s^{-1/2}e^{-isL}$. The Wronskian gives $N_+N_-=1$, and reality gives $\abs{aN_+}=\abs{bN_-}$, so that $\abs{aN_+}^2=\abs{aN_+}\abs{bN_-}=\abs{a}\abs{b}$, and $\sqrt{2\pi}\,\psi_{\rm reg}=2\sqrt{\abs{a}\abs{b}}\,s^{-1/2}\cos(sL+\text{const})$.}
\begin{equation}
    \frac{\pi s}2\,A_s^2=e^{V_-}\,\frac{\big(1+e^{-V_-}\big)\big(1+e^{-V_+}\big)}{1-e^{-V_\infty}} ,
\end{equation}
and \eqref{eq:wkb_rho_def} gives
\begin{equation}
    \rho(s)=2s\,e^{-V_-}\,\frac{1-e^{-V_\infty}}{\big(1+e^{-V_-}\big)\big(1+e^{-V_+}\big)},
\end{equation}
which is \eqref{eq:wkb_main}. Multiplying numerator and denominator by $e^{2\pi\nu}$ one checks directly that this equals $2s\sinh(2\pi s)/(\cosh2\pi\nu+\cosh2\pi s)$.\footnote{For $\delta=0^-$ everything is mirrored and the same conclusion holds.}
\section{Negative modes of wound boundaries at finite cutoff}
\label{app:negmodes}

In this appendix we show that the alternating sign $(-1)^{n-1}$ of the sectors \eqref{eq:np_sectors} is produced by the negative modes of the fluctuations of the boundary curve around the $n$-wound saddle, in the boundary path integral of JT gravity at finite cutoff of \cite{Griguolo:2025kpi}. We write $k$ for the winding number, to keep $n$ for the Fourier modes.

The boundary curve is described by $z(u)$, with $u\in[0,\hat\beta)$, and $\varepsilon$ is the cutoff. A word on conventions is in order. In~\cite{Griguolo:2025kpi} the parameter that we denote here by $\hat\beta$ is the length of the boundary measured in units of the cutoff, $\hat\beta=\varepsilon L$, with $L$ the proper length of the boundary, and $\phi_r=\varepsilon\Phi_b$ is the renormalized boundary dilaton. The inverse temperature of this paper is instead the proper length itself, i.e.\ the inverse Tolman temperature, $\beta=L=\hat\beta/\varepsilon$, in the units of Section \ref{sec:nonpert} in which $\Phi_b=\nu$. We work with $\hat\beta$, $\phi_r$ and $\varepsilon$ in the intermediate steps, and translate to $\beta$ and $\nu$ at the end. The extrinsic curvature of the finite-cutoff curve obeys the Riccati equation~\cite{Griguolo:2025kpi}
\begin{equation}\label{eq:nm_riccati}
    2\varepsilon\,\kappa'-i\kappa^2+2i\varepsilon\Big(\varepsilon\{z(u),u\}+\frac1{2\varepsilon}\Big)=0 .
\end{equation}
After integrating out the dilaton, one is left with a boundary action proportional to $\int_0^{\hat\beta}\dd u\,(\kappa-1)$ with coefficient $\phi_r/\varepsilon^2$, with $\kappa(u)$ given by the solution of \eqref{eq:nm_riccati}. Configurations with constant Schwarzian derivative solve the equations of motion obtained by varying $\kappa$, as can be checked using \eqref{eq:nm_riccati}. The $k$-wound saddle is then given by:
\begin{equation}\label{eq:nm_saddle}
    z(u)=\tan\frac{\pi k u}{\hat\beta},\qquad \{z(u),u\}=\frac{2\pi^2k^2}{\hat\beta^2},\qquad \kappa_0^2=1+\Big(\frac{2\pi k\varepsilon}{\hat\beta}\Big)^2,
\end{equation}
a circle of constant extrinsic curvature traversed $k$ times. It is the saddle associated to a defect of angle $2\pi k$ in the language of~\cite{Mertens:2019tcm}, with a conical excess $2\pi(k-1)$ at the horizon.

On the perturbative branch $\kappa_0>0$, the on-shell action associated to \eqref{eq:nm_saddle} is $-I_0=\frac{\phi_r\hat\beta}{\varepsilon^2}(\kappa_0-1)$. In terms of $\Phi_b=\phi_r/\varepsilon$ and $L=\hat\beta/\varepsilon$,
\begin{equation}
    -I_0=\Phi_b\Big(\sqrt{L^2+4\pi^2k^2}-L\Big).
\end{equation}
\rev{Setting $L=\beta$ and $\Phi_b=\nu$, this is the gravitational on-shell action of the $k$-wound boundary, $\nu\big(\sqrt{4\pi^2k^2+\beta^2}-\beta\big)$. It already contains the Gauss--Bonnet term $2\pi k\nu$ and grows with $k$. The on-shell action of the $k$-th sector of Section~\ref{subsec:np_gravity}, obtained there from the saddle point of the density \cite{Griguolo:2026eoi}, is $-I_k=\nu\big(\sqrt{4\pi^2k^2+\beta^2}-\beta\big)-2\pi k\nu$: the two differ by the turning-number factor $e^{-2\pi k\nu}$, by which the canonical quantization weights the wound sectors relative to the gravitational boundary term, as explained in Section~\ref{subsec:np_gravity}.}

We now consider the one-loop determinant associated to quadratic fluctuations.
We expand $z(u)=\tan\big(\pi k\eta(u)/\hat\beta\big)$ around the classical solution as $\eta(u)=u+\phi_r^{-1/2}\delta\eta(u)$. We then expand in Fourier modes as $\delta\eta(u)=\sum_n e^{-2\pi inu/\hat\beta}\delta\eta_n$ and solve \eqref{eq:nm_riccati} order by order, $\kappa=\kappa_0+\phi_r^{-1/2}\kappa_1+\phi_r^{-1}\kappa_2+\dots$. The only change with respect to the disk is in the Schwarzian derivative
\begin{equation}
    \{z,u\}=\frac{2\pi^2k^2}{\hat\beta^2}+\phi_r^{-1/2}\Big(\frac{4\pi^2k^2}{\hat\beta^2}\,\delta\eta'+\delta\eta'''\Big)+\phi_r^{-1}\Big(\frac{2\pi^2k^2}{\hat\beta^2}\,\delta\eta'^2-\frac32\,\delta\eta''^2-\delta\eta'''\delta\eta'\Big)+\dots .
\end{equation}
At first order one finds\footnote{At first order \eqref{eq:nm_riccati} gives $2i\varepsilon\kappa_1'+2\kappa_0\kappa_1=2\varepsilon^2\big(\frac{4\pi^2k^2}{\hat\beta^2}\delta\eta'+\delta\eta'''\big)$, and at second order $2i\varepsilon\kappa_2'+2\kappa_0\kappa_2+\kappa_1^2=2\varepsilon^2\big(\frac{2\pi^2k^2}{\hat\beta^2}\delta\eta'^2-\frac32\delta\eta''^2-\delta\eta'''\delta\eta'\big)$. The denominator in \eqref{eq:nm_kappa2} arises from $\kappa_{1,n}\kappa_{1,-n}$, through $\kappa_0^2-(2\pi n\varepsilon/\hat\beta)^2$.}
\begin{equation}
    \kappa_{1,n}=i\varepsilon^2\Big(\frac{2\pi}{\hat\beta}\Big)^3\frac{n\,(n^2-k^2)}{\kappa_0+2\pi n\varepsilon/\hat\beta}\,\delta\eta_n ,
\end{equation}
and the zero mode of the second-order equation gives
\begin{equation}\label{eq:nm_kappa2}
    \kappa_{2,0}=-\varepsilon^2\Big(\frac{2\pi}{\hat\beta}\Big)^4\sum_n\frac{n^2\,(n^2-k^2)}{2\kappa_0\Big[1-\big(\frac{2\pi\varepsilon}{\hat\beta}\big)^2(n^2-k^2)\Big]}\,\abs{\delta\eta_n}^2 .
\end{equation}
For $k=1$ these are eqs.~(4.28)--(4.29) of~\cite{Griguolo:2025kpi}. The quadratic action is therefore
\begin{equation}\label{eq:nm_I2}
    I_2=\frac{2\pi}{\kappa_0}\Big(\frac{2\pi}{\hat\beta}\Big)^3\sum_{n\ge1}\frac{n^2\,(n^2-k^2)}{1-\big(\frac{2\pi\varepsilon}{\hat\beta}\big)^2(n^2-k^2)}\,\abs{\delta\eta_n}^2 ,
\end{equation}
which reduces to eq.~(4.30) of~\cite{Griguolo:2025kpi} for $k=1$ and to the Schwarzian result $\propto n^2(n^2-k^2)$ for $\varepsilon\to0$.

The modes $n=0,\pm k$ are zero modes. They are generated by the $PSL(2,\mathbb R)$ transformations of $z$, which leaves $\{z,u\}$ and $\kappa$ invariant, and they are removed as gauge redundancies as for the disk. Instead the modes with $1\le n\le k-1$ have $n^2-k^2<0$, so that both the numerator and the denominator in \eqref{eq:nm_I2} have definite sign,
\begin{equation}
    n^2(n^2-k^2)<0,\qquad 1+\Big(\frac{2\pi\varepsilon}{\hat\beta}\Big)^2\big(k^2-n^2\big)>1 .
\end{equation}
These $k-1$ complex modes, i.e.\ $2(k-1)$ real modes, therefore have the wrong sign for every value of the cutoff. The large modes for which the denominator becomes negative are instead common to all sectors and are treated by analytic continuation in $\varepsilon$ as in~\cite{Griguolo:2025kpi}. We have not recomputed the measure factors, i.e.\ the Pfaffian of the symplectic form and the determinant arising from the constraint \cite{Griguolo:2025kpi}, around the $k$-wound saddle. They affect the one-loop prefactor of each sector but not its phase.

The Gaussian integral over each wrong-sign complex mode is defined by rotating the contours of both $\mathrm{Re}\,\delta\eta_n$ and $\mathrm{Im}\,\delta\eta_n$, which produces a factor $-1$. Therefore the $k$-th sector acquires, relative to the disk,
\begin{equation}
    (-1)^{k-1},
\end{equation}
which is the sign of the sectors in \eqref{eq:np_sectors}. Each additional winding adds one complex negative mode, i.e.\ two real ones, in agreement with the two reflection phases $i\cdot i=-1$ of each bounce in the exact WKB analysis.


\bibliographystyle{ourbst}
\bibliography{references}

\end{document}